\documentclass[aps,prd,10pt,nofootinbib,onecolumn,eqsecnum,superscriptaddress,preprintnumbers,altaffilletter]{revtex4}
\usepackage[T1]{fontenc}
\usepackage[utf8]{inputenc}
\RequirePackage{graphicx}
\usepackage{amsmath}
\usepackage{amssymb}
\usepackage{booktabs}
\usepackage{soul}

\usepackage{adjustbox}

\usepackage{hyperref}
\hypersetup{
    colorlinks=true,
    linkcolor=blue,
    filecolor=magenta,
    urlcolor=cyan,
    citecolor=cyan
}

\newcommand{\simgt}{\lower.5ex\hbox{$\; \buildrel > \over \sim \;$}}
\newcommand{\simlt}{\lower.5ex\hbox{$\; \buildrel < \over \sim \;$}}

\begin{document}

\title{Addressing Cosmological Tensions by Non-Local Gravity}

\author{Filippo Bouchè}
\email{filippo.bouche-ssm@unina.it}
\affiliation{Dipartimento di Fisica ``E. Pancini'', Universit\`a  di Napoli  ``Federico II", Via Cinthia 21, I-80126, Napoli, Italy}
\author{Salvatore Capozziello}
\email{capozziello@na.infn.it}
\affiliation{Dipartimento di Fisica ``E. Pancini'', Universit\`a  di Napoli  ``Federico II", Via Cinthia 21, I-80126, Napoli, Italy}
\affiliation{DipaScuola Superiore Meridionale, Largo S. Marcellino 10, I-80138, Napoli, Italy}
\affiliation{Istituto Nazionale di Fisica Nucleare, Sez. di Napoli,  Via Cinthia 21, I-80126, Napoli, Italy}
\author{V. Salzano}
\email{vincenzo.salzano@usz.edu.pl}
\affiliation{Institute of Physics, University of Szczecin, Wielkopolska 15, 70-451 Szczecin, Poland}

\date{\today}

\begin{abstract}
Alternative cosmological models have been under deep scrutiny in recent years, aiming to address the main shortcomings of the $\Lambda$CDM model. Moreover, as the accuracy of cosmological surveys improved, new tensions have risen between the model-dependent analysis of the Cosmic Microwave Background and lower redshift probes. Within this framework, we review two quantum-inspired non-locally extended theories of gravity, whose main cosmological feature is a geometrically driven accelerated expansion. The models are especially investigated in  light of the Hubble and  growth tension, and promising features emerge for the Deser--Woodard one. 
On the one hand, the cosmological analysis of the phenomenological formulation of the model shows a lowered growth of structures but an equivalent background with respect to $\Lambda$CDM. On the other hand, the study of the lensing features at the galaxy cluster scale of a new formulation of non-local cosmology, based on Noether symmetries, makes room for the possibility of alleviating both the $H_0$ and  $\sigma_8$ tension. However, the urgent need for a screening mechanism arises for this non-local theory of gravity.
\end{abstract}

\maketitle

\section{Introduction}

Recent astrophysical and cosmological surveys, both from ground-based and space experiments, have provided extremely high-quality data. The observations point towards a Universe in which the cosmological principle holds on large scales, namely the Universe appears homogeneous and isotropic if averaged over scales of $\sim$100 h$^{-1}${Mpc}
~or more \cite{Hogg:2004vw,SylosLabini_2009}. Moreover, the Universe is undergoing an accelerated expansion phase \cite{Riess_1998,Perlmutter_1999}, subsequent to a decelerated era in which the structure formation occurred. All these features, together with the nuclei abundances produced in the Big Bang Nucleosynthesis (BBN), the Baryon Acoustic Oscillations (BAO) and many others, are well predicted by the Lambda Cold Dark Matter ($\Lambda$CDM) model, which has been adopted as the standard cosmological model accordingly. Such a model provides an effective description of the Universe, which relies on the assumption that General Relativity (GR) is the final theory of gravitation that governs the cosmic dynamics. As a consequence, two more fluids other than baryonic matter and radiation must be inserted in the matter--energy content of the Universe in order to adjust GR predictions to data: the Dark Energy (DE), responsible for the late time cosmic acceleration, and the Dark Matter (DM), which accounts for the structure formation. Together, they should represent  $\sim$95\% of the matter--energy budget of the Universe \cite{Planck:2018vyg,Alam_2021,Abbott_2022}, thus dominating the cosmic dynamics at all scales. 

Building on its capability to fit the whole cosmological and astrophysical dataset with a relatively small number of parameters, the $\Lambda$CDM model stands as the pillar of our comprehension of the Universe. However, several shortcomings \cite{Bull_2016} and recently risen tensions \cite{Perivolaropoulos_2022,Cosmology_intertwined_II,Cosmology_intertwined_III} affect its reliability. On the one hand, we have a huge assortment of candidates but no final solution for DM \cite{Bertone_2005,https://doi.org/10.48550/arxiv.2110.10074} and DE \cite{Copeland2006wr,Padmanabhan2002ji,Nojiri_2006ri}. On the other hand, the presence of singularities, as well as the inconsistency at quantum level, undermine the credibility of GR as the final theory of gravity. Even more puzzling are the cosmological tensions, which have emerged in recent years as the result of the growing availability of a wide range of extremely precise data. A multitude of independent observations appears indeed to be in a $\gtrsim$2$\sigma$ tension with the reference $\Lambda$CDM estimates by the Planck collaboration \cite{Planck:2018vyg}. Even though systematic experimental errors may account for part of these tensions, their statistical significance and their persistence after several check analyses have thrown up some serious red flags. Since the Planck constraints for the cosmological parameters relies on a strongly $\Lambda$CDM-model-dependent analysis of the Cosmic Microwave Background (CMB), such tensions may be the signature of the brakedown of the concordance model, hence of new physics. In this paper, we  pay particular attention to the two most well-known tensions: the $H_0$ tension \cite{Cosmology_intertwined_II}, which emerges from the comparison between early-time \cite{Planck:2018vyg} and late-time measurements \cite{Riess_2022} of the Hubble constant, and the growth tension \cite{Cosmology_intertwined_III} between the CMB value \cite{Planck:2018vyg} of the cosmological parameters $\Omega_{M}$ and $\sigma_{8}$ and those from lower redshift probes, such as Weak Lensing (WL) \cite{Asgari_2021}, Cluster Counts (CC) \cite{Lesci_2022} and Redshift Space Distortion (RSD) \cite{Chen_2022}. 

In order to meet the challenges posed by $\Lambda$CDM theoretical shortcomings, as well as by its observational tensions, a zoo of alternative cosmological models has been formulated in recent years. The common feature of any proposed model is the introduction of additional degrees of freedom, whether in the gravitational or the matter--energy Lagrangian. Several approaches have been adopted: from the simple generalization of the Hilbert--Einstein action to functions of the curvature scalar, namely $f(R)$ theories \cite{CAPOZZIELLO_2002,Nojiri_2006,Starobinsky_2007,Fay_2007uy,Lazkoz_2018,Bajardi_2022,D_Agostino_2020_H0}, to the addition of further geometric invariants such as the torsion scalar $T$ \cite{Nesseris_2013jea,Cai_2016,Golovnev_2018wbh,Wang_Mota_2020,D_Agostino_2020_H0,Bahamonde_2021gfp} or the Gauss--Bonnet scalar $\mathcal{G}$ \cite{Nojiri_2005jg,Li_2007jm,Myrzakulov_2010gt,Bajardi_2020}. The introduction of scalar/vector fields minimally or non-minimally coupled to gravity \cite{Capozziello_2007iu,Heisenberg_2018vsk,Langlois_2018dxi,D_Agostino_2018,Bajardi_2020_aaa}, as well as the emergence of non-trivial dynamics in the dark sector \cite{Amendola_2007_DE,Wang_2016_DE,Di_Valentino_2017,Asghari_2019_DE,Yang_2019_DE}, also represent intriguing possibilities in the extremely wide framework of the alternatives to GR/$\Lambda$CDM (see \cite{CANTATA_2021ktz,DiValentino_2021izs} for the state of the art). In this paper, we want to inquire into a specific class of alternative cosmological models ruled by non-local gravitational interactions \cite{Capozziello_Review}. Among others, we  investigate the cosmological implications of two non-locally {extended} theories of gravity: the Deser--Woodard (DW) model \cite{Deser_2007} and the Ricci-Transverse (RT) model \cite{Maggiore_2014_RT}. These theories have drawn increasing attention in recent years due to their capability to account for  late-time cosmic acceleration, thus avoiding the introduction of any form of unknown dark energy. 
Moreover, the non-local corrections may provide a viable mechanism to alleviate some of the main cosmological tensions.

The paper is organized as follow: in Section~\ref{NonLocal}, we outline the main motives for formulating non-local theories of gravity, and we present the two ways in which dynamical non-locality can be implemented. Then, we introduce the two chosen models and their theoretical features. In Section~\ref{DE}, we investigate the mechanisms through which the DW and the RT model account for the accelerated expansion of the Universe. In Section~\ref{S8}, we present the non-locally driven evolution of cosmological perturbations for the two models and the resulting impact on the $\sigma_{8}$ tension. Moreover, in Section~\ref{H0}, we assess the $H_{0}$ tension in  light of the {non-local theories}. Finally, in Section~\ref{AstroTest}, we present the main astrophysical tests of {the non-local gravity models}. The conclusions are drawn in Section~\ref{Conclusions}.


\section{Non-Local Gravity}\label{NonLocal}

Non-locality naturally emerges in Quantum Physics, both as a kinematical and a dynamical feature. On the other hand, locality is a key property of classical field theories, and thus represents one of the main obstacles to overcome in order to merge  gravitational interaction formalism with that of Quantum Field Theory (QFT). As a consequence, the introduction of non-locality in our theory of gravitation seems to be an unavoidable step towards the unification of the fundamental interactions. There exist at least two ways to achieve non-locality: at fundamental level, in which kinematical non-locality can be implemented by discretizing  spacetime and introducing a minimal length scale (usually the Planck length); as an effective approach, in which non-local geometrical operators can be added to the gravitational Lagrangian to obtain a non-local dynamics in a continuum background spacetime \cite{Buoninf_thesis}. Here, we want to focus on the latter scenario, which is of great interest for cosmological applications.

Two main classes of non-locally {extended} theories of gravity have been developed in recent years \cite{Capozziello_Review}: Infinite Derivative theories of Gravity (IDGs), involving entire analytic transcendental functions of a differential operator, and Integral Kernel theories of Gravity (IKGs) based on integral kernels of differential operators, such as

\begin{equation}\label{InverseBox}
    \Box^{-1}R(x) = \int d^{4}x' G(x,x') R(x') \, ,
\end{equation}
where $G(x'x')$ is the Green function associated to the inverse d’Alembertian. 
IDGs usually address the ultraviolet (UV) problems of the $\Lambda$CDM model by ensuring classical asymptotic freedom. The gravitational interaction is weakened on small scales and the singularities disappear accordingly. Non-singular black holes \cite{Biswas_2012}, as well as inflationary \cite{Briscese_2013} and bouncing cosmologies \cite{Biswas_2006}, are indeed forecast in the IDG framework. On the other hand, IKGs are introduced to account for the infrared (IR) shortcomings of the concordance model of cosmology. The phenomenology of both  dark fluids can be actually reproduced by non-local corrections that switch on at large scales \cite{Deser_2007,Belgacem_2020,BORKA2022}.

In this paper, we focus on two specific curvature-based IKGs \cite{Deser_2007,Maggiore_2014_RT} and their cosmological features. IKGs indeed have special relevance due to the fact that they combine suitable cosmological behavior with well-justified Lagrangians at the fundamental level. 
GR is actually plagued by {quantum} IR divergences that already appear for pure gravity in flat space \cite{https://doi.org/10.48550/arxiv.1703.05448}. This pathological behavior implies that the long-range dynamics of the gravitational interaction may be non-trivial, and non-perturbative techniques are thus required. Applying such non-perturbative methods to the renormalization of the quantum effective action of the gravity theory, non-local terms emerge both associated \cite{Reuter_2002,Wetterich_2018} or not associated \cite{Barvinsky_2003rx,Barvinsky_2015} to a dynamical mass scale. Analogous results can be recovered through the trace anomaly \cite{https://doi.org/10.48550/arxiv.1606.08784}.

\subsection{The Deser--Woodard Model}

The first model that we want to highlight is an IKG initially proposed in \cite{Deser_2007}. The non-locally {extended} gravitational action of the Deser--Woodard model reads

\begin{equation}\label{DWaction}
    S = \frac{1}{16 \pi G} \int d^{4}x \sqrt{-g} \Big\{ R \big[ 1 + f(\Box^{-1}R) \big] \Big\} \, ,
\end{equation}
where the non-local correction is given by the so-called distortion function, namely a general function of the inverse box of the Ricci scalar, as in Equation~(\ref{InverseBox}).
It is worth noticing that the non-local theory reduces to GR as soon as $f(\Box^{-1}R)$ vanishes. The modified field equations descending from Equation~(\ref{DWaction}) are

\begin{equation}
    G_{\mu\nu} + \Delta G_{\mu\nu} = \kappa T_{\mu\nu}^{(m)} \, ,
\end{equation}
where the non-local correction reads

\begin{equation}\label{DWequationExplicit}
\begin{aligned}
\Delta G_{\mu\nu} =& \bigg( G_{\mu\nu} + g_{\mu\nu} \square - \nabla_{\mu}\nabla_{\nu} \bigg) \bigg\lbrace f \big( \square^{-1}R \big) + \square^{-1} \big[ R f'\big( \square^{-1}R \big)\big] \bigg\rbrace \\
&+ \Bigg[ \frac{1}{2} \bigg( \delta_{\mu}^{\alpha} \delta_{\nu}^{\beta} + \delta_{\mu}^{\beta} \delta_{\nu}^{\alpha} \bigg) - \frac{1}{2} g_{\mu\nu} g^{\alpha\beta} \Bigg] \partial_{\alpha} \big( \square^{-1}R \big) \partial_{\beta} \bigg\lbrace \square^{-1} \big[ R f'\big( \square^{-1}R \big)\big] \bigg\rbrace \, .
\end{aligned}
\end{equation}

Furthermore, the non-local gravitational action in Equation~(\ref{DWaction}) can be easily rewritten under the standard of local scalar--tensor theories by introducing an auxiliary scalar field 

\begin{equation}
    R(x) = \Box \eta(x) \, ,
\end{equation}
which does not carry any independent degree of freedom. The local canonical form of the scalar--tensor action, equivalent to the non-local theory, thus reads \cite{Nojiri_2008}

\begin{equation}\label{NojiriAction}
    S = \frac{1}{2\kappa} \int d^{4}x \sqrt{-g} \Big\{ R \big[ 1 + f\big( \eta \big) \big] - \partial_{\mu}\xi \partial^{\mu}\eta - \xi R \Big\} \, ,
\end{equation}
where $\xi(x)$ is a Lagrangian multiplier which has been promoted to a position- and time-dependent scalar field. In this formulation, the gravitational field equation is

\begin{equation}\label{NLfieldeq}
\begin{aligned}
G_{\mu\nu} =& \frac{1}{1+f(\eta)-\xi} \Bigg[ \kappa T_{\mu\nu}^{(m)} - \frac{1}{2} g_{\mu\nu} \partial^{\alpha}\xi \partial_{\alpha}\eta + \frac{1}{2} \big( \partial_{\mu}\xi \partial_{\nu}\eta + \partial_{\mu}\eta \partial_{\nu}\xi \big) - \big( g_{\mu\nu} \square - \nabla_{\mu} \nabla_{\nu} \big) \big( f(\eta) - \xi \big) \Bigg] \, ,
\end{aligned}
\end{equation}
while the Klein--Gordon equations for the two auxiliary scalar fields are

\begin{equation}\label{KGeqeta}
    \Box \eta = R \, ,
\end{equation}
\begin{equation}\label{KGeqxi}
    \Box \xi = -R \frac{\partial f(\eta)}{\partial \eta} \, .
\end{equation}

\subsection{The Ricci-Transverse Model}

The second non-local model that we investigate through this paper is a metric IKG proposed in \cite{Maggiore_2014_RT}. This is a quantum-inspired model, whose quantum effective action is

\begin{equation}\label{RTaction}
    \Gamma = \frac{1}{64 \pi G} \int d^{4}x \bigg[ h_{\mu\nu} \mathcal{E}^{\mu\nu,\alpha\beta} h_{\alpha\beta} - \frac{2}{3} m^2 \big( P^{\mu\nu} h_{\mu\nu} \big)^2 \bigg] \, ,
\end{equation}
where $g_{\mu\nu}=\eta_{\mu\nu}+\kappa h_{\mu\nu}$ {is the linearized metric tensor}, $\mathcal{E}^{\mu\nu,\alpha\beta}$ is the Lichnerowicz operator\footnote{$\mathcal{E}^{\mu\nu,\alpha\beta} = \frac{1}{2} (\eta^{\mu\rho}\eta^{\nu\sigma} + \eta^{\mu\sigma}\eta^{\nu\rho} - 2 \eta^{\mu\nu}\eta^{\rho\sigma}) \Box + (\eta^{\rho\sigma}\partial^{\mu}\partial^{\nu} + \eta^{\mu\nu}\partial^{\rho}\partial^{\sigma}) - \frac{1}{2} (\eta^{\mu\rho}\partial^{\sigma}\partial^{\nu} + \eta^{\nu\rho}\partial^{\sigma}\partial^{\mu} + \eta^{\mu\sigma}\partial^{\rho}\partial^{\nu} + \eta^{\mu\sigma}\partial^{\rho}\partial^{\mu})$}, $P^{\mu\nu} = \eta^{\mu\nu} - (\partial^{\mu}\partial^{\nu}/\Box)$ is a projector operator and $m$ is the mass of the conformal mode of the gravitational field. Performing the covariantization of Equation~(\ref{RTaction}), the modified gravitational field equation reads

\begin{equation}\label{RTequation}
    G_{\mu\nu} - \frac{1}{3} m^2 (g_{\mu\nu} \Box^{-1}R)^{T} = \kappa T_{\mu\nu} \, ,
\end{equation}
where we take the transverse part of the symmetric non-local tensor $S_{\mu\nu} = g_{\mu\nu} \Box^{-1}R$

\begin{equation}\label{Transverse}
    S_{\mu\nu} = S_{\mu\nu}^{T} + \frac{1}{2} (\nabla_{\mu} S_{\nu} + \nabla_{\nu} S_{\mu}) \, ,
\end{equation}
and $S_{\mu}$ is an associated four-vector. The Bianchi identities are guaranteed accordingly, i.e., $\nabla^{\mu}S_{\mu\nu}^{T}=0$.

In the same way as the DW model, the Ricci-Transverse model can be localized through a scalar--tensor--vector formulation \cite{Maggiore_2014_RT,Kehagias_2014}. 
Here, we introduce two auxiliary objects, namely

\begin{equation}
    U(x)=-\Box^{-1}R(x) \, , \qquad \mathcal{S}_{\mu\nu}(x) = -U(x) g_{\mu\nu}(x) = g_{\mu\nu}(x) \Box^{-1}R(x) \, .
\end{equation}

An auxiliary four-vector field $\mathcal{S}_{\mu}(x)$ therefore enters the localized equations because of Equation~(\ref{Transverse}). The gravitational field equation, Equation~(\ref{RTequation}), turns into

\begin{equation}\label{GFeqRT}
    G_{\mu\nu} + \frac{m^2}{6} \big( 2Ug_{\mu\nu} + \nabla_{\mu} \mathcal{S}_{\nu} + \nabla_{\nu} \mathcal{S}_{\mu} \big) = \kappa T_{\mu\nu}
\end{equation}
plus the two equations of motions of the two auxiliary fields

\begin{equation}\label{KGeqRT}
    \Box U = - R \, ,
\end{equation}

\begin{equation}\label{4VeqRT}
    \big( \delta_{\nu}^{\mu}\Box + \nabla^{\mu}\nabla_{\nu} \big) \mathcal{S}_{\mu} = -2 \partial_{\nu}U \, .
\end{equation}


\section{The Late-Time Cosmic Acceleration}\label{DE} 

The Universe is currently undergoing an accelerated expansion. The first evidence of this peculiar behavior dates back to the end of the twentieth century, when the observation of several Type Ia Supernovae (SNIa) \cite{Riess_1998,Perlmutter_1999} pointed out the unavoidable necessity of a cosmological constant to fit the cosmic expansion history. On the one hand, these results have been corroborated by the observations of all the recent surveys \cite{Planck:2018vyg,Aiola_2020,Tr_ster_2021_DE,Pandey_2022_DE}. On the other hand, the theoretical explanation of this issue has two main drawbacks: the fine tuning problem \cite{RevModPhys_61_1} and the coincidence problem \cite{Velten_2014}. The former is related to the huge discrepancy ($\sim$120 orders of magnitude) between the observed value of the cosmological constant and the vacuum energy density calculated via QFT. The latter is linked with the similar current values of $\Omega_{\Lambda}$ and $\Omega_{M}$, despite their radically different evolution laws. 

The next generation of cosmological surveys should boost the investigation of the nature of the so-called cosmological constant, providing powerful data to discriminate between DE solutions and extended theories of gravity. Within this framework, non-local gravity provides viable mechanisms that could account for the observed accelerated expansion of the Universe.

\subsection{The DW Case: Delayed Response to Cosmic Events}

The main reason why the DW model has been in the spotlight since its formulation is its effective way to explain  late-time cosmic acceleration without the introduction of any form of dark energy. Computing the non-local correction of Equation~(\ref{DWaction}) in the Friedmann--Lemaitre--Robertson--Walker (FLRW) metric, 

\begin{equation}\label{FLRW}
    ds^2 = - dt^2 + a^2(t) d\Vec{x}\cdot d\Vec{x} \, ,
\end{equation}
one obtains a non-negligible geometrical contribution,

\begin{equation}\label{InverseBoxCosmology}
\begin{aligned}
\big[ \square^{-1} R \big](t) &= \int_{0}^{t} dt' \frac{1}{a^{3}(t')} \int_{0}^{t'} dt'' a^{3}(t'') R(t'') = - \frac{6s (2s-1)}{3s-1} \left[ \mathrm{ln}\Bigg(\frac{t}{t_{eq}} \Bigg) - \frac{1}{3s-1} + \frac{1}{3s-1} \Bigg( \frac{t_{eq}}{t} \Bigg)^{3s-1} \right] \, ,
\end{aligned}
\end{equation}
where $a(t) \sim t^{\,s}$, and the integration constant is set to make the non-local correction vanish before the radiation--matter equivalence time, $t_{eq}$. 
Then, for $t>t_{eq}$ the non-local correction starts to grow, becoming non-negligible at late time and driving the accelerated cosmic expansion. Non-locality thus emerges in the cosmological framework as a delayed response to the radiation-to-matter dominance transition, i.e., as a time-like non-local effect.

The introduction of non-locality may therefore have beneficial effects on cosmological scales, both at background level and perturbations level. Two different approaches can be adopted for the DW model: on the one hand, one can exploit the freedom guaranteed by the undetermined form of the distortion function to fit the observed expansion history of the universe. In such a scenario, the DW cosmology is made equivalent to $\Lambda$CDM at the background level. However, different features emerge when perturbations are taken into account, and the physics of structure formation is affected accordingly. On the other hand, one can select the form of the distortion function building on some fundamental principles, such as the Noether symmetries of the system. In this case, the non-local cosmology is also modified at background level and stronger deviations from the concordance model should rise.

In \cite{Deffayet_2009}, the first method has been applied, and the $\Lambda$CDM expansion history of the Universe has been accurately reproduced by matching the data through a non-trivial form of the distortion function

\begin{equation}\label{DistFun}
    f(\Box^{-1}R) = 0.245 \big[ tanh \big( 0.350 X + 0.032 X^{2} + 0.003 X^{3} \big) -1 \big] \, ,
\end{equation}
where $X = \Box^{-1}R + 16.5$.

\subsection{The RT Case: Dynamical Dark Energy}

Considering a spatially flat FLRW metric, Equation~(\ref{FLRW}), the equations of motion of the RT model, Equations~(\ref{GFeqRT})--(\ref{4VeqRT}), become \cite{Foffa_2014}

\begin{equation}\label{RTcosm1}
    {H^2 - \frac{m^2}{9} (U - \Dot{\mathcal{S}}_{0}) = \frac{8\pi G}{3} \rho \, ,}
\end{equation}

\begin{equation}\label{RTcosm2}
    \Ddot{U} + 3H\Dot{U} = 6 \Dot{H} + 12 H^2 \, ,
\end{equation}

\begin{equation}\label{RTcosm3}
    \Ddot{\mathcal{S}}_{0} + 3H\Dot{\mathcal{S}}_{0} - 3 H^2 \mathcal{S}_{0} = \Dot{U} \, ,
\end{equation}
where the spatial components of the vector field $\mathcal{S}_{\mu}$ vanish to preserve the rotational invariance of the FLRW metric, and the stress--energy tensor is taken to be $T^{\mu}_{\nu} =$diag$(-\rho, p, p, p)$. 
Defining $Y=U-\Dot{\mathcal{S}}_{0}$, $\Tilde{h} = H/H_0$ and the dimensionless variable $x \equiv \ln{a(t)}$, then the modified Friedmann equation, Equation~(\ref{RTcosm1}), reads

\begin{equation}
    \Tilde{h}^{2}(x) = \Omega_{M} e^{-3x} + \gamma Y(x) \, ,
\end{equation}
with $\gamma \equiv m^2 / (9 H_{0}^{2})$. An effective dark energy thus appears

\begin{equation}
    \rho_{DE}(t) = \rho_{0} \gamma Y(t) \, ,
\end{equation}
where $\rho_{0} = 3 H_{0}^{2} / (8\pi G)$. Once the initial conditions for the auxiliary fields are set (see \cite{Belgacem_2020} for details), the evolution of $\rho_{DE}(t)/\rho_{TOT}(t)$ can be studied: the non-local effective dark energy is actually negligible until recent time and then starts to dominate the cosmic expansion. 
Moreover, it is possible to study the DE equation of state

\begin{equation}
    \Dot{\rho}_{DE} + 3H (1 + \omega_{DE})\rho_{DE} = 0 \, ,
\end{equation}
and different evolutions for $\rho_{DE}(z)$ follow from different choices for the initial conditions of the auxiliary fields. For small values of the initial conditions, one obtains a fully phantom DE, namely $\omega_{DE}(z)$ is always less than $-1$. For large values of the initial conditions, $\rho_{DE}(z)$ has a "phantom crossing" behavior, i.e., there is a transition from the phantom regime to $-1<\omega_{DE}<0$ of about $z \simeq 0.3$. 
Regardless of the initial conditions, therefore, the non-local model provides a dynamical DE density which drives the accelerated expansion of the Universe.


\section{The Growth of Perturbations and the \texorpdfstring{\boldmath{$\sigma_{8}$}}{S8} Tension}\label{S8}

Building on the primordial density fluctuations emerged from the inflation,  cosmic structures have formed due to  gravitational instability. 
Studying the large-scale structure of the Universe and its evolution through the cosmic epochs, it is possible to trace the growth of the so-called cosmological perturbations.

Associated to this observable, one of the main cosmological tensions has risen: the growth tension. It has come about as the result of the discrepancy between the Planck value of the cosmological parameters $\Omega_{M}$ and $\sigma_{8}$ and those from WL measurements, CC and RSD data. The former dynamical probes point towards lower values of the amplitude ($\sigma_{8}$) or the rate ($f\sigma_{8} = [\Omega_{M}(z=0)]^{0.55}\sigma_{8}$) of growth of structures with respect to the CMB experiments, giving rise to a $2-3 \sigma$ tension \cite{Perivolaropoulos_2022,Cosmology_intertwined_III}. Moreover, the Planck 2018 value of the joint parameter $S_{8} = \sigma_{8}\sqrt{\Omega_{M}/0.3}\,$ ($S_{8} = 0.834 \pm 0.016$ \cite{Planck:2018vyg}) is confirmed by another recent CMB analysis by the ACT + WMAP collaboration ($S_{8} = 0.840 \pm 0.030$ \cite{Aiola_2020}), thus erasing the possibility of a systematic error related to the excess of lensing amplitude measured by Planck \cite{Calabrese_2008}. A $2.3\sigma$ tension emerges accordingly with both the original WL analysis of KiDS-450 \cite{Hildebrandt_2016} and KiDS-450 + VIKING data \cite{Hildebrandt_2020}, while updated constraints from the same datasets \cite{K_hlinger_2017,Wright_2020} show greater discrepancies. 
The same $2.3\sigma$ tension also occurs with the data from DES's first year release (DESY1) \cite{Troxel_2018}, while the combination of KiDS-450 + VIKING + DESY1 weak lensing datasets results in a $2.5\sigma$ \cite{Joudaki_2020} or $3.2\sigma$ \cite{Asgari_2020} tension depending on the analysis. The most recent cosmic
shear data release from both KiDS-1000 and DESY3 confirms the previous estimates \cite{Asgari_2021,Loureiro_2022,https://doi.org/10.48550/arxiv.2203.12440,Amon_2022,Secco_2022} ($S_{8} = 0.759^{+0.024}_{-0.021}$ from KiDS-1000 \cite{Asgari_2021}). Analogous results have been obtained with the $3 \times 2$ {pt}
~correlation function analysis (cosmic shear correlation function, galaxy clustering angular auto-correlation function and galaxy--galaxy lensing cross-correlation function) of KiDS-1000 + BOSS + 2dFLenS datset \cite{Heymans_2021}. Additional results, in agreement with those from WL surveys, have been achieved by number counting of galaxy clusters, using multiwavelength datasets \cite{Mantz_2014,Salvati_2018,Costanzi_2019,Bocquet_2019,Abbott_2020_CC,Abdullah_2020,Lesci_2022}. 
Supplementary observational evidence for the weaker growth of structures is also given by the exploitation of RSD data \cite{Kazantzidis_2018,Benisty_2021,Nunes_2021,Philcox_2022,Chen_2022}.

\subsection{The Deser--Woodard Evolution of Scalar Perturbations}

To investigate the growth of structures in the non-local DW model, we select the phenomenological form of the distortion function given by Equation~(\ref{DistFun}). As a consequence, the non-local background evolution is made equivalent to that of $\Lambda$CDM, and any deviation is enclosed in the cosmological perturbations.

Consider the field equations of the scalar--tensor equivalent of the DW model, Equations~(\ref{NLfieldeq})--(\ref{KGeqxi}). Specializing to the cosmological case by assuming the FLRW metric, Equation~(\ref{FLRW}), the field equations now read

\begin{equation}
        {H^2 \big[ 1 + f - \xi \big] + H \big[ f' \Dot{\eta} - \Dot{\xi} \big] - \frac{1}{6} \Dot{\eta} \Dot{\xi} =\frac{8 \pi G}{3} \rho \, ,}
\end{equation}

\begin{equation}
        {\Dot{H} \big[ 1 + f - \xi \big] - \frac{H}{2} \big[ f' \Dot{\eta} - \Dot{\xi} \big] + \frac{1}{2} \Dot{\eta} \Dot{\xi} + \frac{1}{2} \big[ f'' \Dot{\eta}^2 + f' \Ddot{\eta} - \Ddot{\xi} \big] = - 4 \pi G (p + \rho) \, ,}
\end{equation}

\begin{equation}
    {\Ddot{\eta} + 3H\Dot{\eta} = -6 \big( \Dot{H} + 2 H^2 \big) \, ,}
\end{equation}

\begin{equation}
    {\Ddot{\xi} + 3H\Dot{\xi} = 6 f' \big( \Dot{H} + 2 H^2 \big) \, ,}
\end{equation}
where the former two are the (0,0) and the (1,1) component of Equation~(\ref{NLfieldeq}), while the latter two are the cosmological formulation of Equations~(\ref{KGeqeta}) and~(\ref{KGeqxi}).

The linear perturbation equations have been derived in \cite{Nersisyan_2017} for the scalar--tensor equivalent of the non-local theory, and then analogous results have been found in \cite{Park_2018} for the original formulation. Using the perturbed FLRW metric in the Newtonian gauge,

\begin{equation}\label{NewtGauge}
    ds^2 = - (1+2\Psi)dt^2 + a^2(t) (1+2\Phi)\delta_{ij}dx^i dx^j \, ,
\end{equation}
the growth equation reads

\begin{equation}
  {\Ddot{\delta}_{M} + (2-\xi) \Dot{\delta}_{M} = \frac{3 H_{0}^{2} \big[ 1-\xi - 8 f'(\eta) + f(\eta) \big] \Omega_{M}^{0}}{2 a^3 H^2 \big[ 1-\xi - 6 f'(\eta) + f(\eta) \big] \big[ 1 + f(\eta) - \xi \big]} \, \delta_{M} \, ,}
\end{equation}
where $\delta_{M} = \delta \rho_{M}/\rho_{M}$ is the matter density perturbation in the sub-horizon limit.

\begin{figure}
\includegraphics[width=9.35cm]{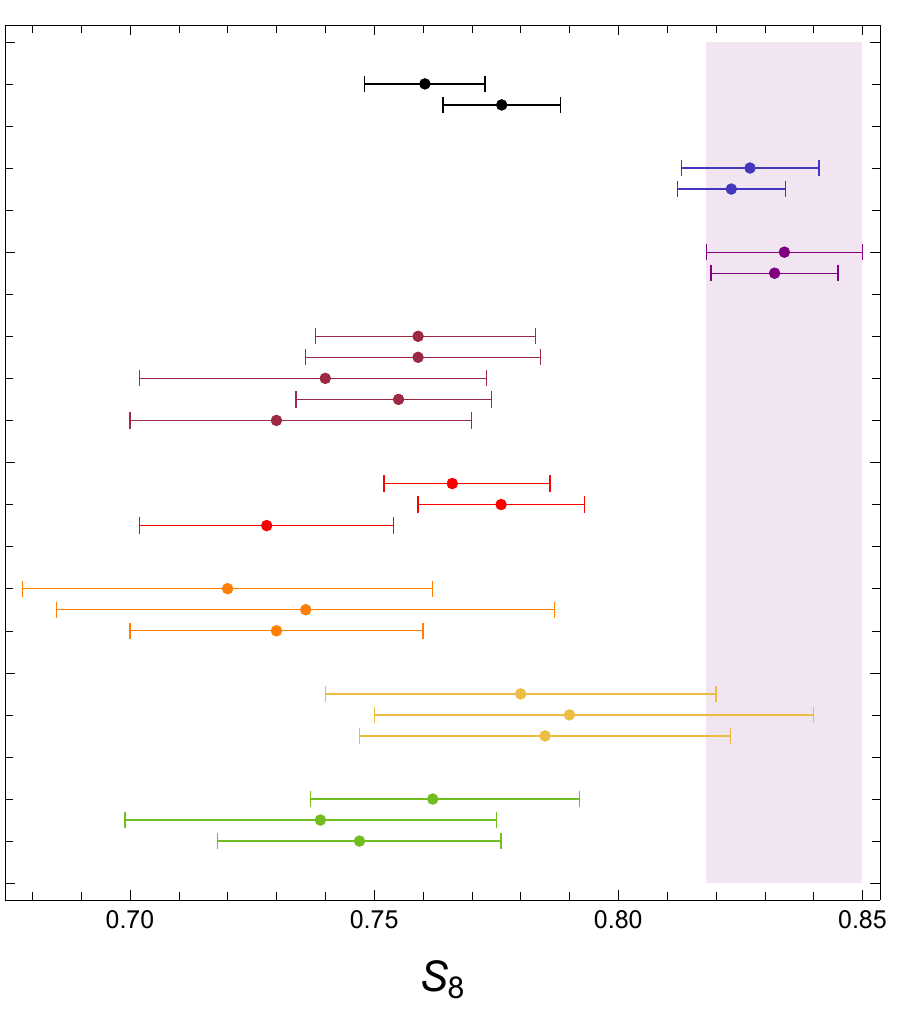}\includegraphics[width=4.5cm,trim=0cm -1.16cm 0cm 0cm]{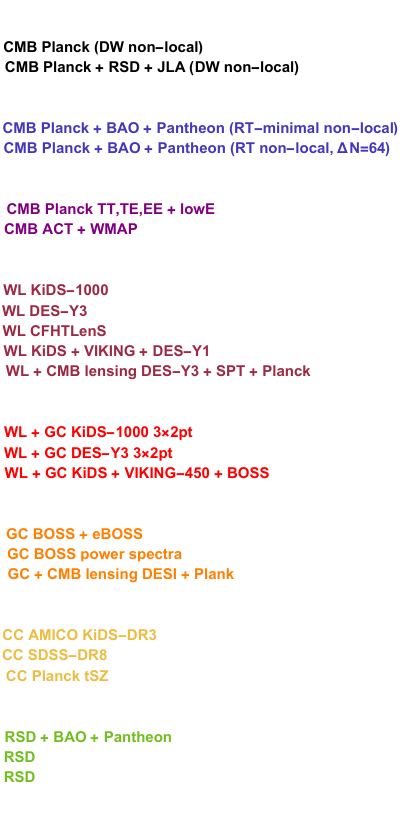}
\caption{{Estimates of} $S_{8}$ provided by the two non-local cosmological analyses \cite{Amendola_2019,Belgacem_2020}, and the $\Lambda$CDM fit of the CMB \cite{Planck:2018vyg,Aiola_2020}, the WL data \cite{Asgari_2021,Amon_2022,Secco_2022,Joudaki_2017,Asgari_2020,Chen_2022}, the combination of WL and galaxy clustering observations \cite{Ivanov_2021,Ivanov_2020,White_2022}, cluster counting \cite{Lesci_2022,Costanzi_2019,Salvati_2018} and RSD surveys \cite{Nunes_2021,Kazantzidis_2018}. The colored band corresponds to the $S_8$ value derived by the analysis of the Planck--CMB data in the $\Lambda$CDM framework \cite{Planck:2018vyg}.}\label{FigureBars1}
\end{figure}

Numerical results for the growth rate $f \sigma_{8} \equiv \sigma_{8} \delta_{M}'/\delta_{M}$ have been obtained in \cite{Nersisyan_2017} and \cite{Park_2018} for both the formulations of the Deser--Woodard model, and good agreement with the Redshift Space Distortion (RSD) data has emerged ($\sigma_{8}^{NL}=0.78$). Moreover, when the DW cosmological parameters are inferred by matching the CMB data \cite{Amendola_2019}, a lower growth amplitude with respect to that of $\Lambda$CDM turns out. The non-local model thus alleviates the growth tension, predicting compatible values of $\sigma_{8}$ both from Planck--CMB and the other dynamical probes, as shown in Figure~\ref{FigureBars1}. 

However, even though the non-local clustering of linear structures is weakened with respect to $\Lambda$CDM, and the DW prediction for the matter power spectrum is about $10\%$ lower \cite{Amendola_2019}, the non-local lensing response is counterintuitively enhanced due to a severe increase in the lensing potential. This peculiar behavior results in a slight tension between CMB and RSD \cite{Amendola_2019}: performing the joint fit, the RSD dataset tends to push the DW predictions for the CMB lensing potential $C^{\phi\phi}_{\ell}$ out of the $1\sigma$ error bars at low-$\ell$. Applying the Bayesian tools for the model selection, a ``weak evidence'' \cite{JeffreysScale} for the $\Lambda$CDM model consequently emerges.

\subsection{The Ricci-Transverse Evolution of Scalar Perturbations}

The scalar perturbations of the RT model have been investigated in \cite{Nesseris_2014_pert,Belgacem_2020}, using the FLRW metric in the Newtonian gauge, Equation~(\ref{NewtGauge}), and perturbing the auxiliary fields as

\begin{equation}
    U(t,\mathbf{x}) = \Bar{U}(t) + \delta U(t,\mathbf{x}) \, ,
\end{equation}

\begin{equation}
    \mathcal{S}_{\mu}(t,\mathbf{x}) = \Bar{\mathcal{S}}_{0}(t) + \delta \mathcal{S}_{\mu}(t,\mathbf{x}) = \Bar{\mathcal{S}}_{0}(t) + \delta \mathcal{S}_{0}(t,\mathbf{x}) +  \partial_{i} \big[ \delta \mathcal{S}(t,\mathbf{x}) \big] \, ,
\end{equation}
where the spatial part of the vector perturbation does not vanish and, for scalar perturbations, only depends on $\delta \mathcal{S}$. Building on the RT cosmological equations Equations~(\ref{RTcosm1})--(\ref{RTcosm3}), the growth equation for the matter density perturbation in the sub-horizon limit reads \cite{Nesseris_2014_pert}

\begin{equation}
     {\Ddot{\delta}_{M} + 2H \Dot{\delta}_{M} = \frac{3}{2} \frac{G_{eff}}{G} H_{0}^{2} \Omega_{M} \delta_{M} \, ,}
\end{equation}
where in $G_{eff} \big( \Psi, \Phi, \delta U, \delta \mathcal{S}_0 ,  \mathcal{S} \big)$ is encoded the deviation of the non-local theory from GR. In the sub-horizon modes, namely $\hat{k} \gg 1$, such deviation is

\begin{equation}\label{Geff}
    1- \frac{G_{eff}}{G} = \mathcal{O}\bigg( \frac{1}{\hat{k}^2} \bigg) \, ,
\end{equation}
and the RT model is thus safe regarding the time variation of the effective Newton's constant. $G_{eff}$ indeed reduces to $G$ at the Solar System scale, while a deviation of $\sim$1\% rises at cosmological scales.

In \cite{Belgacem_2020}, the growth rate $f(z, k) \equiv d \ln{\delta_{M}}/d\ln{a}$ is also derived. The results do not differ from those of $\Lambda$CDM cosmology: $f(z, k)$ can be fitted with a $k$-independent function $f(z) = [\Omega_{M}(z)]^{\gamma}$, where $\gamma \simeq 0.55$ is roughly constant. 
Accordingly, any possible deviation in the growth of perturbations should be due to the amplitude $\sigma_{8}$. In order to find any signature of the non-local model at perturbation level, which could account for the growth tension, the theory was compared with cosmological observations: Planck--CMB, Pantheon SNIa and SDSS-BAO. The Bayesian parameter estimation shows a full equivalence between the RT non-local cosmology and the $\Lambda$CDM one. No statistically significant deviation in the $\sigma_{8}$ parameter emerges for any of the tested versions of the Ricci-Transverse model. Eventually, this theory cannot alleviate the growth tension, as shown in Figure~\ref{FigureBars1}.


\section{Hubble Tension in Light of the Non-Local Models}\label{H0} 

Hubble tension is certainly the most renowned and significant tension of the $\Lambda$CDM model. It emerges from the comparison between early-time and late-time measurements of the Hubble constant. From one side, CMB analysis \cite{Planck:2018vyg,Dutcher_2021,Aiola_2020,Wang_2020,Balkenhol_2021,Addison_2021}, BAO surveys \cite{Ivanov_2020,d_Amico_2020,Colas_2020,Alam_2021} with standard BBN constraints \cite{Cooke_2018} and combinations of CMB, BAO, SNIa \cite{Scolnic_2018}, RSD and cosmic shear data \cite{Abbott_2018,Troxel_2018,https://doi.org/10.48550/arxiv.1706.09359} point towards lower values of $H_0$ ($H_0 = 67.4 \pm 0.5$ {km s}$^{-1}$Mpc$^{-1}$
~from Planck 2018 \cite{Planck:2018vyg}). On the other side, the local measurements based on standard candles prefer higher values for the Hubble constant \cite{Riess_2019_review} ($H_0 = 73.04 \pm 1.04$ {km s}$^{-1}$Mpc$^{-1}$ from SH0ES 2022 \cite{Riess_2022}). The main results are achieved by the SH0ES collaboration using Hubble Space Telescope observations: on the one hand, they analyzed SNIa data with distance calibration by Cepheid variables in the host galaxies \cite{Riess_2011,Riess_2016}; on the other hand, they targeted long-period pulsating Cepheid variables \cite{Riess_2021,Riess_2022}, calibrating the geometric distance to the Large Magellanic Cloud, both from eclipsing {binaries}
and parallaxes from the Gaia satellite \cite{GAIA,refId0}. Moreover, other independent	local measurement of $H_0$ have been performed by using time delays	between	multiple images	of strong lensed quasars \cite{Wong_2019aa,Shajib_2020}, the tip of the Red Giant Branch \cite{Yuan_2019} and Miras (variable red giant stars) \cite{Huang_2020_miras} with water megamaser as distance indicator \cite{Reid_2019}. All these measurements agree on higher values of the Hubble constant, thus generating a $4-5 \sigma$ tension with early-time model-dependent estimates.

\subsection{The Deser--Woodard Expansion History}

\begin{figure}
\includegraphics[width=9.35cm]{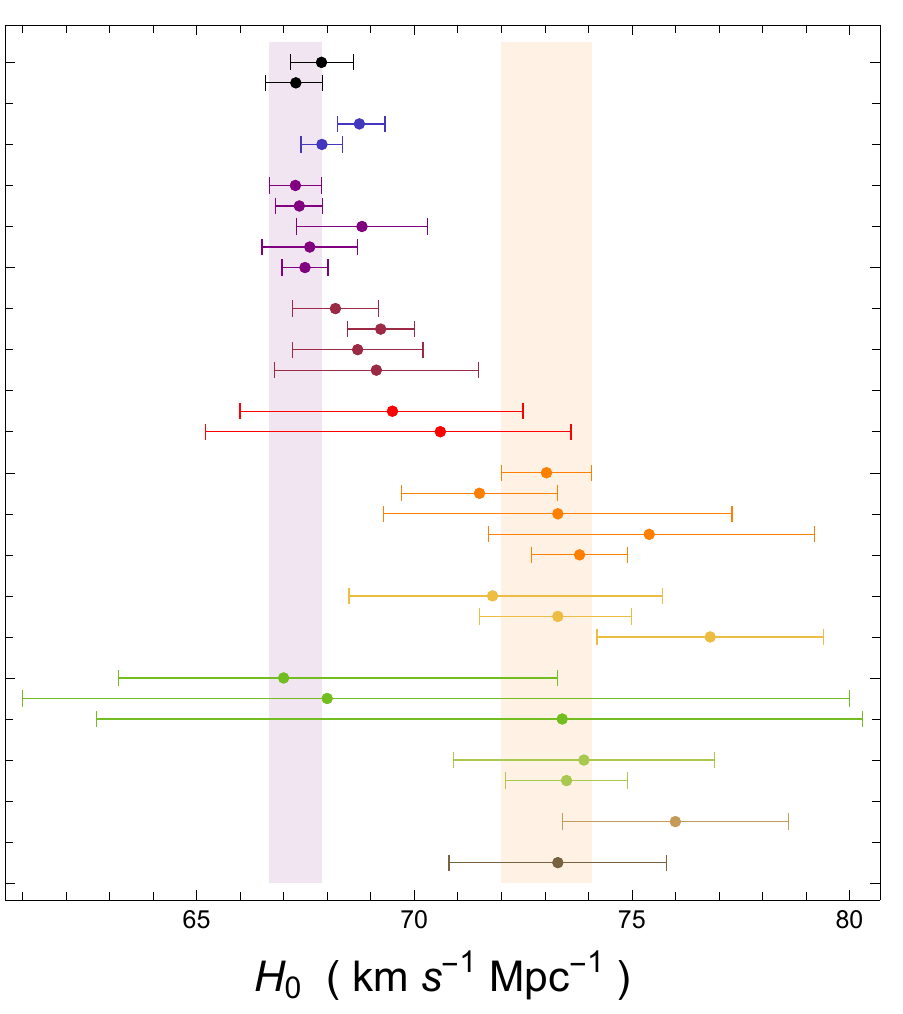}\includegraphics[width=4.37cm,trim=0cm -1.21cm 0cm 0cm]{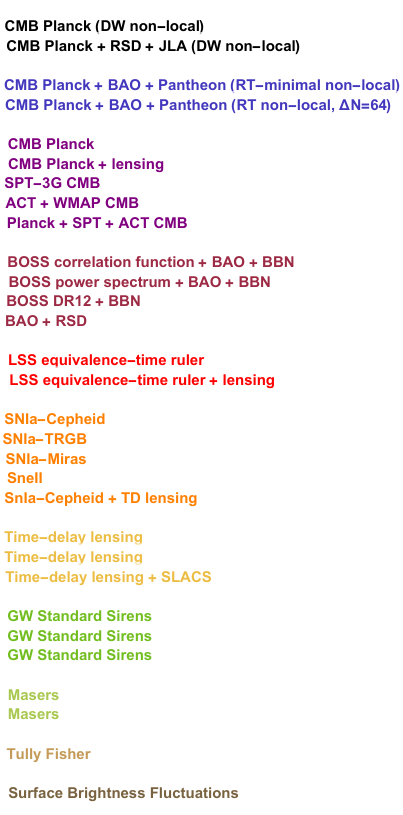}
\caption{Estimates of $H_0$ provided by the two non-local cosmological analyses \cite{Amendola_2019,Belgacem_2020}, and the $\Lambda$CDM fit of the CMB \cite{Planck:2018vyg,Dutcher_2021,Aiola_2020,Balkenhol_2021}, the matter power spectrum combined with BAO \cite{Zhang_2022,Chen_2022,Colas_2020} and RSD \cite{Wang_2017}, the Large Scale Structure $t_{eq}$ standard ruler \cite{Farren_2022,Philcox_2021}, the supernovae \cite{Riess_2022,Anand_2022,Huang_2020_miras,de_Jaeger_2022,Wong_2019aa}, the time-delay lensing \cite{Denzel_2020,Wong_2019aa,Chen_2019}, the gravitational waves \cite{Muk_2022,Abb_2021_b,Gayat_GW}, the water megamasers \cite{Pesce_2020,Reid_2019}, the Tully--Fisher relation \cite{Kourkchi_2020} and the SBF \cite{Blakeslee_2021}. The colored bands correspond to the $H_0$ estimates derived by the Planck--CMB analysis in the $\Lambda$CDM framework (purple) \cite{Planck:2018vyg} and the SNIa--Cepheids analysis by SH0ES (orange) \cite{Riess_2022}.}\label{FigureBarsH0} 
\end{figure}

The stat-of-the-art investigation of the DW non-local model does not allow any attempt to address  $H_0$ tension. 
To make the model predictive, the form of the distortion function needs to be specified, and most of the analyses have been focused on the phenomenological $\Lambda$CDM form of Equation~(\ref{DistFun}), until now (see \cite{Amendola_2019} for the latest results). This choice implies that the DW cosmology is made equivalent to that of the concordance model at background level, and the same expansion history, as well as the same $H_0$, are thus predicted (see Figure~\ref{FigureBarsH0}).  

However, another option is also available for the selection of the distortion function. In \cite{Dialektopoulos_2018iph,Bouche:2022jts,BORKA2022}, a specific form of $f(\eta)$ has been derived by exploiting the Noether symmetries \cite{Capozziello:1996bi} of a spherically symmetric background spacetime

\begin{equation}\label{ExpDistFun}
    f(\eta) = 1 + e^{\,\eta} \, .
\end{equation}

The accurate cosmological analysis of this form of the DW model has yet to be carried out, but some results have already been achieved, such as exact solutions \cite{Capozziello_Review} and a phase-space view of solutions \cite {Capozziello:2022rac}. {Furthermore}, several astrophysical tests have been performed on very different scales, and viable results turned out. In \cite{Bouche:2022jts}, the lensing properties of the galaxy clusters have been investigated in  light of the exponential DW model, and a fully non-local regime with enhanced lensing strength has been highlighted. This feature clearly resembles the improved lensing response of the phenomenological form of the DW model, which is co-responsible for the lowered estimate of the $\sigma_{8}$ parameter. A promising insight upon the cosmological behavior of the non-local theory based on Equation~(\ref{ExpDistFun}) thus emerges. Therefore, this model may alleviate not only the growth tension but also the Hubble tension, since it also deviates from GR  at the background level {\cite{NLicecube}}. 
Further analysis of the exponential DW model should be carried out accordingly.

\subsection{The Ricci-Transverse Expansion History}

For what concerns the RT model, the most updated cosmological analysis is due to Belgacem {et al.} \cite{Belgacem_2020}. As we saw in Section~\ref{S8}, different versions of the non-local model, relying on different choices for initial conditions of the auxiliary fields, have been compared against Planck--CMB, Pantheon SNIa and SDSS-BAO data. Since the RT model has no freedom with regard to the functional form of the action, the theory cannot be adjusted to data, and no background equivalence to the $\Lambda$CDM cosmology can be established a priori. 
In such a scenario, the solution to the Hubble tension may thus be possible. However, the estimator tool defined in Equation~(\ref{Geff}), which accounts for the deviation from GR of the non-local theory, only shows little discrepancies. This manifests in the values of the cosmological parameters inferred via Markov Chain Monte Carlo (MCMC), which are almost equivalent to those of the concordance model. The only model that exhibits some slight discrepancies is the so-called "RT-minimal", which relies on the assumption that the auxiliary scalar field $U(x)$ starts its evolution during the radiation dominance era. On the one hand, this model predict a non-vanishing value for the sum of neutrino masses, while the $\Lambda$CDM model and the other versions of the RT model show a marginalized posterior which is peaked in zero. On the other hand, the RT-minimal model provides a barely higher estimate of the Hubble constant, i.e., $H_0 = 68.74^{+0.59}_{-0.51}$ {km s}$^{-1}$Mpc$^{-1}$. Accordingly, the Hubble tension is just reduced to $\sim$4$\sigma$, as shown in Figure~\ref{FigureBarsH0}.

The RT model in its minimal setup thus provides a viable mechanism to account for the late-time cosmic acceleration, as well as for the inclusion of non-zero neutrino masses. However, the predictions for both the background evolution and the linear perturbations are too similar to that of the $\Lambda$CDM model, hence the cosmological tensions cannot be alleviated.


\section{Astrophysical Tests of  Non-Local Gravity}\label{AstroTest}

As we saw, good cosmological behavior emerges for both of the non-local models, thus enabling the possibility to avoid the introduction of any form of unknown dark energy. In order to further investigate the viability of such models as alternatives to GR, it is then necessary to test the non-local predictions down to astrophysical scales. Moreover, an accurate investigation of the possible screening mechanisms should be performed, if necessary. 

\subsection{Testing the Deser--Woodard Model by Galaxy Clusters, Elliptical Galaxies and the S2 Star}

The DW model has been tested on a wide range of astrophysical scales, from the galaxy clusters \cite{Bouche:2022jts} to the stellar orbits around Sagittarius A* \cite{Dialektopoulos_2018iph}. Most of the tests have been carried out for the exponential form of the non-local model, namely

\begin{equation}
    S = \frac{1}{2\kappa} \int d^{4}x \sqrt{-g} \Big[ R \big( 2 + e^{\,\eta} \big) - \partial_{\mu}\xi \partial^{\mu}\eta - \xi R \Big] \, ,
\end{equation}
where the distortion function has been picked out by exploiting the Noether Symmetry Approach \cite{Dialektopoulos_2018_NS}. The analyses  presented are all performed in the post-Newtonian limit, hence the non-local gravitational and metric potential are 

\begin{equation}\label{PhiNL}
    \begin{aligned}
     \phi(r) =& - \frac{GM\eta_{c}}{r} + \frac{G^{2}M^{2}}{2c^{2}r^{2}} \Bigg[ \frac{14}{9}\eta_{c}^{2} + \frac{18 r_{\xi} - 11 r_{\eta}}{6 r_{\eta} r_{\xi}} \, r \Bigg] - \frac{G^{3}M^{3}}{2c^{4}r^{3}} \Bigg[ \frac{50 r_{\xi} - 7 r_{\eta}}{12 r_{\eta} r_{\xi}} \, \eta_{c} r + \frac{16}{27} \eta_{c}^{3} - \frac{2 r_{\xi}^{2} - r_{\eta}^{2}}{r_{\eta}^{2} r_{\xi}^{2}} \, r^{2} \Bigg] \, ,
\end{aligned}
\end{equation}

\begin{equation}\label{PsiNL}
     \psi(r) = - \frac{GM\eta_{c}}{3r} - \frac{G^{2}M^{2}}{2c^{2}r^{2}} \Bigg[ \frac{2}{9}\eta_{c}^{2} + \frac{3 r_{\eta} - 2 r_{\xi}}{2 r_{\eta} r_{\xi}} \, r \Bigg] \, ,
\end{equation}
where $\eta_{c}$ is set to 1 so as to recover GR in the limit of $\phi(r)$. The two length scales $r_{\eta}$ and $r_{\xi}$ are the characteristic non-local parameters that define the scale at which the non-local gravity corrections become effective.

The first test of this form of the DW model has been carried out in \cite{Dialektopoulos_2018iph}, where the weak field non-local predictions have been compared against the NTT/VLT observations of S2 star orbit \cite{Gillessen_2009}. Exploiting a modified Marquardt--Levenberg algorithm, the fit between the simulated orbit and the observed one has shown a slightly better agreement for the non-local model with respect to the Keplerian orbit. Moreover, some constraints have been set on the non-local length scales.

A further test has been subsequently performed in \cite{Bouche:2022jts}, exploiting the CLASH lensing data from 19 massive clusters \cite{Postman_2012,Umetsu_2016}. The point-mass potentials of Equations~(\ref{PhiNL}) and (\ref{PsiNL}) were extended to a spherically symmetric mass distribution, i.e., the Navarro--Frenk--White density profile, and the non-local predictions for the lensing convergence were achieved. Therefore, the MCMC analysis has highlighted two effective regimes in which the non-local model is able to match the observations at the same level of statistical significance as GR. In the high-value limit of the non-local parameters, the non-local model reduces to a GR-like theory, whose lensing strength is $2/3$ of the standard one. In this scenario, the DW theory is thus able to fit the data at the cost of increased cluster mass estimates. On the other hand, approaching the low-value limit of the non-local length scales, the non-local corrections to the lensing potential become larger and comparable to the zeroth-order terms. In this regime, the non-local model is able to reproduce GR phenomenology, neither affecting the mass estimates nor the statistical viability of the model. Furthermore, when the non-local contributions becomes completely dominant, the non-local theory seems to be able to fit the lensing observations with extremely low cluster masses. Accordingly,  an intriguing possibility to fit data with no dark matter emerges. Additional constraints on the non-local parameters were also derived. 

The most recent astrophysical test of the exponential-DW model was carried out in \cite{BORKA2022}, using the velocity distribution of elliptical galaxies \cite{Burstein_1997}. Computing the non-local velocity dispersion as a function of the galaxy effective radius, the empirical relation of the so-called Fundamental Plane has been recovered so as to constrain the non-local gravity parameters. The results of the fit highlight the possibility to recover the fundamental plane without the dark matter hypothesis, setting new constraints for $r_{\eta}$ and $r_{\xi}$.

It is worth noticing, however, that the non-local Deser--Woodard model exhibits worrisome features at the scale of the Solar System. Indeed, in \cite{Belgacem_2019_LLR}, it was demonstrated that the screening mechanism proposed by the same authors of the non-local model does not work. As a consequence, the DW model would show a time dependence of the effective Newton constant in the small-scale limit, and it would be ruled out by Lunar Laser Ranging (LLR) observations. This conclusion, however, seems to be too strong, since it is still not clear how an FLRW background quantity behaves when evaluated from cosmological scales down to Solar System ones, where the system decouples from the Hubble flow. In fact, a full non-linear time- and scale-dependent solution around a non-linear structure would be necessary. A number of proposals go in this direction, and the so-called Vainshtein mechanism \cite{VAINSHTEIN1972393} can be regarded as the paradigm to realize the screening. Basically, any screening mechanism requires a scalar field coupled to matter and mediating a fifth force which might span from Solar System up to cosmological scales. Since non-local terms can be “localized”, thus resulting in effective scalar fields depending on the scale, some screening mechanism could naturally emerge so as to solve the above reported problems.

\subsection{Testing the Ricci-Transverse Model by Solar System Observations and Gravitational Waves Detection}

The main astrophysical tests of the RT model are related to Solar System observations. As we saw in the previous sub-section, any theory that extends GR has to reduce to Einstein's theory at small scales. However, this is highly non-trivial when additional degrees of freedom are included, and screening mechanisms involving non-linear features are required. The RT model, instead, smoothly reduces to GR already at linear level, and no vDVZ discontinuity arises when $m \rightarrow 0$ \cite{Belgacem_2020}. Note that such results are valid both in the flat and the Schwarzschild spacetime. Moreover, the non-local model passes the LLR test about the time variation of the effective Newton constant \cite{Belgacem_2019_LLR}. As stated in Equation~(\ref{Geff}), indeed, the deviation parameter $G_{eff}$ reduces to $G_{N}$ as soon as the system's characteristic scale decreases.

Another non-local feature of the RT model that has been investigated is the deviation from GR of the predicted gravitational radiation \cite{Belgacem_2018_DW1,Belgacem_2018_DW2}. The RT model, similarly to some other {extended theories of gravity} such as $f(R)$ gravity and DHOST theories, has survived  the GW170817 event, which set a stringent constraint on the Gravitational Waves (GW) speed \cite{Abbott_2017_GW}. 
Indeed, this non-local model only modifies the friction term in the GW propagation equation, thus predicting a massless graviton. Moreover, neither the coupling with matter nor the gravitational interaction between the coalescing binaries are affected (the RT model reduces to GR at short distances), and the only difference will  therefore be due to the free propagation of the GW from the source to the observer. The GW amplitude indeed undergoes a modified dampening in the non-local model

\begin{equation}\label{GWnonlocal}
    \Tilde{h}_{A}(\eta , \mathbf{k}) \sim \frac{1}{d_{L}^{gw}(z)} = \Bigg[ d_{L}^{em}(z) \exp{\bigg\{  -\int_{0}^{z} \frac{dz'}{1+z'} \delta(z') \bigg\}} \Bigg]^{-1} \, ,
\end{equation}
where

\begin{equation}
    \delta(\eta) = \frac{m^2 \bar{\mathcal{S}_0}(\eta)}{6H(\eta)} \, ,
\end{equation}
and $\Tilde{h}_{A}(\eta , \mathbf{k})$ are the Fourier modes of the GW amplitude, with $A=\times,+$ labeling the polarization. Computing the ratio between the non-local behavior given by Equation~(\ref{GWnonlocal}) and the GR behavior, $\Tilde{h}_{A}(\eta , \mathbf{k}) \sim 1/d_{L}^{em}(z)$, little deviation emerges for the RT-minimal model, while a $20-80\%$ deviation manifests at large $z$ for the RT formulations in which the auxiliary fields start their evolution during the de Sitter inflation. The more e-folds we consider between the onset of the auxiliary fields and the end of the inflation, the greater  the deviation from GR. We can use a simple parametrization for the considered ratio

\begin{equation}
    \frac{d_{L}^{gw}(z)}{d_{L}^{em}(z)} = \Xi_0 + \frac{1 - \Xi_0}{(1+z)^n} \, ,
\end{equation}
where $\Xi_0$ is the asymptotic value reached by the ratio and $n$ is the rate at which $\Xi_0$ is approached. Then,

\begin{equation}
    \delta(z) = \frac{\delta(0)}{1 - \Xi_0 + \Xi_0 (1+z)^n} \, ,
\end{equation}
with $\delta(0) = n(1- \Xi_0)$, and the event GW170817 provided the following constraint for such a parameter \cite{Belgacem_2018_DW2}: $\delta(0) = -7.8^{+9.7}_{-18.4}$. More stringent constraints will  certainly be set with the next generation of GW detectors by exploiting the observations of GW events with electromagnetic counterparts. 


\section{Conclusions and Perspectives}\label{Conclusions}

In this paper, we  reviewed the cosmological properties of two of the main proposals in the framework of the non-locally {extended} theories of gravity. In particular, we  considered two metric IKGs that are inspired by quantum corrections and manifest a suitable cosmological behavior as well. Both the DW and  RT models are able to reproduce the expansion history of the Universe, exhibiting a late-time accelerated expansion driven by the onset of the non-local corrections. The non-local extensions of the Hilbert--Einstein Lagrangian thus provide a viable mechanism to avoid the introduction of any form of unknown dark energy. Building on these appealing properties, we  inquired into the chance of addressing the two main cosmological tensions, namely the $\sigma_8$ and  $H_0$ tensions. 

On the one hand, the non-local DW model has shown suitable features towards this aim. The phenomenological formulation of the model indeed predicts a lowered amplitude of growth of perturbations, therefore solving the $\sigma_8$ tension. However, this model is made equivalent to the $\Lambda$CDM cosmology at the background level, hence no chance to account for the Hubble tension arises. Another formulation of the DW theory, based on the Noether symmetries of the system, may address both the tensions. This model lacks  a proper cosmological analysis, but the investigation of its lensing properties at the galaxy clusters scale has shown the same features that, in the phenomenological DW model, allow the weakening of the growth of structures. Moreover, this formulation of the non-local theory deviates from GR also at background level, thus enabling the possibility to alleviate the Hubble tension as well. The model has been also tested on astrophysical scales, and substantial statistical equivalence to GR has emerged in very different systems, such as the S2 star, the elliptical galaxies and the galaxy clusters. The main drawback of the DW model, however, is the absence of an effective screening mechanism on small scales, which has to be further investigated.

On the other hand, the non-local RT model perfectly reduces to GR at the Solar System scale, thus avoiding the necessity of non-trivial screening mechanisms. Accordingly, the model is not ruled out by the LLR test. Moreover, the non-minimal formulations of the RT model show a strong deviation from GR for what concerns the GW propagation at large redshift. A powerful tool to test the model with the next generation of GW detectors thus emerges. However, this non-local model is not able to address any of the cosmological tensions, as it mimics the $\Lambda$CDM evolution both at the background and linear perturbations level.

In  view of the fact that the next generation of cosmological surveys are expected to provide sufficiently accurate data to reach a turning point in our comprehension of the Universe, it is of great interest to further investigate the main alternatives to GR. A complete cosmological analysis should  especially be carried out for the non-local DW model in its formulation based on the Noether Symmetry Approach. This model indeed provides one of the most promising windows towards the solution of both the cosmological tensions and the dark energy problem. The large-scale structure especially appears as a privileged environment for testing the non-local models, since one of their main features is the emergence of characteristic length scales. However, it must be stressed that as long as no screening mechanism will be found for the DW model, its reliability will be compromised.




\acknowledgments{
This article is based upon work from COST Action CA21136 Addressing
observational tensions in cosmology with systematic and fundamental
physics (CosmoVerse) supported by COST (European Cooperation in Science and
Technology). FB and SC acknowledge the support of {\it Istituto Nazionale di Fisica Nucleare} (INFN), {\it iniziative specifiche} QGSKY and MOONLIGHT2.}

\appendix

\section{Abbreviations}

The following abbreviations are used in this manuscript:\\
\noindent
\begin{tabular}{@{}ll}
BBN & Big Bang Nucleosynthesis\\
BAO & Baryon Acoustic Oscillations\\
$\Lambda$CDM & Lambda Cold Dark Matter\\
GR & General Relativity\\
DE & Dark Energy\\
DM & Dark Matter\\
CMB & Cosmic Microwave Background\\
WL & Weak Lensing\\
CC & Cluster Counts\\
RSD & Redshift Space Distortion\\
DW & Deser--Woodard\\
RT & Ricci-Transverse\\
QFT & Quantum Field Theory\\
IDG & Infinite Derivative Theory of Gravity\\
IKG & Integral Kernel Theory of Gravity\\
UV & UltraViolet\\
IR & InfraRed\\
SNIa & Type Ia Supernovae\\
FLRW & Friedmann--Lemaitre--Robertson--Walker\\
MCMC & Markov Chain Monte Carlo\\
LLR & Lunar Laser Ranging\\
GW & Gravitational Waves
\end{tabular}




\begin{thebibliography}{999}

\bibitem[Hogg \em{et~al.}(2005)Hogg, Eisenstein, Blanton, Bahcall, Brinkmann,
  Gunn, and Schneider]{Hogg:2004vw}
Hogg, D.W.; Eisenstein, D.J.; Blanton, M.R.; Bahcall, N.A.; Brinkmann, J.;
  Gunn, J.E.; Schneider, D.P.
\newblock {Cosmic homogeneity demonstrated with luminous red galaxies}.
\newblock {\em Astrophys. J.} {\bf 2005}, {\em 624}, 54.
\newblock {\url{https://doi.org/10.1086/429084}}.

\bibitem[Labini \em{et~al.}(2009)Labini, Vasilyev, Pietronero, and
  Baryshev]{SylosLabini_2009}
Labini, F.S.; Vasilyev, N.L.; Pietronero, L.; Baryshev, Y.V.
\newblock Absence of self-averaging and of homogeneity in the large-scale
  galaxy distribution.
\newblock {\em Europhys. Lett.} {\bf 2009}, {\em 86}, 49001.
\newblock {\url{https://doi.org/10.1209/0295-5075/86/49001}}.

\bibitem[Riess \em{et~al.}(1998)Riess, Filippenko, Challis, Clocchiatti,
  Diercks, Garnavich, Gilliland, Hogan, Jha, Kirshner, Leibundgut, Phillips,
  Reiss, Schmidt, Schommer, Smith, Spyromilio, Stubbs, Suntzeff, and
  Tonry]{Riess_1998}
Riess, A.G.; Filippenko, A.V.; Challis, P.; Clocchiatti, A.; Diercks, A.;
  Garnavich, P.M.; Gilliland, R.L.; Hogan, C.J.; Jha, S.; Kirshner, R.P.;
  et~al.
\newblock Observational Evidence from Supernovae for an Accelerating Universe
  and a Cosmological Constant.
\newblock {\em  Astron. J.} {\bf 1998}, {\em 116}, 1009.
\newblock {\url{https://doi.org/10.1086/300499}}.

\bibitem[Perlmutter \em{et~al.}(1999)Perlmutter, Aldering, Goldhaber, Knop,
  Nugent, Castro, Deustua, Fabbro, Goobar, Groom, Hook, Kim, Kim, Lee, Nunes,
  Pain, Pennypacker, Quimby, Lidman, Ellis, Irwin, McMahon, Ruiz-Lapuente,
  Walton, Schaefer, Boyle, Filippenko, Matheson, Fruchter, Panagia, Newberg,
  Couch, and Project]{Perlmutter_1999}
Perlmutter, S.; Aldering, G.; Goldhaber, G.; Knop, R.A.; Nugent, P.; Castro,
  P.G.; Deustua, S.; Fabbro, S.; Goobar, A.; Groom, D.E.;  et~al.
\newblock Measurements of $Omega$ and $\Lambda$ from 42 High-Redshift
  Supernovae.
\newblock {\em  Astrophys. J.} {\bf 1999}, {\em 517}, 565.
\newblock {\url{https://doi.org/10.1086/307221}}.

\bibitem[Aghanim \em{et~al.}(2020)Aghanim et~al.]{Planck:2018vyg}
{Aghanim, N.;} Akrami, Y.; Ashdown, M.; Aumont, J.; Baccigalupi, C.; Ballardini, M.; Banday, A.J.; Barreiro, R.B.; Bartolo, N.; Basak, S.; et~al.
\newblock {Planck 2018 results. VI. Cosmological parameters}.
\newblock {\em Astron. Astrophys.} {\bf 2020}, {\em 641},~A6; Erratum in \emph{Astron. Astrophys.} \textbf{2021}, \emph{652}, C4. 
  {\url{https://doi.org/10.1051/0004-6361/201833910}}.

\bibitem[Alam \em{et~al.}(2021)Alam, Aubert, Avila, Balland, Bautista,
  Bershady, Bizyaev, Blanton, Bolton, Bovy, Brinkmann, Brownstein, Burtin,
  Chabanier, Chapman, Choi, Chuang, Comparat, Cousinou, Cuceu, Dawson, de~la
  Torre, de~Mattia, de~Sainte~Agathe, du~Mas~des Bourboux, Escoffier,
  Etourneau, Farr, Font-Ribera, Frinchaboy, Fromenteau, Gil-Mar{\'{\i}}n, Goff,
  Gonzalez-Morales, Gonzalez-Perez, Grabowski, Guy, Hawken, Hou, Kong, Parker,
  Klaene, Kneib, Lin, Long, Lyke, de~la Macorra, Martini, Masters, Mohammad,
  Moon, Mueller, Mu{\~{n}}oz-Guti{\'{e}}rrez, Myers, Nadathur, Neveux, Newman,
  Noterdaeme, Oravetz, Oravetz, Palanque-Delabrouille, Pan, Paviot, Percival,
  P{\'{e}}rez-R{\`{a}}fols, Petitjean, Pieri, Prakash, Raichoor, Ravoux,
  Rezaie, Rich, Ross, Rossi, Ruggeri, Ruhlmann-Kleider, S{\'{a}}nchez,
  S{\'{a}}nchez, S{\'{a}}nchez-Gallego, Sayres, Schneider, Seo, Shafieloo,
  Slosar, Smith, Stermer, Tamone, Tinker, Tojeiro, Vargas-Maga{\~{n}}a, Variu,
  Wang, Weaver, Weijmans, Y{\`{e}}che, Zarrouk, Zhao, Zhao, and
  Zheng]{Alam_2021}
Alam, S.; Aubert, M.; Avila, S.; Balland, C.; Bautista, J.E.; Bershady, M.A.;
  Bizyaev, D.; Blanton, M.R.; Bolton, A.S.; Bovy, J.;  et~al.
\newblock Completed {SDSS}-{IV} extended Baryon Oscillation Spectroscopic
  Survey: Cosmological implications from two decades of spectroscopic surveys
  at the Apache Point Observatory.
\newblock {\em Phys. Rev. D} {\bf 2021}, {\em 103}, 083533.
\newblock {\url{https://doi.org/10.1103/physrevd.103.083533}}.

\bibitem[Abbott \em{et~al.}(2022)Abbott, Aguena, Alarcon, Allam, Alves, Amon,
  Andrade-Oliveira, Annis, Avila, Bacon, Baxter, Bechtol, Becker, Bernstein,
  Bhargava, Birrer, Blazek, Brandao-Souza, Bridle, Brooks, Buckley-Geer, Burke,
  Camacho, Campos, Rosell, Kind, Carretero, Castander, Cawthon, Chang, Chen,
  Chen, Choi, Conselice, Cordero, Costanzi, Crocce, da~Costa, da~Silva~Pereira,
  Davis, Davis, Vicente, DeRose, Desai, Valentino, Diehl, Dietrich, Dodelson,
  Doel, Doux, Drlica-Wagner, Eckert, Eifler, Elsner, Elvin-Poole, Everett,
  Evrard, Fang, Farahi, Fernandez, Ferrero, Fert{\'{e}}, Fosalba, Friedrich,
  Frieman, Garc{\'{\i}}a-Bellido, Gatti, Gaztanaga, Gerdes, Giannantonio,
  Giannini, Gruen, Gruendl, Gschwend, Gutierrez, Harrison, Hartley, Herner,
  Hinton, Hollowood, Honscheid, Hoyle, Huff, Huterer, Jain, James, Jarvis,
  Jeffrey, Jeltema, Kovacs, Krause, Kron, Kuehn, Kuropatkin, Lahav, Leget,
  Lemos, Liddle, Lidman, Lima, Lin, MacCrann, Maia, Marshall, Martini,
  McCullough, Melchior, Mena-Fern{\'{a}}ndez, Menanteau, Miquel, Mohr, Morgan,
  Muir, Myles, Nadathur, Navarro-Alsina, Nichol, Ogando, Omori, Palmese,
  Pandey, Park, Paz-Chinch{\'{o}}n, Petravick, Pieres, Malag{\'{o}}n, Porredon,
  Prat, Raveri, Rodriguez-Monroy, Rollins, Romer, Roodman, Rosenfeld, Ross,
  Rykoff, Samuroff, S{\'{a}}nchez, Sanchez, Sanchez, Cid, Scarpine, Schubnell,
  Scolnic, Secco, Serrano, Sevilla-Noarbe, Sheldon, Shin, Smith, Soares-Santos,
  Suchyta, Swanson, Tabbutt, Tarle, Thomas, To, Troja, Troxel, Tucker,
  Tutusaus, Varga, Walker, Weaverdyck, Wechsler, Weller, Yanny, Yin, Zhang, and
  and]{Abbott_2022}
Abbott, T.; Aguena, M.; Alarcon, A.; Allam, S.; Alves, O.; Amon, A.;
  Andrade-Oliveira, F.; Annis, J.; Avila, S.; Bacon, D.;  et~al.
\newblock Dark Energy Survey Year 3 results: Cosmological constraints from
  galaxy clustering and weak lensing.
\newblock {\em Phys. Rev. D} {\bf 2022}, {\em 105}, 023520.
\newblock {\url{https://doi.org/10.1103/physrevd.105.023520}}.

\bibitem[Bull \em{et~al.}(2016)Bull, Akrami, Adamek, Baker, Bellini,
  Jim{\'{e}}nez, Bentivegna, Camera, Clesse, Davis, Dio, Enander, Heavens,
  Heisenberg, Hu, Llinares, Maartens, Mörtsell, Nadathur, Noller, Pasechnik,
  Pawlowski, Pereira, Quartin, Ricciardone, Riemer-S{\o}rensen, Rinaldi,
  Sakstein, Saltas, Salzano, Sawicki, Solomon, Spolyar, Starkman, Steer,
  Tereno, Verde, Villaescusa-Navarro, von Strauss, and Winther]{Bull_2016}
Bull, P.; Akrami, Y.; Adamek, J.; Baker, T.; Bellini, E.; Jim{\'{e}}nez, J.B.;
  Bentivegna, E.; Camera, S.; Clesse, S.; Davis, J.H.;  et~al.
\newblock Beyond $\Lambda$CDM: Problems, solutions, and the road ahead.
\newblock {\em Phys. Dark Universe} {\bf 2016}, {\em 12}, 56--99.
\newblock {\url{https://doi.org/10.1016/j.dark.2016.02.001}}.

\bibitem[Perivolaropoulos and Skara(2022)]{Perivolaropoulos_2022}
Perivolaropoulos, L.; Skara, F.
\newblock Challenges for $\Lambda$CDM: An update.
\newblock {\em New Astron. Rev.} {\bf 2022}, {\em 95},~101659.
\newblock {\url{https://doi.org/10.1016/j.newar.2022.101659}}.

\bibitem[{Di Valentino} \em{et~al.}(2021{\natexlab{a}}){Di Valentino},
  Anchordoqui, Özgür Akarsu, Ali-Haimoud, Amendola, Arendse, Asgari,
  Ballardini, Basilakos, Battistelli, Benetti, Birrer, Bouchet, Bruni,
  Calabrese, Camarena, Capozziello, Chen, Chluba, Chudaykin, Colg{\'{a}}in,
  Cyr-Racine, de~Bernardis, de~Cruz~P{\'{e}}rez, Delabrouille, Dunkley,
  Escamilla-Rivera, Fert{\'{e}}, Finelli, Freedman, Frusciante, Giusarma,
  G{\'{o}}mez-Valent, Guy, Handley, Harrison, Hart, Heavens, Hildebrandt, Holz,
  Huterer, Ivanov, Joudaki, Kamionkowski, Karwal, Knox, Kumar, Lamagna,
  Lesgourgues, Lucca, Marra, Masi, Matarrese, Mazumdar, Melchiorri, Mena,
  Mersini-Houghton, Miranda, Moreno-Pulido, Mota, Muir, Mukherjee, Niedermann,
  Notari, Nunes, Pace, Paliathanasis, Palmese, Pan, Paoletti, Pettorino,
  Piacentini, Poulin, Raveri, Riess, Salzano, Saridakis, Sen, Shafieloo,
  Shajib, Silk, Silvestri, Sloth, Smith, Peracaula, van~de Bruck, Verde,
  Visinelli, Wandelt, Wang, Wang, Yadav, and Yang]{Cosmology_intertwined_II}
{Di Valentino}, E.; Anchordoqui, L.A.; Özgür Akarsu.; Ali-Haimoud, Y.;
  Amendola, L.; Arendse, N.; Asgari, M.; Ballardini, M.; Basilakos, S.;
  Battistelli, E.;  et~al.
\newblock Snowmass2021---Letter of interest cosmology intertwined {II}: The
  hubble constant tension.
\newblock {\em Astropart. Phys.} {\bf 2021}, {\em 131}, 102605.
\newblock {\url{https://doi.org/10.1016/j.astropartphys.2021.102605}}.

\bibitem[{Di Valentino} \em{et~al.}(2021{\natexlab{b}}){Di Valentino},
  Anchordoqui, Özgür Akarsu, Ali-Haimoud, Amendola, Arendse, Asgari,
  Ballardini, Basilakos, Battistelli, Benetti, Birrer, Bouchet, Bruni,
  Calabrese, Camarena, Capozziello, Chen, Chluba, Chudaykin, Colg{\'{a}}in,
  Cyr-Racine, de~Bernardis, de~Cruz~P{\'{e}}rez, Delabrouille, Dunkley,
  Escamilla-Rivera, Fert{\'{e}}, Finelli, Freedman, Frusciante, Giusarma,
  G{\'{o}}mez-Valent, Handley, Harrison, Hart, Heavens, Hildebrandt, Holz,
  Huterer, Ivanov, Joudaki, Kamionkowski, Karwal, Knox, Kumar, Lamagna,
  Lesgourgues, Lucca, Marra, Masi, Matarrese, Mazumdar, Melchiorri, Mena,
  Mersini-Houghton, Miranda, Moreno-Pulido, Mota, Muir, Mukherjee, Niedermann,
  Notari, Nunes, Pace, Paliathanasis, Palmese, Pan, Paoletti, Pettorino,
  Piacentini, Poulin, Raveri, Riess, Salzano, Saridakis, Sen, Shafieloo,
  Shajib, Silk, Silvestri, Sloth, Smith, Peracaula, van~de Bruck, Verde,
  Visinelli, Wandelt, Wang, Wang, Yadav, and Yang]{Cosmology_intertwined_III}
{Di Valentino}, E.; Anchordoqui, L.A.; Özgür Akarsu.; Ali-Haimoud, Y.;
  Amendola, L.; Arendse, N.; Asgari, M.; Ballardini, M.; Basilakos, S.;
  Battistelli, E.;  et~al. Cosmology intertwined {III}: $f\sigma_{8}$ and $S_{8}$. {\em Astropart. Phys.} {\bf 2021}, {\em 131}, 102604. {\url{https://doi.org/10.1016/j.astropartphys.2021.102604}}.

\bibitem[Bertone \em{et~al.}(2005)Bertone, Hooper, and Silk]{Bertone_2005}
Bertone, G.; Hooper, D.; Silk, J.
\newblock Particle dark matter: Evidence, candidates and constraints.
\newblock {\em Phys. Rep.} {\bf 2005}, {\em 405}, 279--390.
\newblock {\url{https://doi.org/10.1016/j.physrep.2004.08.031}}.

\bibitem[Batista \em{et~al.}(2021)Batista, Amin, Barenboim, Bartolo, Baumann,
  Bauswein, Bellini, Benisty, Bertone, Blasi, Böhmer, Bošnjak, Bringmann,
  Burrage, Bustamante, Bustillo, Byrnes, Calore, Catena, Cerdeño, Cerri,
  Chianese, Clough, Cole, Coloma, Coogan, Covi, Cutting, Davis, de~Rham,
  di~Matteo, Domènech, Drewes, Dietrich, Edwards, Esteban, Erdem, Evoli,
  Fasiello, Feeney, Ferreira, Fialkov, Fornengo, Gabici, Galatyuk, Gaggero,
  Grasso, Guépin, Harz, Herrero-Valea, Hinderer, Hogg, Hooper, Iocco, Isern,
  Karchev, Kavanagh, Korsmeier, Kotera, Koyama, Krishnan, Lesgourgues, Said,
  Lombriser, Lorenz, Manconi, Mapelli, Marcowith, Markoff, Marsh, Martinelli,
  Martins, Matthews, Meli, Mena, Mifsud, Bertolami, Millington, Moesta, Nippel,
  Niro, O'Connor, Oikonomou, Paganini, Pagliaroli, Pani, Pfrommer, Pascoli,
  Pinol, Pizzuti, Porto, Pound, Quevedo, Raffelt, Raccanelli, Ramirez-Ruiz,
  Raveri, Renaux-Petel, Ricciardone, Khalifeh, Riotto, Roiban, Rubio, Sahlén,
  Sabti, Sagunski, Šarčević, Schmitz, Schwaller, Schwetz, Sedrakian,
  Sellentin, Serenelli, Serpico, Sfakianakis, Shalgar, Silvestri, Tamborra,
  Tanidis, Teresi, Tokareva, Tolos, Trojanowski, Trotta, Uhlemann, Urban,
  Vernizzi, van Vliet, Villante, Vincent, Vink, Vitagliano, Weniger,
  Wickenbrock, Winter, Zell, and
  Zeng]{https://doi.org/10.48550/arxiv.2110.10074}
Batista, R.A.; Amin, M.A.; Barenboim, G.; Bartolo, N.; Baumann, D.; Bauswein,
  A.; Bellini, E.; Benisty, D.; Bertone, G.; Blasi, P.;  et~al.
\newblock EuCAPT White Paper: Opportunities and Challenges for Theoretical
  Astroparticle Physics in the Next Decade. \emph{arXiv Prepr.} \textbf{2021}, arXiv:2110.10074.
\newblock {\url{https://doi.org/10.48550/ARXIV.2110.10074}}.

\bibitem[Copeland \em{et~al.}(2006)Copeland, Sami, and
  Tsujikawa]{Copeland2006wr}
Copeland, E.J.; Sami, M.; Tsujikawa, S.
\newblock {Dynamics of dark energy}.
\newblock {\em Int. J. Mod. Phys. D} {\bf 2006}, {\em 15}, 1753--1935.
\newblock {\url{https://doi.org/10.1142/S021827180600942X}}.

\bibitem[Padmanabhan(2003)]{Padmanabhan2002ji}
Padmanabhan, T.
\newblock {Cosmological constant: The Weight of the vacuum}.
\newblock {\em Phys. Rept.} {\bf 2003}, {\em 380}, 235--320.
\newblock {\url{https://doi.org/10.1016/S0370-1573(03)00120-0}}.

\bibitem[Nojiri and Odintsov(2006)]{Nojiri_2006ri}
Nojiri, S.; Odintsov, S.D.
\newblock {Introduction to modified gravity and gravitational alternative for
  dark energy}.
\newblock \emph{{Int. J. Geom. Methods Mod. Phys.}} {\bf 2007}, {\em 4}, 115--145.
~\newblock {\url{https://doi.org/10.1142/S0219887807001928}}.

\bibitem[Riess \em{et~al.}(2022)Riess, Yuan, Macri, Scolnic, Brout, Casertano,
  Jones, Murakami, Anand, Breuval, Brink, Filippenko, Hoffmann, Jha, Kenworthy,
  Mackenty, Stahl, and Zheng]{Riess_2022}
Riess, A.G.; Yuan, W.; Macri, L.M.; Scolnic, D.; Brout, D.; Casertano, S.;
  Jones, D.O.; Murakami, Y.; Anand, G.S.; Breuval, L.;  et~al.
\newblock A Comprehensive Measurement of the Local Value of the Hubble Constant
  with 1 km s$^{-1}$ Mpc$^{-1}$ Uncertainty from the Hubble Space Telescope and
  the SH0ES Team.
\newblock {\em  Astrophys. J. Lett.} {\bf 2022}, {\em 934}, L7.
\newblock {\url{https://doi.org/10.3847/2041-8213/ac5c5b}}.

\bibitem[Asgari \em{et~al.}(2021)Asgari, Lin, Joachimi, Giblin, Heymans,
  Hildebrandt, Kannawadi, Stölzner, Tröster, van~den Busch, Wright, Bilicki,
  Blake, de~Jong, Dvornik, Erben, Getman, Hoekstra, Köhlinger, Kuijken,
  Miller, Radovich, Schneider, Shan, and Valentijn]{Asgari_2021}
Asgari, M.; Lin, C.A.; Joachimi, B.; Giblin, B.; Heymans, C.; Hildebrandt, H.;
  Kannawadi, A.; Stölzner, B.; Tröster, T.; van~den Busch, J.L.;  et~al.
\newblock {KiDS}-1000 cosmology: Cosmic shear constraints and comparison
  between two point statistics.
\newblock {\em Astron. Astrophys.} {\bf 2021}, {\em 645},  A104.
\newblock {\url{https://doi.org/10.1051/0004-6361/202039070}}.

\bibitem[Lesci \em{et~al.}(2022)Lesci, Marulli, Moscardini, Sereno,
  Veropalumbo, Maturi, Giocoli, Radovich, Bellagamba, Roncarelli, Bardelli,
  Contarini, Covone, Ingoglia, Nanni, and Puddu]{Lesci_2022}
Lesci, G.F.; Marulli, F.; Moscardini, L.; Sereno, M.; Veropalumbo, A.; Maturi,
  M.; Giocoli, C.; Radovich, M.; Bellagamba, F.; Roncarelli, M.;  et~al.
\newblock {AMICO} galaxy clusters in {KiDS}-{DR}3: Cosmological constraints
  from counts and stacked weak lensing.
\newblock {\em Astron. Astrophys.} {\bf 2022}, {\em 659}, A88.
\newblock {\url{https://doi.org/10.1051/0004-6361/202040194}}.

\bibitem[Chen \em{et~al.}(2022)Chen, Vlah, and White]{Chen_2022}
Chen, S.F.; Vlah, Z.; White, M.
\newblock A new analysis of galaxy 2-point functions in the {BOSS} survey,
  including full-shape information and post-reconstruction {BAO}.
\newblock {\em J. Cosmol. Astropart. Phys.} {\bf 2022}, {\em
  2022}, 008.
\newblock {\url{https://doi.org/10.1088/1475-7516/2022/02/008}}.

\bibitem[Capozziello(2002)]{CAPOZZIELLO_2002}
Capozziello, S.
\newblock {Curvature} {Quintessence}.
\newblock {\em Int. J. Mod. Phys. D} {\bf 2002}, {\em 11}, 483--491.
\newblock {\url{https://doi.org/10.1142/s0218271802002025}}.

\bibitem[Nojiri and Odintsov(2006)]{Nojiri_2006}
Nojiri, S.; Odintsov, S.D.
\newblock Modified $f(R)$ gravity consistent with realistic cosmology: From a
  matter dominated epoch to a dark energy universe.
\newblock {\em Phys. Rev. D} {\bf 2006}, {\em 74}, 086005.
\newblock {\url{https://doi.org/10.1103/physrevd.74.086005}}.

\bibitem[Starobinsky(2007)]{Starobinsky_2007}
Starobinsky, A.A.
\newblock Disappearing cosmological constant in $f(R)$ gravity.
\newblock {\em {JETP} Lett.} {\bf 2007}, {\em 86}, 157--163.
\newblock {\url{https://doi.org/10.1134/s0021364007150027}}.

\bibitem[Fay \em{et~al.}(2007)Fay, Nesseris, and Perivolaropoulos]{Fay_2007uy}
Fay, S.; Nesseris, S.; Perivolaropoulos, L.
\newblock {Can $f(R)$ Modified Gravity Theories Mimic a $\Lambda$CDM Cosmology?}
\newblock {\em Phys. Rev. D} {\bf 2007}, {\em 76}, 063504.
\newblock {\url{https://doi.org/10.1103/PhysRevD.76.063504}}.

\bibitem[Lazkoz \em{et~al.}(2018)Lazkoz, Ortiz-Ba{\~{n}}os, and
  Salzano]{Lazkoz_2018}
Lazkoz, R.; Ortiz-Ba{\~{n}}os, M.; Salzano, V.
\newblock $f(R)$ gravity modifications: From the action to the data.
\newblock {\em  Eur. Phys. J. C} {\bf 2018}, {\em 78}, 213.
\newblock {\url{https://doi.org/10.1140/epjc/s10052-018-5711-6}}.

\bibitem[Bajardi \em{et~al.}(2022)Bajardi, D'Agostino, Benetti, {De Falco}, and
  Capozziello]{Bajardi_2022}
Bajardi, F.; D'Agostino, R.; Benetti, M.; {De Falco}, V.; Capozziello, S.
\newblock Early and late time cosmology: The $f(R)$ gravity perspective.
\newblock {\em  Eur. Phys. J. Plus} {\bf 2022}, {\em 137}, 1239.
\newblock {\url{https://doi.org/10.1140/epjp/s13360-022-03418-8}}.

\bibitem[D'Agostino and Nunes(2020)]{D_Agostino_2020_H0}
D'Agostino, R.; Nunes, R.C.
\newblock Measurements of $H_0$ in modified gravity theories: The role of
  lensed quasars in the late-time Universe.
\newblock {\em Phys. Rev. D} {\bf 2020}, {\em 101}, 103505.
\newblock {\url{https://doi.org/10.1103/physrevd.101.103505}}.

\bibitem[Nesseris \em{et~al.}(2013)Nesseris, Basilakos, Saridakis, and
  Perivolaropoulos]{Nesseris_2013jea}
Nesseris, S.; Basilakos, S.; Saridakis, E.N.; Perivolaropoulos, L.
\newblock {Viable $f(T)$ models are practically indistinguishable from
  $\Lambda$CDM}.
\newblock {\em Phys. Rev. D} {\bf 2013}, {\em 88}, 103010.
\newblock {\url{https://doi.org/10.1103/PhysRevD.88.103010}}.

\bibitem[Cai \em{et~al.}(2016)Cai, Capozziello, {De Laurentis}, and
  Saridakis]{Cai_2016}
Cai, Y.F.; Capozziello, S.; {De Laurentis}, M.; Saridakis, E.N.
\newblock $f(T)$ teleparallel gravity and cosmology.
\newblock {\em Rep. Prog. Phys.} {\bf 2016}, {\em 79}, 106901.
\newblock {\url{https://doi.org/10.1088/0034-4885/79/10/106901}}.

\bibitem[Golovnev and Koivisto(2018)]{Golovnev_2018wbh}
Golovnev, A.; Koivisto, T.
\newblock {Cosmological perturbations in modified teleparallel gravity models}.
\newblock {\em J. Cosmol. Astropart. Phys.} {\bf 2018}, {\em 11}, 012.
\newblock {\url{https://doi.org/10.1088/1475-7516/2018/11/012}}.

\bibitem[Wang and Mota(2020)]{Wang_Mota_2020}
Wang, D.; Mota, D.
\newblock Can $f(T)$ gravity resolve the $H_0$ tension?
\newblock {\em Phys. Rev. D} {\bf 2020}, {\em 102}, 063530.
\newblock {\url{https://doi.org/10.1103/physrevd.102.063530}}.

\bibitem[Bahamonde \em{et~al.}(2021)Bahamonde, Dialektopoulos,
  Escamilla-Rivera, Farrugia, Gakis, Hendry, Hohmann, Said, Mifsud, and
  Di~Valentino]{Bahamonde_2021gfp}
Bahamonde, S.; Dialektopoulos, K.F.; Escamilla-Rivera, C.; Farrugia, G.; Gakis,
  V.; Hendry, M.; Hohmann, M.; Said, J.L.; Mifsud, J.; Di~Valentino, E.
\newblock {Teleparallel Gravity: From Theory to Cosmology.}
\emph{{Rep. Prog. Phys.}} {\bf 2022}, doi:10.1088/1361-6633/ac9cef.

\bibitem[Nojiri and Odintsov(2005)]{Nojiri_2005jg}
Nojiri, S.; Odintsov, S.D.
\newblock {Modified Gauss-Bonnet theory as gravitational alternative for dark
  energy}.
\newblock {\em Phys. Lett. B} {\bf 2005}, {\em 631}, 1--6.
\newblock {\url{https://doi.org/10.1016/j.physletb.2005.10.010}}.

\bibitem[Li \em{et~al.}(2007)Li, Barrow, and Mota]{Li_2007jm}
Li, B.; Barrow, J.D.; Mota, D.F.
\newblock {The Cosmology of Modified Gauss-Bonnet Gravity}.
\newblock {\em Phys. Rev. D} {\bf 2007}, {\em 76}, 044027.
\newblock {\url{https://doi.org/10.1103/PhysRevD.76.044027}}.

\bibitem[Myrzakulov \em{et~al.}(2011)Myrzakulov, Saez-Gomez, and
  Tureanu]{Myrzakulov_2010gt}
Myrzakulov, R.; Saez-Gomez, D.; Tureanu, A.
\newblock {On the $\Lambda$CDM Universe in $f(G)$ gravity}.
\newblock {\em Gen. Relativ. Gravit.} {\bf 2011}, {\em 43}, 1671--1684.
\newblock {\url{https://doi.org/10.1007/s10714-011-1149-y}}.

\bibitem[Bajardi and Capozziello(2020)]{Bajardi_2020}
Bajardi, F.; Capozziello, S.
\newblock $f(\mathcal{G})$ Noether cosmology.
\newblock {\em  Eur. Phys. J. C} {\bf 2020}, {\em 80}, 704.
\newblock {\url{https://doi.org/10.1140/epjc/s10052-020-8258-2}}.

\bibitem[Capozziello \em{et~al.}(2007)Capozziello, Nesseris, and
  Perivolaropoulos]{Capozziello_2007iu}
Capozziello, S.; Nesseris, S.; Perivolaropoulos, L.
\newblock {Reconstruction of the Scalar-Tensor Lagrangian from a LCDM
  Background and Noether Symmetry}.
\newblock {\em J. Cosmol. Astropart. Phys.} {\bf 2007}, {\em 12}, 009.
\newblock {\url{https://doi.org/10.1088/1475-7516/2007/12/009}}.

\bibitem[Heisenberg(2019)]{Heisenberg_2018vsk}
Heisenberg, L.
\newblock {A systematic approach to generalisations of General Relativity and
  their cosmological implications}.
\newblock {\em Phys. Rept.} {\bf 2019}, {\em 796}, 1--113.
\newblock {\url{https://doi.org/10.1016/j.physrep.2018.11.006}}.

\bibitem[Langlois(2019)]{Langlois_2018dxi}
Langlois, D.
\newblock {Dark energy and modified gravity in degenerate higher-order
  scalar--tensor (DHOST) theories: A review}.
\newblock {\em Int. J. Mod. Phys. D} {\bf 2019}, {\em 28}, 1942006.
\newblock {\url{https://doi.org/10.1142/S0218271819420069}}.

\bibitem[D'Agostino and Luongo(2018)]{D_Agostino_2018}
D'Agostino, R.; Luongo, O.
\newblock Growth of matter perturbations in nonminimal teleparallel dark
  energy.
\newblock {\em Phys. Rev. D} {\bf 2018}, {\em 98}, 124013.
\newblock {\url{https://doi.org/10.1103/physrevd.98.124013}}.

\bibitem[Bajardi and Capozziello(2020)]{Bajardi_2020_aaa}
Bajardi, F.; Capozziello, S.
\newblock Equivalence of nonminimally coupled cosmologies by Noether
  symmetries.
\newblock {\em Int. J. Mod. Phys. D} {\bf 2020}, {\em 29}, 2030015.
\newblock {\url{https://doi.org/10.1142/s0218271820300153}}.

\bibitem[Amendola \em{et~al.}(2007)Amendola, Campos, and
  Rosenfeld]{Amendola_2007_DE}
Amendola, L.; Campos, G.C.; Rosenfeld, R.
\newblock Consequences of dark matter-dark energy interaction on cosmological
  parameters derived from type~Ia supernova data.
\newblock {\em Phys. Rev. D} {\bf 2007}, {\em 75}, 083506.
\newblock {\url{https://doi.org/10.1103/physrevd.75.083506}}.

\bibitem[Wang \em{et~al.}(2016)Wang, Abdalla, Atrio-Barandela, and Pav{\'{o}
  }n]{Wang_2016_DE}
Wang, B.; Abdalla, E.; Atrio-Barandela, F.; Pav{\'{o}}n, D.
\newblock Dark matter and dark energy interactions: Theoretical challenges,
  cosmological implications and observational signatures.
\newblock {\em Rep. Prog. Phys.} {\bf 2016}, {\em 79}, 096901.
\newblock {\url{https://doi.org/10.1088/0034-4885/79/9/096901}}.

\bibitem[{Di Valentino} \em{et~al.}(2017){Di Valentino}, Melchiorri, and
  Mena]{Di_Valentino_2017}
{Di Valentino}, E.; Melchiorri, A.; Mena, O.
\newblock Can interacting dark energy solve the $H_0$ tension?
\newblock {\em Phys. Rev. D} {\bf 2017}, {\em 96}, 043503.
\newblock {\url{https://doi.org/10.1103/physrevd.96.043503}}.

\bibitem[Asghari \em{et~al.}(2019)Asghari, Jim{\'{e}}nez, Khosravi, and
  Mota]{Asghari_2019_DE}
Asghari, M.; Jim{\'{e}}nez, J.B.; Khosravi, S.; Mota, D.F.
\newblock On structure formation from a small-scales-interacting dark sector.
\newblock {\em J. Cosmol. Astropart. Phys.} {\bf 2019}, {\em
  2019}, 042.
\newblock {\url{https://doi.org/10.1088/1475-7516/2019/04/042}}.

\bibitem[Yang \em{et~al.}(2019)Yang, Vagnozzi, {Di Valentino}, Nunes, Pan, and
  Mota]{Yang_2019_DE}
Yang, W.; Vagnozzi, S.; {Di Valentino}, E.; Nunes, R.C.; Pan, S.; Mota, D.F.
\newblock Listening to the sound of dark sector interactions with gravitational
  wave standard sirens.
\newblock {\em J. Cosmol. Astropart. Phys.} {\bf 2019}, {\em
  2019}, 037.
\newblock {\url{https://doi.org/10.1088/1475-7516/2019/07/037}}.

\bibitem[Saridakis \em{et~al.}(2021)Saridakis et~al.]{CANTATA_2021ktz}
Saridakis, E.N.; Lazkoz, R.; Salzano, V.; Moniz, P.V.; Capozziello, S.; Jiménez, J.B.; De Laurentis, M.; Olmo, G.J.
\newblock {Modified Gravity and Cosmology: An Update by the CANTATA Network}; {Springer: Cham, Switzerland,}
  {2021}. 

\bibitem[{Di Valentino} \em{et~al.}(2021){Di Valentino}, Mena, Pan, Visinelli,
  Yang, Melchiorri, Mota, Riess, and Silk]{DiValentino_2021izs}
{Di Valentino}, E.; Mena, O.; Pan, S.; Visinelli, L.; Yang, W.; Melchiorri, A.;
  Mota, D.F.; Riess, A.G.; Silk, J.
\newblock {In the realm of the Hubble tension---A review of
  solutions}.
\newblock {\em Class. Quant. Grav.} {\bf 2021}, {\em 38}, 153001.
\newblock {\url{https://doi.org/10.1088/1361-6382/ac086d}}.

\bibitem[Capozziello and Bajardi(2021)]{Capozziello_Review}
Capozziello, S.; Bajardi, F.
\newblock Nonlocal gravity cosmology: An overview.
\newblock {\em Int. J. Mod. Phys. D} {\bf 2021}, {\em 31}, 2230009.
\newblock {\url{https://doi.org/10.1142/s0218271822300099}}.

\bibitem[Deser and Woodard(2007)]{Deser_2007}
Deser, S.; Woodard, R.P.
\newblock Nonlocal Cosmology.
\newblock {\em Phys. Rev. Lett.} {\bf 2007}, {\em 99}, 111301.
\newblock {\url{https://doi.org/10.1103/physrevlett.99.111301}}.

\bibitem[Maggiore(2014)]{Maggiore_2014_RT}
Maggiore, M.
\newblock Phantom dark energy from nonlocal infrared modifications of general
  relativity.
\newblock {\em Phys. Rev. D} {\bf 2014}, {\em 89}, 043008.
\newblock {\url{https://doi.org/10.1103/physrevd.89.043008}}.

\bibitem[Buoninfante(2019)]{Buoninf_thesis}
Buoninfante, L.
\newblock Nonlocal Field Theories: Theoretical and Phenomenological Aspects.
\newblock Ph.D. Thesis, University of Groningen, {Groningen, The Netherlands}, 2019. 
~\newblock {\url{https://doi.org/10.33612/diss.99349099}}.

\bibitem[Biswas \em{et~al.}(2012)Biswas, Gerwick, Koivisto, and
  Mazumdar]{Biswas_2012}
Biswas, T.; Gerwick, E.; Koivisto, T.; Mazumdar, A.
\newblock Towards Singularity- and Ghost-Free Theories of Gravity.
\newblock {\em Phys. Rev. Lett.} {\bf 2012}, {\em 108}, 031101.
\newblock {\url{https://doi.org/10.1103/physrevlett.108.031101}}.

\bibitem[Briscese \em{et~al.}(2013)Briscese, Marcian{\`{o} }, Modesto, and
  Saridakis]{Briscese_2013}
Briscese, F.; Marcian{\`{o} }, A.; Modesto, L.; Saridakis, E.N.
\newblock Inflation in (super-)renormalizable gravity.
\newblock {\em Phys. Rev. D} {\bf 2013}, {\em 87}, 083507.
\newblock {\url{https://doi.org/10.1103/physrevd.87.083507}}.

\bibitem[Biswas \em{et~al.}(2006)Biswas, Mazumdar, and Siegel]{Biswas_2006}
Biswas, T.; Mazumdar, A.; Siegel, W.
\newblock Bouncing universes in string-inspired gravity.
\newblock {\em J. Cosmol. Astropart. Phys.} {\bf 2006}, {\em
  2006}, 009.
\newblock {\url{https://doi.org/10.1088/1475-7516/2006/03/009}}.

\bibitem[Belgacem \em{et~al.}(2020)Belgacem, Dirian, Finke, Foffa, and
  Maggiore]{Belgacem_2020}
Belgacem, E.; Dirian, Y.; Finke, A.; Foffa, S.; Maggiore, M.
\newblock Gravity in the infrared and effective nonlocal models.
\newblock {\em J. Cosmol. Astropart. Phys.} {\bf 2020}, {\em
  2020}, 010.
\newblock {\url{https://doi.org/10.1088/1475-7516/2020/04/010}}.

\bibitem[Borka \em{et~al.}(2022)Borka, {Borka Jovanović}, Capozziello, and
  Jovanović]{BORKA2022}
Borka, D.; {Borka Jovanović}, V.; Capozziello, S.; Jovanović, P.
\newblock Velocity distribution of elliptical galaxies in the framework of
  Non-local Gravity model.
\newblock {\em Adv. Space Res.} {\bf 2022}, in press.
\newblock {\url{https://doi.org/https://doi.org/10.1016/j.asr.2022.08.060}}.

\bibitem[Strominger(2017)]{https://doi.org/10.48550/arxiv.1703.05448}
Strominger, A.
\newblock \emph{Lectures on the Infrared Structure of Gravity and Gauge Theory}; {Princeton University Press: Princeton, NJ, USA,  2018}. 
~\newblock {\url{https://doi.org/10.48550/ARXIV.1703.05448}}.

\bibitem[Reuter and Saueressig(2002)]{Reuter_2002}
Reuter, M.; Saueressig, F.
\newblock A class of nonlocal truncations in quantum Einstein gravity and its
  renormalization group behavior.
\newblock {\em Phys. Rev. D} {\bf 2002}, {\em 66}, 125001.
\newblock {\url{https://doi.org/10.1103/physrevd.66.125001}}.

\bibitem[Wetterich(2018)]{Wetterich_2018}
Wetterich, C.
\newblock Infrared limit of quantum gravity.
\newblock {\em Phys. Rev. D} {\bf 2018}, {\em 98}, 026028.
\newblock {\url{https://doi.org/10.1103/physrevd.98.026028}}.

\bibitem[Barvinsky \em{et~al.}(2003)Barvinsky, Gusev, Mukhanov, and
  Nesterov]{Barvinsky_2003rx}
Barvinsky, A.O.; Gusev, Y.V.; Mukhanov, V.F.; Nesterov, D.V.
\newblock {Nonperturbative late time asymptotics for heat kernel in gravity
  theory}.
\newblock {\em Phys. Rev. D} {\bf 2003}, {\em 68}, 105003.
\newblock {\url{https://doi.org/10.1103/PhysRevD.68.105003}}.

\bibitem[Barvinsky(2015)]{Barvinsky_2015}
Barvinsky, A.O.
\newblock Aspects of nonlocality in quantum field theory, quantum gravity and
  cosmology.
\newblock {\em Mod. Phys. Lett. A} {\bf 2015}, {\em 30}, 1540003.
\newblock {\url{https://doi.org/10.1142/s0217732315400039}}.

\bibitem[Maggiore(2016)]{https://doi.org/10.48550/arxiv.1606.08784}
Maggiore, M.
\newblock Nonlocal Infrared Modifications of Gravity. A Review.  In \emph{{Gravity and the Quantum}}; Springer: Cham, Switzerland, 2017; pp. 221--281. 
\newblock {\url{https://doi.org/10.48550/ARXIV.1606.08784}}.

\bibitem[Nojiri and Odintsov(2008)]{Nojiri_2008}
Nojiri, S.; Odintsov, S.D.
\newblock Modified non-local-$F(R)$ gravity as the key for the inflation and dark
  energy.
\newblock {\em Phys. Lett. B} {\bf 2008}, {\em 659}, 821--826.
\newblock {\url{https://doi.org/10.1016/j.physletb.2007.12.001}}.

\bibitem[Kehagias and Maggiore(2014)]{Kehagias_2014}
Kehagias, A.; Maggiore, M.
\newblock Spherically symmetric static solutions in a nonlocal infrared
  modification of General Relativity.
\newblock {\em J. High Energy Phys.} {\bf 2014}, {\em 2014}, 29.
\newblock {\url{https://doi.org/10.1007/jhep08(2014)029}}.

\bibitem[Aiola \em{et~al.}(2020)Aiola, Calabrese, Maurin, Naess, Schmitt,
  Abitbol, Addison, Ade, Alonso, Amiri, Amodeo, Angile, Austermann, Baildon,
  Battaglia, Beall, Bean, Becker, Bond, Bruno, Calafut, Campusano, Carrero,
  Chesmore, mei Cho, Choi, Clark, Cothard, Crichton, Crowley, Darwish, Datta,
  Denison, Devlin, Duell, Duff, Duivenvoorden, Dunkley, Dünner,
  Essinger-Hileman, Fankhanel, Ferraro, Fox, Fuzia, Gallardo, Gluscevic, Golec,
  Grace, Gralla, Guan, Hall, Halpern, Han, Hargrave, Hasselfield, Helton,
  Henderson, Hensley, Hill, Hilton, Hilton, Hincks, Hlo{\v{z}}ek, Ho, Hubmayr,
  Huffenberger, Hughes, Infante, Irwin, Jackson, Klein, Knowles, Koopman,
  Kosowsky, Lakey, Li, Li, Li, Lokken, Louis, Lungu, MacInnis, Madhavacheril,
  Maldonado, Mallaby-Kay, Marsden, McMahon, Menanteau, Moodley, Morton,
  Namikawa, Nati, Newburgh, Nibarger, Nicola, Niemack, Nolta, Orlowski-Sherer,
  Page, Pappas, Partridge, Phakathi, Pisano, Prince, Puddu, Qu, Rivera,
  Robertson, Rojas, Salatino, Schaan, Schillaci, Sehgal, Sherwin, Sierra,
  Sievers, Sifon, Sikhosana, Simon, Spergel, Staggs, Stevens, Storer, Sunder,
  Switzer, Thorne, Thornton, Trac, Treu, Tucker, Vale, Engelen, Lanen,
  Vavagiakis, Wagoner, Wang, Ward, Wollack, Xu, Zago, and Zhu]{Aiola_2020}
Aiola, S.; Calabrese, E.; Maurin, L.; Naess, S.; Schmitt, B.L.; Abitbol, M.H.;
  Addison, G.E.; Ade, P.A.R.; Alonso, D.; Amiri, M.;  et~al.
\newblock The Atacama Cosmology Telescope: {DR}4 maps and cosmological
  parameters.
\newblock {\em J. Cosmol. Astropart. Phys.} {\bf 2020}, {\em
  2020}, 47.
\newblock {\url{https://doi.org/10.1088/1475-7516/2020/12/047}}.

\bibitem[Tröster \em{et~al.}(2021)Tröster, Asgari, Blake, Cataneo, Heymans,
  Hildebrandt, Joachimi, Lin, S{\'{a} }nchez, Wright, Bilicki, Bose, Crocce,
  Dvornik, Erben, Giblin, Glazebrook, Hoekstra, Joudaki, Kannawadi, Köhlinger,
  Kuijken, Lidman, Lombriser, Mead, Parkinson, Shan, Wolf, and
  Xia]{Tr_ster_2021_DE}
Tröster, T.; Asgari, M.; Blake, C.; Cataneo, M.; Heymans, C.; Hildebrandt, H.;
  Joachimi, B.; Lin, C.A.; S{\'{a} }nchez, A.G.; Wright, A.H.;  et~al.
\newblock {KiDS}-1000 Cosmology: Constraints beyond flat $Lambda${CDM}.
\newblock {\em Astron. Astrophys.} {\bf 2021}, {\em 649}, A88.
\newblock {\url{https://doi.org/10.1051/0004-6361/202039805}}.

\bibitem[Pandey \em{et~al.}(2022)Pandey, Krause, DeRose, MacCrann, Jain,
  Crocce, Blazek, Choi, Huang, To, Fang, Elvin-Poole, Prat, Porredon, Secco,
  Rodriguez-Monroy, Weaverdyck, Park, Raveri, Rozo, Rykoff, Bernstein,
  S{\'{a}}nchez, Jarvis, Troxel, Zacharegkas, Chang, Alarcon, Alves, Amon,
  Andrade-Oliveira, Baxter, Bechtol, Becker, Camacho, Campos, Rosell, Kind,
  Cawthon, Chen, Chintalapati, Davis, Valentino, Diehl, Dodelson, Doux,
  Drlica-Wagner, Eckert, Eifler, Elsner, Everett, Farahi, Fert{\'{e}}, Fosalba,
  Friedrich, Gatti, Giannini, Gruen, Gruendl, Harrison, Hartley, Huff, Huterer,
  Kovacs, Leget, McCullough, Muir, Myles, Navarro-Alsina, Omori, Rollins,
  Roodman, Rosenfeld, Sevilla-Noarbe, Sheldon, Shin, Troja, Tutusaus, Varga,
  Wechsler, Yanny, Yin, Zhang, Zuntz, Abbott, Aguena, Allam, Annis, Bacon,
  Bertin, Brooks, Burke, Carretero, Conselice, Costanzi, da~Costa, Pereira,
  Vicente, Dietrich, Doel, Evrard, Ferrero, Flaugher, Frieman,
  Garc{\'{\i}}a-Bellido, Gaztanaga, Gerdes, Giannantonio, Gschwend, Gutierrez,
  Hinton, Hollowood, Honscheid, James, Jeltema, Kuehn, Kuropatkin, Lahav, Lima,
  Lin, Maia, Marshall, Melchior, Menanteau, Miller, Miquel, Mohr, Morgan,
  Palmese, Paz-Chinch{\'{o}}n, Petravick, Pieres, Malag{\'{o}}n, Sanchez,
  Scarpine, Serrano, Smith, Soares-Santos, Suchyta, Tarle, Thomas, and
  and]{Pandey_2022_DE}
Pandey, S.; Krause, E.; DeRose, J.; MacCrann, N.; Jain, B.; Crocce, M.; Blazek,
  J.; Choi, A.; Huang, H.; To, C.;  et~al.
\newblock Dark Energy Survey year 3 results: Constraints on cosmological
  parameters and galaxy-bias models from galaxy clustering and galaxy-galaxy
  lensing using the {redMaGiC} sample.
\newblock {\em Phys. Rev. D} {\bf 2022}, {\em 106}, 043520.
\newblock {\url{https://doi.org/10.1103/physrevd.106.043520}}.

\bibitem[Weinberg(1989)]{RevModPhys_61_1}
Weinberg, S.
\newblock The cosmological constant problem.
\newblock {\em Rev. Mod. Phys.} {\bf 1989}, {\em 61}, 1.
\newblock {\url{https://doi.org/10.1103/RevModPhys.61.1}}.

\bibitem[Velten \em{et~al.}(2014)Velten, vom Marttens, and
  Zimdahl]{Velten_2014}
Velten, H.E.S.; vom Marttens, R.F.; Zimdahl, W.
\newblock Aspects of the cosmological {\textquotedblleft}coincidence
  problem{\textquotedblright}.
\newblock {\em  Eur. Phys. J. C} {\bf 2014}, {\em 74}, 3160.
\newblock {\url{https://doi.org/10.1140/epjc/s10052-014-3160-4}}.

\bibitem[Deffayet and Woodard(2009)]{Deffayet_2009}
Deffayet, C.; Woodard, R.
\newblock Reconstructing the distortion function for nonlocal cosmology.
\newblock {\em J. Cosmol. Astropart. Phys.} {\bf 2009}, {\em
  2009},~023.
\newblock {\url{https://doi.org/10.1088/1475-7516/2009/08/023}}.

\bibitem[Foffa \em{et~al.}(2014)Foffa, Maggiore, and Mitsou]{Foffa_2014}
Foffa, S.; Maggiore, M.; Mitsou, E.
\newblock Cosmological dynamics and dark energy from nonlocal infrared
  modifications of gravity.
\newblock {\em Int. J. Mod. Phys. A} {\bf 2014}, {\em 29}, 1450116.
\newblock {\url{https://doi.org/10.1142/s0217751x14501164}}.

\bibitem[Calabrese \em{et~al.}(2008)Calabrese, Slosar, Melchiorri, Smoot, and
  Zahn]{Calabrese_2008}
Calabrese, E.; Slosar, A.; Melchiorri, A.; Smoot, G.F.; Zahn, O.
\newblock Cosmic microwave weak lensing data as a test for the dark universe.
\newblock {\em Phys. Rev. D} {\bf 2008}, {\em 77}, 123531.
\newblock {\url{https://doi.org/10.1103/physrevd.77.123531}}.

\bibitem[Hildebrandt \em{et~al.}(2016)Hildebrandt, Viola, Heymans, Joudaki,
  Kuijken, Blake, Erben, Joachimi, Klaes, Miller, Morrison, Nakajima, Kleijn,
  Amon, Choi, Covone, de~Jong, Dvornik, Conti, Grado, Harnois-D{\'{e} }raps,
  Herbonnet, Hoekstra, Köhlinger, McFarland, Mead, Merten, Napolitano,
  Peacock, Radovich, Schneider, Simon, Valentijn, van~den Busch, van Uitert,
  and Waerbeke]{Hildebrandt_2016}
Hildebrandt, H.; Viola, M.; Heymans, C.; Joudaki, S.; Kuijken, K.; Blake, C.;
  Erben, T.; Joachimi, B.; Klaes, D.; Miller, L.;  et~al.
\newblock {KiDS}-450: Cosmological parameter constraints from tomographic weak
  gravitational lensing.
\newblock {\em Mon. Not. R. Astron. Soc.} {\bf 2016},
  {\em 465}, 1454--1498.
\newblock {\url{https://doi.org/10.1093/mnras/stw2805}}.

\bibitem[Hildebrandt \em{et~al.}(2020)Hildebrandt, Köhlinger, van~den Busch,
  Joachimi, Heymans, Kannawadi, Wright, Asgari, Blake, Hoekstra, Joudaki,
  Kuijken, Miller, Morrison, Tröster, Amon, Archidiacono, Brieden, Choi,
  de~Jong, Erben, Giblin, Mead, Peacock, Radovich, Schneider, Sif{\'{o} }n, and
  Tewes]{Hildebrandt_2020}
Hildebrandt, H.; Köhlinger, F.; van~den Busch, J.L.; Joachimi, B.; Heymans,
  C.; Kannawadi, A.; Wright, A.H.; Asgari, M.; Blake, C.; Hoekstra, H.;  et~al.
\newblock {KiDS}+{VIKING}-450: Cosmic shear tomography with optical and
  infrared data.
\newblock {\em Astron. Astrophys.} {\bf 2020}, {\em 633}, A69.
\newblock {\url{https://doi.org/10.1051/0004-6361/201834878}}.

\bibitem[Köhlinger \em{et~al.}(2017)Köhlinger, Viola, Joachimi, Hoekstra, van
  Uitert, Hildebrandt, Choi, Erben, Heymans, Joudaki, Klaes, Kuijken, Merten,
  Miller, Schneider, and Valentijn]{K_hlinger_2017}
Köhlinger, F.; Viola, M.; Joachimi, B.; Hoekstra, H.; van Uitert, E.;
  Hildebrandt, H.; Choi, A.; Erben, T.; Heymans, C.; Joudaki, S.;  et~al.
\newblock {KiDS}-450: The tomographic weak lensing power spectrum and
  constraints on cosmological parameters.
\newblock {\em Mon. Not. R. Astron. Soc.} {\bf 2017},
  {\em 471}, 4412--4435.
\newblock {\url{https://doi.org/10.1093/mnras/stx1820}}.

\bibitem[Wright \em{et~al.}(2020)Wright, Hildebrandt, van~den Busch, Heymans,
  Joachimi, Kannawadi, and Kuijken]{Wright_2020}
Wright, A.H.; Hildebrandt, H.; van~den Busch, J.L.; Heymans, C.; Joachimi, B.;
  Kannawadi, A.; Kuijken, K.
\newblock {KiDS}+{VIKING}-450: Improved cosmological parameter constraints from
  redshift calibration with self-organising maps.
\newblock {\em Astron. Astrophys.} {\bf 2020}, {\em 640}, L14.
\newblock {\url{https://doi.org/10.1051/0004-6361/202038389}}.

\bibitem[Troxel \em{et~al.}(2018)Troxel, MacCrann, Zuntz, Eifler, Krause,
  Dodelson, Gruen, Blazek, Friedrich, Samuroff, Prat, Secco, Davis,
  Fert{\'{e}}, DeRose, Alarcon, Amara, Baxter, Becker, Bernstein, Bridle,
  Cawthon, Chang, Choi, Vicente, Drlica-Wagner, Elvin-Poole, Frieman, Gatti,
  Hartley, Honscheid, Hoyle, Huff, Huterer, Jain, Jarvis, Kacprzak, Kirk,
  Kokron, Krawiec, Lahav, Liddle, Peacock, Rau, Refregier, Rollins, Rozo,
  Rykoff, S{\'{a}}nchez, Sevilla-Noarbe, Sheldon, Stebbins, Varga, Vielzeuf,
  Wang, Wechsler, Yanny, Abbott, Abdalla, Allam, Annis, Bechtol,
  Benoit-L{\'{e}}vy, Bertin, Brooks, Buckley-Geer, Burke, Rosell, Kind,
  Carretero, Castander, Crocce, Cunha, D'Andrea, da~Costa, DePoy, Desai, Diehl,
  Dietrich, Doel, Fernandez, Flaugher, Fosalba, Garc{\'{\i}}a-Bellido,
  Gaztanaga, Gerdes, Giannantonio, Goldstein, Gruendl, Gschwend, Gutierrez,
  James, Jeltema, Johnson, Johnson, Kent, Kuehn, Kuhlmann, Kuropatkin, Li,
  Lima, Lin, Maia, March, Marshall, Martini, Melchior, Menanteau, Miquel, Mohr,
  Neilsen, Nichol, Nord, Petravick, Plazas, Romer, Roodman, Sako, Sanchez,
  Scarpine, Schindler, Schubnell, Smith, Smith, Soares-Santos, Sobreira,
  Suchyta, Swanson, Tarle, Thomas, Tucker, Vikram, Walker, Weller, and
  and]{Troxel_2018}
Troxel, M.; MacCrann, N.; Zuntz, J.; Eifler, T.; Krause, E.; Dodelson, S.;
  Gruen, D.; Blazek, J.; Friedrich, O.; Samuroff, S.;  et~al.
\newblock Dark Energy Survey Year 1 results: Cosmological constraints from
  cosmic shear.
\newblock {\em Phys. Rev. D} {\bf 2018}, {\em 98}, 043528.
\newblock {\url{https://doi.org/10.1103/physrevd.98.043528}}.

\bibitem[Joudaki \em{et~al.}(2020)Joudaki, Hildebrandt, Traykova, Chisari,
  Heymans, Kannawadi, Kuijken, Wright, Asgari, Erben, Hoekstra, Joachimi,
  Miller, Tröster, and van~den Busch]{Joudaki_2020}
Joudaki, S.; Hildebrandt, H.; Traykova, D.; Chisari, N.E.; Heymans, C.;
  Kannawadi, A.; Kuijken, K.; Wright, A.H.; Asgari, M.; Erben, T.;  et~al.
\newblock {KiDS}+{VIKING}-450 and {DES}-Y1 combined: Cosmology with cosmic
  shear.
\newblock {\em Astron. Astrophys.} {\bf 2020}, {\em 638}, L1.
\newblock {\url{https://doi.org/10.1051/0004-6361/201936154}}.

\bibitem[Asgari \em{et~al.}(2020)Asgari, Tröster, Heymans, Hildebrandt,
  van~den Busch, Wright, Choi, Erben, Joachimi, Joudaki, Kannawadi, Kuijken,
  Lin, Schneider, and Zuntz]{Asgari_2020}
Asgari, M.; Tröster, T.; Heymans, C.; Hildebrandt, H.; van~den Busch, J.L.;
  Wright, A.H.; Choi, A.; Erben, T.; Joachimi, B.; Joudaki, S.;  et~al.
\newblock {KiDS}+{VIKING}-450 and {DES}-Y1 combined: Mitigating baryon feedback
  uncertainty with {COSEBIs}.
\newblock {\em Astron. Astrophys.} {\bf 2020}, {\em 634}, A127.
\newblock {\url{https://doi.org/10.1051/0004-6361/201936512}}.

\bibitem[Loureiro \em{et~al.}(2022)Loureiro, Whittaker, Mancini, Joachimi,
  Cuceu, Asgari, Stölzner, Tröster, Wright, Bilicki, Dvornik, Giblin,
  Heymans, Hildebrandt, Shan, Amara, Auricchio, Bodendorf, Bonino, Branchini,
  Brescia, Capobianco, Carbone, Carretero, Castellano, Cavuoti, Cimatti,
  Cledassou, Congedo, Conversi, Copin, Corcione, Cropper, Silva, Douspis,
  Dubath, Duncan, Dupac, Dusini, Farrens, Ferriol, Fosalba, Frailis,
  Franceschi, Fumana, Garilli, Gillis, Giocoli, Grazian, Grupp, Haugan, Holmes,
  Hormuth, Jahnke, Kümmel, Kermiche, Kiessling, Kilbinger, Kitching, Kuijken,
  Kunz, Kurki-Suonio, Ligori, Lilje, Lloro, Mansutti, Marggraf, Markovic,
  Marulli, Massey, Meneghetti, Meylan, Moresco, Morin, Moscardini, Munari,
  Niemi, Padilla, Paltani, Pasian, Pedersen, Pettorino, Pires, Poncet, Popa,
  Raison, Rhodes, Rix, Roncarelli, Saglia, Schneider, Secroun, Serrano,
  Sirignano, Sirri, Stanco, Starck, Tallada-Cresp{\'{\i} }, Taylor, Tereno,
  Toledo-Moreo, Torradeflot, Valentijn, Wang, Welikala, Weller, Zamorani,
  Zoubian, Andreon, Baldi, Camera, Farinelli, Polenta, and
  Tessore]{Loureiro_2022}
Loureiro, A.; Whittaker, L.; Mancini, A.S.; Joachimi, B.; Cuceu, A.; Asgari,
  M.; Stölzner, B.; Tröster, T.; Wright, A.H.; Bilicki, M.;  et~al.
\newblock {KiDS} and Euclid: Cosmological implications of a pseudo angular
  power spectrum analysis of {KiDS}-1000 cosmic shear tomography.
\newblock {\em Astron. Astrophys.} {\bf 2022}, {\em 665}.
\newblock {\url{https://doi.org/10.1051/0004-6361/202142481}}.

\bibitem[Chang \em{et~al.}(2022)Chang, Omori, Baxter, Doux, Choi, Pandey,
  Alarcon, Alves, Amon, Andrade-Oliveira, Bechtol, Becker, Bernstein,
  Bianchini, Blazek, Bleem, Camacho, Campos, Rosell, Kind, Cawthon, Chen,
  Cordero, Crawford, Crocce, Davis, DeRose, Dodelson, Drlica-Wagner, Eckert,
  Eifler, Elsner, Elvin-Poole, Everett, Fang, Ferté, Fosalba, Friedrich,
  Gatti, Giannini, Gruen, Gruendl, Harrison, Herner, Huang, Huff, Huterer,
  Jarvis, Kovacs, Krause, Kuropatkin, Leget, Lemos, Liddle, MacCrann,
  McCullough, Muir, Myles, Navarro-Alsina, Park, Porredon, Prat, Raveri,
  Rollins, Roodman, Rosenfeld, Ross, Rykoff, Sánchez, Sanchez, Secco,
  Sevilla-Noarbe, Sheldon, Shin, Troxel, Tutusaus, Varga, Weaverdyck, Wechsler,
  Wu, Yanny, Yin, Zhang, Zuntz, Abbott, Aguena, Allam, Annis, Bacon, Benson,
  Bertin, Bocquet, Brooks, Burke, Carlstrom, Carretero, Chang, Chown, Costanzi,
  da~Costa, Crites, Pereira, de~Haan, De~Vicente, Desai, Diehl, Dobbs, Doel,
  Everett, Ferrero, Flaugher, Friedel, Frieman, García-Bellido, Gaztanaga,
  George, Giannantonio, Halverson, Hinton, Holder, Hollowood, Holzapfel,
  Honscheid, Hrubes, James, Knox, Kuehn, Lahav, Lee, Lima, Luong-Van, March,
  McMahon, Melchior, Menanteau, Meyer, Miquel, Mocanu, Mohr, Morgan, Natoli,
  Padin, Palmese, Paz-Chinchón, Pieres, Malagón, Pryke, Reichardt,
  Rodríguez-Monroy, Romer, Ruhl, Sanchez, Schaffer, Schubnell, Serrano,
  Shirokoff, Smith, Staniszewski, Stark, Suchyta, Tarle, Thomas, To, Vieira,
  Weller, and Williamson]{https://doi.org/10.48550/arxiv.2203.12440}
Chang, C.; Omori, Y.; Baxter, E.J.; Doux, C.; Choi, A.; Pandey, S.; Alarcon,
  A.; Alves, O.; Amon, A.; Andrade-Oliveira, F.;  et~al.
\newblock Joint analysis of DES Year 3 data and CMB lensing from SPT and Planck
  II: Cross-correlation measurements and cosmological constraints. \emph{arXiv Prepr.} \textbf{2022}, arXiv:2203.12440.
\newblock {\url{https://doi.org/10.48550/ARXIV.2203.12440}}.

\bibitem[Amon \em{et~al.}(2022)Amon, Gruen, Troxel, MacCrann, Dodelson, Choi,
  Doux, Secco, Samuroff, Krause, Cordero, Myles, DeRose, Wechsler, Gatti,
  Navarro-Alsina, Bernstein, Jain, Blazek, Alarcon, Fert{\'{e}}, Lemos, Raveri,
  Campos, Prat, S{\'{a}}nchez, Jarvis, Alves, Andrade-Oliveira, Baxter,
  Bechtol, Becker, Bridle, Camacho, Rosell, Kind, Cawthon, Chang, Chen,
  Chintalapati, Crocce, Davis, Diehl, Drlica-Wagner, Eckert, Eifler,
  Elvin-Poole, Everett, Fang, Fosalba, Friedrich, Gaztanaga, Giannini, Gruendl,
  Harrison, Hartley, Herner, Huang, Huff, Huterer, Kuropatkin, Leget, Liddle,
  McCullough, Muir, Pandey, Park, Porredon, Refregier, Rollins, Roodman,
  Rosenfeld, Ross, Rykoff, Sanchez, Sevilla-Noarbe, Sheldon, Shin, Troja,
  Tutusaus, Tutusaus, Varga, Weaverdyck, Yanny, Yin, Zhang, Zuntz, Aguena,
  Allam, Annis, Bacon, Bertin, Bhargava, Brooks, Buckley-Geer, Burke,
  Carretero, Costanzi, da~Costa, Pereira, Vicente, Desai, Dietrich, Doel,
  Ferrero, Flaugher, Frieman, Garc{\'{\i}}a-Bellido, Gaztanaga, Gerdes,
  Giannantonio, Gschwend, Gutierrez, Hinton, Hollowood, Honscheid, Hoyle,
  James, Kron, Kuehn, Lahav, Lima, Lin, Maia, Marshall, Martini, Melchior,
  Menanteau, Miquel, Mohr, Morgan, Ogando, Palmese, Paz-Chinch{\'{o}}n,
  Petravick, Pieres, Romer, Sanchez, Scarpine, Schubnell, Serrano, Smith,
  Soares-Santos, Tarle, Thomas, To, and and]{Amon_2022}
Amon, A.; Gruen, D.; Troxel, M.; MacCrann, N.; Dodelson, S.; Choi, A.; Doux,
  C.; Secco, L.; Samuroff, S.; Krause, E.;  et~al.
\newblock Dark Energy Survey Year 3 results: Cosmology from cosmic shear and
  robustness to data calibration.
\newblock {\em Phys. Rev. D} {\bf 2022}, {\em 105}, 023514.
\newblock {\url{https://doi.org/10.1103/physrevd.105.023514}}.

\bibitem[Secco \em{et~al.}(2022)Secco, Samuroff, Krause, Jain, Blazek, Raveri,
  Campos, Amon, Chen, Doux, Choi, Gruen, Bernstein, Chang, DeRose, Myles,
  Fert{\'{e}}, Lemos, Huterer, Prat, Troxel, MacCrann, Liddle, Kacprzak, Fang,
  S{\'{a}}nchez, Pandey, Dodelson, Chintalapati, Hoffmann, Alarcon, Alves,
  Andrade-Oliveira, Baxter, Bechtol, Becker, Brandao-Souza, Camacho, Rosell,
  Kind, Cawthon, Cordero, Crocce, Davis, Valentino, Drlica-Wagner, Eckert,
  Eifler, Elidaiana, Elsner, Elvin-Poole, Everett, Fosalba, Friedrich, Gatti,
  Giannini, Gruendl, Harrison, Hartley, Herner, Huang, Huff, Jarvis, Jeffrey,
  Kuropatkin, Leget, Muir, Mccullough, Alsina, Omori, Park, Porredon, Rollins,
  Roodman, Rosenfeld, Ross, Rykoff, Sanchez, Sevilla-Noarbe, Sheldon, Shin,
  Troja, Tutusaus, Varga, Weaverdyck, Wechsler, Yanny, Yin, Zhang, Zuntz,
  Abbott, Aguena, Allam, Annis, Bacon, Bertin, Bhargava, Bridle, Brooks,
  Buckley-Geer, Burke, Carretero, Costanzi, da~Costa, Vicente, Diehl, Dietrich,
  Doel, Ferrero, Flaugher, Frieman, Garc{\'{\i}}a-Bellido, Gaztanaga, Gerdes,
  Giannantonio, Gschwend, Gutierrez, Hinton, Hollowood, Honscheid, Hoyle,
  James, Jeltema, Kuehn, Lahav, Lima, Lin, Maia, Marshall, Martini, Melchior,
  Menanteau, Miquel, Mohr, Morgan, Ogando, Palmese, Paz-Chinch{\'{o}}n,
  Petravick, Pieres, Malag{\'{o}}n, Rodriguez-Monroy, Romer, Sanchez, Scarpine,
  Schubnell, Scolnic, Serrano, Smith, Soares-Santos, Suchyta, Swanson, Tarle,
  Thomas, and and]{Secco_2022}
Secco, L.; Samuroff, S.; Krause, E.; Jain, B.; Blazek, J.; Raveri, M.; Campos,
  A.; Amon, A.; Chen, A.; Doux, C.;  et~al.
\newblock Dark Energy Survey Year 3 results: Cosmology from cosmic shear and
  robustness to modeling uncertainty.
\newblock {\em Phys. Rev. D} {\bf 2022}, {\em 105}, 023515.
\newblock {\url{https://doi.org/10.1103/physrevd.105.023515}}.

\bibitem[Heymans \em{et~al.}(2021)Heymans, Tröster, Asgari, Blake,
  Hildebrandt, Joachimi, Kuijken, Lin, S{\'{a} }nchez, van~den Busch, Wright,
  Amon, Bilicki, de~Jong, Crocce, Dvornik, Erben, Fortuna, Getman, Giblin,
  Glazebrook, Hoekstra, Joudaki, Kannawadi, Köhlinger, Lidman, Miller,
  Napolitano, Parkinson, Schneider, Shan, Valentijn, Kleijn, and
  Wolf]{Heymans_2021}
Heymans, C.; Tröster, T.; Asgari, M.; Blake, C.; Hildebrandt, H.; Joachimi,
  B.; Kuijken, K.; Lin, C.A.; S{\'{a} }nchez, A.G.; van~den Busch, J.L.;
  et~al.
\newblock {KiDS}-1000 Cosmology: Multi-probe weak gravitational lensing and
  spectroscopic galaxy clustering constraints.
\newblock {\em Astron. Astrophys.} {\bf 2021}, {\em 646}, A140.
\newblock {\url{https://doi.org/10.1051/0004-6361/202039063}}.

\bibitem[Mantz \em{et~al.}(2014)Mantz, von~der Linden, Allen, Applegate, Kelly,
  Morris, Rapetti, Schmidt, Adhikari, Allen, Burchat, Burke, Cataneo, Donovan,
  Ebeling, Shandera, and Wright]{Mantz_2014}
Mantz, A.B.; von~der Linden, A.; Allen, S.W.; Applegate, D.E.; Kelly, P.L.;
  Morris, R.G.; Rapetti, D.A.; Schmidt, R.W.; Adhikari, S.; Allen, M.T.;
  et~al.
\newblock Weighing the giants {\textendash} {IV}. Cosmology and neutrino mass.
\newblock {\em Mon. Not. R. Astron. Soc.} {\bf 2014},
  {\em 446}, 2205--2225.
\newblock {\url{https://doi.org/10.1093/mnras/stu2096}}.

\bibitem[Salvati \em{et~al.}(2018)Salvati, Douspis, and Aghanim]{Salvati_2018}
Salvati, L.; Douspis, M.; Aghanim, N.
\newblock Constraints from thermal Sunyaev-Zel'dovich cluster counts and power
  spectrum combined with {CMB}.
\newblock {\em Astron. Astrophys.} {\bf 2018}, {\em 614}, A13.
\newblock {\url{https://doi.org/10.1051/0004-6361/201731990}}.

\bibitem[Costanzi \em{et~al.}(2019)Costanzi, Rozo, Simet, Zhang, Evrard, Mantz,
  Rykoff, Jeltema, Gruen, Allen, McClintock, Romer, von~der Linden, Farahi,
  DeRose, Varga, Weller, Giles, Hollowood, Bhargava, Bermeo-Hernandez, Chen,
  Abbott, Abdalla, Avila, Bechtol, Brooks, Buckley-Geer, Burke, Rosell, Kind,
  Carretero, Crocce, Cunha, da~Costa, Davis, Vicente, Diehl, Dietrich, Doel,
  Eifler, Estrada, Flaugher, Fosalba, Frieman, Garc{\'{\i} }a-Bellido,
  Gaztanaga, Gerdes, Giannantonio, Gruendl, Gschwend, Gutierrez, Hartley,
  Honscheid, Hoyle, James, Krause, Kuehn, Kuropatkin, Lima, Lin, Maia, March,
  Marshall, Martini, Menanteau, Miller, Miquel, Mohr, Ogando, Plazas, Roodman,
  Sanchez, Scarpine, Schindler, Schubnell, Serrano, Sevilla-Noarbe, Sheldon,
  Smith, Soares-Santos, Sobreira, Suchyta, Swanson, Tarle, Thomas, and
  Wechsler]{Costanzi_2019}
Costanzi, M.; Rozo, E.; Simet, M.; Zhang, Y.; Evrard, A.E.; Mantz, A.; Rykoff,
  E.S.; Jeltema, T.; Gruen, D.; Allen, S.;  et~al.
\newblock Methods for cluster cosmology and application to the {SDSS} in
  preparation for {DES} Year 1 release.
\newblock {\em Mon. Not. R. Astron. Soc.} {\bf 2019},
  {\em 488}, 4779--4800.
\newblock {\url{https://doi.org/10.1093/mnras/stz1949}}.

\bibitem[Bocquet \em{et~al.}(2019)Bocquet, Dietrich, Schrabback, Bleem, Klein,
  Allen, Applegate, Ashby, Bautz, Bayliss, Benson, Brodwin, Bulbul, Canning,
  Capasso, Carlstrom, Chang, Chiu, Cho, Clocchiatti, Crawford, Crites, de~Haan,
  Desai, Dobbs, Foley, Forman, Garmire, George, Gladders, Gonzalez, Grandis,
  Gupta, Halverson, Hlavacek-Larrondo, Hoekstra, Holder, Holzapfel, Hou,
  Hrubes, Huang, Jones, Khullar, Knox, Kraft, Lee, von~der Linden, Luong-Van,
  Mantz, Marrone, McDonald, McMahon, Meyer, Mocanu, Mohr, Morris, Padin, Patil,
  Pryke, Rapetti, Reichardt, Rest, Ruhl, Saliwanchik, Saro, Sayre, Schaffer,
  Shirokoff, Stalder, Stanford, Staniszewski, Stark, Story, Strazzullo, Stubbs,
  Vanderlinde, Vieira, Vikhlinin, Williamson, and Zenteno]{Bocquet_2019}
Bocquet, S.; Dietrich, J.P.; Schrabback, T.; Bleem, L.E.; Klein, M.; Allen,
  S.W.; Applegate, D.E.; Ashby, M.L.N.; Bautz, M.; Bayliss, M.;  et~al.
\newblock Cluster Cosmology Constraints from the 2500 deg$^2$ {SPT}-{SZ}
  Survey: Inclusion of Weak Gravitational Lensing Data from Magellan and the
  Hubble Space Telescope.
\newblock {\em  Astrophys. J.} {\bf 2019}, {\em 878},  55.
\newblock {\url{https://doi.org/10.3847/1538-4357/ab1f10}}.

\bibitem[Abbott \em{et~al.}(2020)Abbott, Aguena, Alarcon, Allam, Allen, Annis,
  Avila, Bacon, Bechtol, Bermeo, Bernstein, Bertin, Bhargava, Bocquet, Brooks,
  Brout, Buckley-Geer, Burke, Rosell, Kind, Carretero, Castander, Cawthon,
  Chang, Chen, Choi, Costanzi, Crocce, da~Costa, Davis, Vicente, DeRose, Desai,
  Diehl, Dietrich, Dodelson, Doel, Drlica-Wagner, Eckert, Eifler, Elvin-Poole,
  Estrada, Everett, Evrard, Farahi, Ferrero, Flaugher, Fosalba, Frieman,
  Garc{\'{\i}}a-Bellido, Gatti, Gaztanaga, Gerdes, Giannantonio, Giles,
  Grandis, Gruen, Gruendl, Gschwend, Gutierrez, Hartley, Hinton, Hollowood,
  Honscheid, Hoyle, Huterer, James, Jarvis, Jeltema, Johnson, Johnson, Kent,
  Krause, Kron, Kuehn, Kuropatkin, Lahav, Li, Lidman, Lima, Lin, MacCrann,
  Maia, Mantz, Marshall, Martini, Mayers, Melchior, Mena-Fern{\'{a}}ndez,
  Menanteau, Miquel, Mohr, Nichol, Nord, Ogando, Palmese, Paz-Chinch{\'{o}}n,
  Plazas, Prat, Rau, Romer, Roodman, Rooney, Rozo, Rykoff, Sako, Samuroff,
  S{\'{a}}nchez, Sanchez, Saro, Scarpine, Schubnell, Scolnic, Serrano,
  Sevilla-Noarbe, Sheldon, Smith, Smith, Suchyta, Swanson, Tarle, Thomas, To,
  Troxel, Tucker, Varga, von~der Linden, Walker, Wechsler, Weller, Wilkinson,
  Wu, Yanny, Zhang, Zhang, and and]{Abbott_2020_CC}
Abbott, T.; Aguena, M.; Alarcon, A.; Allam, S.; Allen, S.; Annis, J.; Avila,
  S.; Bacon, D.; Bechtol, K.; Bermeo, A.;  et~al.
\newblock Dark Energy Survey Year 1 Results: Cosmological constraints from
  cluster abundances and weak lensing.
\newblock {\em Phys. Rev. D} {\bf 2020}, {\em 102}, 023509.
\newblock {\url{https://doi.org/10.1103/physrevd.102.023509}}.

\bibitem[Abdullah \em{et~al.}(2020)Abdullah, Klypin, and Wilson]{Abdullah_2020}
Abdullah, M.H.; Klypin, A.; Wilson, G.
\newblock Cosmological Constraints on $\Omega_{M}$ and $\sigma_{8}$ from
  Cluster Abundances Using the {GalWCat}19 Optical-spectroscopic {SDSS}
  Catalog.
\newblock {\em  Astrophys. J.} {\bf 2020}, {\em 901}, 90.
\newblock {\url{https://doi.org/10.3847/1538-4357/aba619}}.

\bibitem[Kazantzidis and Perivolaropoulos(2018)]{Kazantzidis_2018}
Kazantzidis, L.; Perivolaropoulos, L.
\newblock Evolution of the $f\sigma_{8}$ tension with the Planck15/$\Lambda$CDM
  determination and implications for modified gravity theories.
\newblock {\em Phys. Rev. D} {\bf 2018}, {\em 97}, 103503.
\newblock {\url{https://doi.org/10.1103/physrevd.97.103503}}.

\bibitem[Benisty(2021)]{Benisty_2021}
Benisty, D.
\newblock Quantifying the $S_{8}$ tension with the Redshift Space Distortion
  data set.
\newblock {\em Phys. Dark Universe} {\bf 2021}, {\em 31}, 100766.
\newblock {\url{https://doi.org/10.1016/j.dark.2020.100766}}.

\bibitem[Nunes and Vagnozzi(2021)]{Nunes_2021}
Nunes, R.C.; Vagnozzi, S.
\newblock Arbitrating the $S_{8}$ discrepancy with growth rate measurements
  from redshift-space distortions.
\newblock {\em Mon. Not. R. Astron. Soc.} {\bf 2021},
  {\em 505}, 5427--5437.
\newblock {\url{https://doi.org/10.1093/mnras/stab1613}}.

\bibitem[Philcox and Ivanov(2022)]{Philcox_2022}
Philcox, O.H.; Ivanov, M.M.
\newblock {BOSS} {DR}12 full-shape cosmology: $\Lambda$CDM constraints from the
  large-scale galaxy power spectrum and bispectrum monopole.
\newblock {\em Phys. Rev. D} {\bf 2022}, {\em 105}, 043517.
\newblock {\url{https://doi.org/10.1103/physrevd.105.043517}}.

\bibitem[Nersisyan \em{et~al.}(2017)Nersisyan, Cid, and
  Amendola]{Nersisyan_2017}
Nersisyan, H.; Cid, A.F.; Amendola, L.
\newblock Structure formation in the Deser-Woodard nonlocal gravity model: A
  reappraisal.
\newblock {\em J. Cosmol. Astropart. Phys.} {\bf 2017}, {\em
  2017}, 046.
\newblock {\url{https://doi.org/10.1088/1475-7516/2017/04/046}}.

\bibitem[Park(2018)]{Park_2018}
Park, S.
\newblock Revival of the Deser-Woodard nonlocal gravity model: Comparison of
  the original nonlocal form and a localized formulation.
\newblock {\em Phys. Rev. D} {\bf 2018}, {\em 97}, 044006.
\newblock {\url{https://doi.org/10.1103/physrevd.97.044006}}.

\bibitem[Amendola \em{et~al.}(2019)Amendola, Dirian, Nersisyan, and
  Park]{Amendola_2019}
Amendola, L.; Dirian, Y.; Nersisyan, H.; Park, S.
\newblock Observational constraints in nonlocal gravity: The Deser-Woodard
  case.
\newblock {\em J. Cosmol. Astropart. Phys.} {\bf 2019}, {\em
  2019}, 045.
\newblock {\url{https://doi.org/10.1088/1475-7516/2019/03/045}}.

\bibitem[Joudaki \em{et~al.}(2016)Joudaki, Blake, Heymans, Choi,
  Harnois-Deraps, Hildebrandt, Joachimi, Johnson, Mead, Parkinson, Viola, and
  van Waerbeke]{Joudaki_2017}
Joudaki, S.; Blake, C.; Heymans, C.; Choi, A.; Harnois-Deraps, J.; Hildebrandt,
  H.; Joachimi, B.; Johnson, A.; Mead, A.; Parkinson, D.;  et~al.
\newblock {CFHTLenS} revisited: Assessing concordance with Planck including
  astrophysical systematics.
\newblock {\em Mon. Not. R. Astron. Soc.} {\bf 2016},
  {\em 465}, 2033--2052.
\newblock {\url{https://doi.org/10.1093/mnras/stw2665}}.

\bibitem[Ivanov(2021)]{Ivanov_2021}
Ivanov, M.M.
\newblock Cosmological constraints from the power spectrum of {eBOSS} emission
  line galaxies.
\newblock {\em Phys. Rev. D} {\bf 2021}, {\em 104}, 103514.
\newblock {\url{https://doi.org/10.1103/physrevd.104.103514}}.

\bibitem[Ivanov \em{et~al.}(2020)Ivanov, Simonovi{\'{c} }, and
  Zaldarriaga]{Ivanov_2020}
Ivanov, M.M.; Simonovi{\'{c} }, M.; Zaldarriaga, M.
\newblock Cosmological parameters from the {BOSS} galaxy power spectrum.
\newblock {\em J. Cosmol. Astropart. Phys.} {\bf 2020}, {\em
  2020}, 042.
\newblock {\url{https://doi.org/10.1088/1475-7516/2020/05/042}}.

\bibitem[White \em{et~al.}(2022)White, Zhou, DeRose, Ferraro, Chen, Kokron,
  Bailey, Brooks, Garc{\'{\i} }a-Bellido, Guy, Honscheid, Kehoe, Kremin, Levi,
  Palanque-Delabrouille, Poppett, Schlegel, and Tarle]{White_2022}
White, M.; Zhou, R.; DeRose, J.; Ferraro, S.; Chen, S.F.; Kokron, N.; Bailey,
  S.; Brooks, D.; Garc{\'{\i} }a-Bellido, J.; Guy, J.;  et~al.
\newblock Cosmological constraints from the tomographic cross-correlation of
  {DESI} Luminous Red Galaxies and Planck {CMB} lensing.
\newblock {\em J. Cosmol. Astropart. Phys.} {\bf 2022}, {\em
  2022}, 007.
\newblock {\url{https://doi.org/10.1088/1475-7516/2022/02/007}}.

\bibitem[Jeffreys(1998)]{JeffreysScale}
Jeffreys, H.
\newblock {\em {The Theory of Probability}}; {Oxford University Press}: Oxford, UK, 1998.

\bibitem[Nesseris and Tsujikawa(2014)]{Nesseris_2014_pert}
Nesseris, S.; Tsujikawa, S.
\newblock Cosmological perturbations and observational constraints on nonlocal
  massive gravity.
\newblock {\em Phys. Rev. D} {\bf 2014}, {\em 90}, 024070.
\newblock {\url{https://doi.org/10.1103/physrevd.90.024070}}.

\bibitem[Dutcher \em{et~al.}(2021)Dutcher, Balkenhol, Ade, Ahmed, Anderes,
  Anderson, Archipley, Avva, Aylor, Barry, Thakur, Benabed, Bender, Benson,
  Bianchini, Bleem, Bouchet, Bryant, Byrum, Carlstrom, Carter, Cecil, Chang,
  Chaubal, Chen, Cho, Chou, Cliche, Crawford, Cukierman, Daley, de~Haan,
  Denison, Dibert, Ding, Dobbs, Everett, Feng, Ferguson, Foster, Fu, Galli,
  Gambrel, Gardner, Goeckner-Wald, Gualtieri, Guns, Gupta, Guyser, Halverson,
  Harke-Hosemann, Harrington, Henning, Hilton, Hivon, Holder, Holzapfel, Hood,
  Howe, Huang, Irwin, Jeong, Jonas, Jones, Khaire, Knox, Kofman, Korman, Kubik,
  Kuhlmann, Kuo, Lee, Leitch, Lowitz, Lu, Meyer, Michalik, Millea, Montgomery,
  Nadolski, Natoli, Nguyen, Noble, Novosad, Omori, Padin, Pan, Paschos,
  Pearson, Posada, Prabhu, Quan, Raghunathan, Rahlin, Reichardt, Riebel,
  Riedel, Rouble, Ruhl, Sayre, Schiappucci, Shirokoff, Smecher, Sobrin, Stark,
  Stephen, Story, Suzuki, Thompson, Thorne, Tucker, Umilta, Vale, Vanderlinde,
  Vieira, Wang, Whitehorn, Wu, Yefremenko, Yoon, and and]{Dutcher_2021}
Dutcher, D.; Balkenhol, L.; Ade, P.; Ahmed, Z.; Anderes, E.; Anderson, A.;
  Archipley, M.; Avva, J.; Aylor, K.; Barry, P.;  et~al.
\newblock Measurements of the E-mode polarization and temperature-E-mode
  correlation of the CMB from SPT-3G 2018 data.
\newblock {\em Phys. Rev. D} {\bf 2021}, {\em 104}, 022003.
\newblock {\url{https://doi.org/10.1103/physrevd.104.022003}}.

\bibitem[Wang and Huang(2020)]{Wang_2020}
Wang, K.; Huang, Q.G.
\newblock Implications for cosmology from ground-based Cosmic Microwave
  Background observations.
\newblock {\em J. Cosmol. Astropart. Phys.} {\bf 2020}, {\em
  2020}, 045.
\newblock {\url{https://doi.org/10.1088/1475-7516/2020/06/045}}.

\bibitem[Balkenhol \em{et~al.}(2021)Balkenhol, Dutcher, Ade, Ahmed, Anderes,
  Anderson, Archipley, Avva, Aylor, Barry, Thakur, Benabed, Bender, Benson,
  Bianchini, Bleem, Bouchet, Bryant, Byrum, Carlstrom, Carter, Cecil, Chang,
  Chaubal, Chen, Cho, Chou, Cliche, Crawford, Cukierman, Daley, de~Haan,
  Denison, Dibert, Ding, Dobbs, Everett, Feng, Ferguson, Foster, Fu, Galli,
  Gambrel, Gardner, Goeckner-Wald, Gualtieri, Guns, Gupta, Guyser, Halverson,
  Harke-Hosemann, Harrington, Henning, Hilton, Hivon, Holder, Holzapfel, Hood,
  Howe, Huang, Irwin, Jeong, Jonas, Jones, Khaire, Knox, Kofman, Korman, Kubik,
  Kuhlmann, Kuo, Lee, Leitch, Lowitz, Lu, Meyer, Michalik, Millea, Montgomery,
  Nadolski, Natoli, Nguyen, Noble, Novosad, Omori, Padin, Pan, Paschos,
  Pearson, Posada, Prabhu, Quan, Rahlin, Reichardt, Riebel, Riedel, Rouble,
  Ruhl, Sayre, Schiappucci, Shirokoff, Smecher, Sobrin, Stark, Stephen, Story,
  Suzuki, Thompson, Thorne, Tucker, Umilta, Vale, Vanderlinde, Vieira, Wang,
  Whitehorn, Wu, Yefremenko, Yoon, and and]{Balkenhol_2021}
Balkenhol, L.; Dutcher, D.; Ade, P.; Ahmed, Z.; Anderes, E.; Anderson, A.;
  Archipley, M.; Avva, J.; Aylor, K.; Barry, P.;  et~al.
\newblock Constraints on $\Lambda$CDM extensions from the SPT-3G 2018 EE and TE
  power spectra.
\newblock {\em Phys. Rev. D} {\bf 2021}, {\em 104}, 083509.
\newblock {\url{https://doi.org/10.1103/physrevd.104.083509}}.

\bibitem[Addison(2021)]{Addison_2021}
Addison, G.E.
\newblock High $H_0$ values from CMB E-mode Data: A Clue for Resolving the
  Hubble Tension?
\newblock {\em  Astrophys. J. Lett.} {\bf 2021}, {\em 912}, L1.
\newblock {\url{https://doi.org/10.3847/2041-8213/abf56e}}.

\bibitem[D. Amico \em{et~al.}(2020)D.Amico,
  Gleyzes, Kokron, Markovic, Senatore, Zhang, Beutler, and
  Gil-Mar{\'{\i}}n]{d_Amico_2020}
D'Amico, G.; Gleyzes, J.; Kokron, N.; Markovic, K.; Senatore,
  L.; Zhang, P.; Beutler, F.; Gil-Mar{\'{\i}}n, H.
\newblock The cosmological analysis of the {SDSS}/{BOSS} data from the
  Effective Field Theory of Large-Scale Structure.
\newblock {\em J. Cosmol. Astropart. Phys.} {\bf 2020}, {\em
  2020}, 005.
\newblock {\url{https://doi.org/10.1088/1475-7516/2020/05/005}}.

\bibitem[Colas \em{et~al.}(2020)Colas, d'Amico, Senatore,
  Zhang, and Beutler]{Colas_2020}
Colas, T.; D'Amico, G.; Senatore, L.; Zhang, P.; Beutler, F.
\newblock Efficient cosmological analysis of the {SDSS}/{BOSS} data from the
  Effective Field Theory of Large-Scale Structure.
\newblock {\em J. Cosmol. Astropart. Phys.} {\bf 2020}, {\em
  2020}, 001.
\newblock {\url{https://doi.org/10.1088/1475-7516/2020/06/001}}.

\bibitem[Cooke \em{et~al.}(2018)Cooke, Pettini, and Steidel]{Cooke_2018}
Cooke, R.J.; Pettini, M.; Steidel, C.C.
\newblock One Percent Determination of the Primordial Deuterium Abundance.
\newblock {\em  Astrophys. J.} {\bf 2018}, {\em 855}, 102.
\newblock {\url{https://doi.org/10.3847/1538-4357/aaab53}}.

\bibitem[Scolnic \em{et~al.}(2018)Scolnic, Jones, Rest, Pan, Chornock, Foley,
  Huber, Kessler, Narayan, Riess, Rodney, Berger, Brout, Challis, Drout,
  Finkbeiner, Lunnan, Kirshner, Sanders, Schlafly, Smartt, Stubbs, Tonry,
  Wood-Vasey, Foley, Hand, Johnson, Burgett, Chambers, Draper, Hodapp, Kaiser,
  Kudritzki, Magnier, Metcalfe, Bresolin, Gall, Kotak, McCrum, and
  Smith]{Scolnic_2018}
Scolnic, D.M.; Jones, D.O.; Rest, A.; Pan, Y.C.; Chornock, R.; Foley, R.J.;
  Huber, M.E.; Kessler, R.; Narayan, G.; Riess, A.G.;  et~al.
\newblock The Complete Light-curve Sample of Spectroscopically Confirmed {SNe}
  Ia from Pan-{STARRS}1 and Cosmological Constraints from the Combined Pantheon
  Sample.
\newblock {\em  Astrophys. J.} {\bf 2018}, {\em 859}, 101.
\newblock {\url{https://doi.org/10.3847/1538-4357/aab9bb}}.

\bibitem[Abbott \em{et~al.}(2018)Abbott, Abdalla, Alarcon, Aleksi{\'{c}},
  Allam, Allen, Amara, Annis, Asorey, Avila, Bacon, Balbinot, Banerji, Banik,
  Barkhouse, Baumer, Baxter, Bechtol, Becker, Benoit-L{\'{e}}vy, Benson,
  Bernstein, Bertin, Blazek, Bridle, Brooks, Brout, Buckley-Geer, Burke, Busha,
  Campos, Capozzi, Rosell, Kind, Carretero, Castander, Cawthon, Chang, Chen,
  Childress, Choi, Conselice, Crittenden, Crocce, Cunha, D'Andrea, da~Costa,
  Das, Davis, Davis, Vicente, DePoy, DeRose, Desai, Diehl, Dietrich, Dodelson,
  Doel, Drlica-Wagner, Eifler, Elliott, Elsner, Elvin-Poole, Estrada, Evrard,
  Fang, Fernandez, Fert{\'{e}}, Finley, Flaugher, Fosalba, Friedrich, Frieman,
  Garc{\'{\i}}a-Bellido, Garcia-Fernandez, Gatti, Gaztanaga, Gerdes,
  Giannantonio, Gill, Glazebrook, Goldstein, Gruen, Gruendl, Gschwend,
  Gutierrez, Hamilton, Hartley, Hinton, Honscheid, Hoyle, Huterer, Jain, James,
  Jarvis, Jeltema, Johnson, Johnson, Kacprzak, Kent, Kim, King, Kirk, Kokron,
  Kovacs, Krause, Krawiec, Kremin, Kuehn, Kuhlmann, Kuropatkin, Lacasa, Lahav,
  Li, Liddle, Lidman, Lima, Lin, MacCrann, Maia, Makler, Manera, March,
  Marshall, Martini, McMahon, Melchior, Menanteau, Miquel, Miranda, Mudd, Muir,
  Möller, Neilsen, Nichol, Nord, Nugent, Ogando, Palmese, Peacock, Peiris,
  Peoples, Percival, Petravick, Plazas, Porredon, Prat, Pujol, Rau, Refregier,
  Ricker, Roe, Rollins, Romer, Roodman, Rosenfeld, Ross, Rozo, Rykoff, Sako,
  Salvador, Samuroff, S{\'{a}}nchez, Sanchez, Santiago, Scarpine, Schindler,
  Scolnic, Secco, Serrano, Sevilla-Noarbe, Sheldon, Smith, Smith, Smith,
  Soares-Santos, Sobreira, Suchyta, Tarle, Thomas, Troxel, Tucker, Tucker,
  Uddin, Varga, Vielzeuf, Vikram, Vivas, Walker, Wang, Wechsler, Weller,
  Wester, Wolf, Yanny, Yuan, Zenteno, Zhang, Zhang, and and]{Abbott_2018}
Abbott, T.; Abdalla, F.; Alarcon, A.; Aleksi{\'{c}}, J.; Allam, S.; Allen, S.;
  Amara, A.; Annis, J.; Asorey, J.; Avila, S.;  et~al.
\newblock Dark Energy Survey year 1 results: Cosmological constraints from
  galaxy clustering and weak lensing.
\newblock {\em Phys. Rev. D} {\bf 2018}, {\em 98}, 043526.
\newblock {\url{https://doi.org/10.1103/physrevd.98.043526}}.

\bibitem[Krause \em{et~al.}(2017)Krause, Eifler, Zuntz, Friedrich, Troxel,
  Dodelson, Blazek, Secco, MacCrann, Baxter, Chang, Chen, Crocce, DeRose,
  Ferte, Kokron, Lacasa, Miranda, Omori, Porredon, Rosenfeld, Samuroff, Wang,
  Wechsler, Abbott, Abdalla, Allam, Annis, Bechtol, Benoit-Levy, Bernstein,
  Brooks, Burke, Capozzi, Kind, Carretero, D'Andrea, da~Costa, Davis, DePoy,
  Desai, Diehl, Dietrich, Evrard, Flaugher, Fosalba, Frieman, Garcia-Bellido,
  Gaztanaga, Giannantonio, Gruen, Gruendl, Gschwend, Gutierrez, Honscheid,
  James, Jeltema, Kuehn, Kuhlmann, Lahav, Lima, Maia, March, Marshall, Martini,
  Menanteau, Miquel, Nichol, Plazas, Romer, Rykoff, Sanchez, Scarpine,
  Schindler, Schubnell, Sevilla-Noarbe, Smith, Soares-Santos, Sobreira,
  Suchyta, Swanson, Tarle, Tucker, Vikram, Walker, and
  Weller]{https://doi.org/10.48550/arxiv.1706.09359}
Krause, E.; Eifler, T.F.; Zuntz, J.; Friedrich, O.; Troxel, M.A.; Dodelson, S.;
  Blazek, J.; Secco, L.F.; MacCrann, N.; Baxter, E.;  et~al.
\newblock Dark Energy Survey Year 1 Results: Multi-Probe Methodology and
  Simulated Likelihood Analyses. \emph{arXiv Prepr.}  \textbf{2017}, arXiv:1706.09359.
\newblock {\url{https://doi.org/10.48550/ARXIV.1706.09359}}.

\bibitem[Riess(2019)]{Riess_2019_review}
Riess, A.G.
\newblock The expansion of the Universe is faster than expected.
\newblock {\em Nat. Rev. Phys.} {\bf 2019}, {\em 2}, 10--12.
\newblock {\url{https://doi.org/10.1038/s42254-019-0137-0}}.

\bibitem[Riess \em{et~al.}(2011)Riess, Macri, Casertano, Lampeitl, Ferguson,
  Filippenko, Jha, Li, and Chornock]{Riess_2011}
Riess, A.G.; Macri, L.; Casertano, S.; Lampeitl, H.; Ferguson, H.C.;
  Filippenko, A.V.; Jha, S.W.; Li, W.; Chornock, R.
\newblock A 3{\%} solution: Determination of the Hubble constant with the
  Hubble Space Telescope and the Wide Field Camera 3.
\newblock {\em  Astrophys. J.} {\bf 2011}, {\em 730}, 119.
\newblock {\url{https://doi.org/10.1088/0004-637x/730/2/119}}.

\bibitem[Riess \em{et~al.}(2016)Riess, Macri, Hoffmann, Scolnic, Casertano,
  Filippenko, Tucker, Reid, Jones, Silverman, Chornock, Challis, Yuan, Brown,
  and Foley]{Riess_2016}
Riess, A.G.; Macri, L.M.; Hoffmann, S.L.; Scolnic, D.; Casertano, S.;
  Filippenko, A.V.; Tucker, B.E.; Reid, M.J.; Jones, D.O.; Silverman, J.M.;
  et~al.
\newblock A 2.4{\%} determination of the local value of the Hubble constant.
\newblock {\em  Astrophys. J.} {\bf 2016}, {\em 826}, 56.
\newblock {\url{https://doi.org/10.3847/0004-637x/826/1/56}}.

\bibitem[Riess \em{et~al.}(2021)Riess, Casertano, Yuan, Bowers, Macri, Zinn,
  and Scolnic]{Riess_2021}
Riess, A.G.; Casertano, S.; Yuan, W.; Bowers, J.B.; Macri, L.; Zinn, J.C.;
  Scolnic, D.
\newblock Cosmic Distances Calibrated to 1{\%} Precision with Gaia {EDR}3
  Parallaxes and Hubble Space Telescope Photometry of 75 Milky Way Cepheids
  Confirm Tension with $\Lambda$CDM.
\newblock {\em  Astrophys. J. Lett.} {\bf 2021}, {\em 908}, L6.
\newblock {\url{https://doi.org/10.3847/2041-8213/abdbaf}}.

\bibitem[{Gaia Collaboration} \em{et~al.}(2021){Gaia Collaboration}, {Brown, A.
  G. A.}, {Vallenari, A.}, {Prusti, T.}, {de Bruijne, J. H. J.}, {Babusiaux,
  C.}, {Biermann, M.}, {Creevey, O. L.}, {Evans, D. W.}, {Eyer, L.}, {Hutton,
  A.}, {Jansen, F.}, {Jordi, C.}, {Klioner, S. A.}, {Lammers, U.}, {Lindegren,
  L.}, {Luri, X.}, {Mignard, F.}, {Panem, C.}, {Pourbaix, D.}, {Randich, S.},
  {Sartoretti, P.}, {Soubiran, C.}, {Walton, N. A.}, {Arenou, F.},
  {Bailer-Jones, C. A. L.}, {Bastian, U.}, {Cropper, M.}, {Drimmel, R.}, {Katz,
  D.}, {Lattanzi, M. G.}, {van Leeuwen, F.}, {Bakker, J.}, {Cacciari, C.},
  {Casta\~neda, J.}, {De Angeli, F.}, {Ducourant, C.}, {Fabricius, C.},
  {Fouesneau, M.}, {Fr\'emat, Y.}, {Guerra, R.}, {Guerrier, A.}, {Guiraud, J.},
  {Jean-Antoine Piccolo, A.}, {Masana, E.}, {Messineo, R.}, {Mowlavi, N.},
  {Nicolas, C.}, {Nienartowicz, K.}, {Pailler, F.}, {Panuzzo, P.}, {Riclet,
  F.}, {Roux, W.}, {Seabroke, G. M.}, {Sordo, R.}, {Tanga, P.}, {Th\'evenin,
  F.}, {Gracia-Abril, G.}, {Portell, J.}, {Teyssier, D.}, {Altmann, M.},
  {Andrae, R.}, {Bellas-Velidis, I.}, {Benson, K.}, {Berthier, J.}, {Blomme,
  R.}, {Brugaletta, E.}, {Burgess, P. W.}, {Busso, G.}, {Carry, B.}, {Cellino,
  A.}, {Cheek, N.}, {Clementini, G.}, {Damerdji, Y.}, {Davidson, M.},
  {Delchambre, L.}, {Dell\'{}Oro, A.}, {Fern\'andez-Hern\'andez, J.},
  {Galluccio, L.}, {Garc\'{\i}a-Lario, P.}, {Garcia-Reinaldos, M.},
  {Gonz\'alez-N\'u\~nez, J.}, {Gosset, E.}, {Haigron, R.}, {Halbwachs, J.-L.},
  {Hambly, N. C.}, {Harrison, D. L.}, {Hatzidimitriou, D.}, {Heiter, U.},
  {Hern\'andez, J.}, {Hestroffer, D.}, {Hodgkin, S. T.}, {Holl, B.},
  {Jan\ss{}en, K.}, {Jevardat de Fombelle, G.}, {Jordan, S.}, {Krone-Martins,
  A.}, {Lanzafame, A. C.}, {L\"offler, W.}, {Lorca, A.}, {Manteiga, M.},
  {Marchal, O.}, {Marrese, P. M.}, {Moitinho, A.}, {Mora, A.}, {Muinonen, K.},
  {Osborne, P.}, {Pancino, E.}, {Pauwels, T.}, {Petit, J.-M.}, {Recio-Blanco,
  A.}, {Richards, P. J.}, {Riello, M.}, {Rimoldini, L.}, {Robin, A. C.},
  {Roegiers, T.}, {Rybizki, J.}, {Sarro, L. M.}, {Siopis, C.}, {Smith, M.},
  {Sozzetti, A.}, {Ulla, A.}, {Utrilla, E.}, {van Leeuwen, M.}, {van Reeven,
  W.}, {Abbas, U.}, {Abreu Aramburu, A.}, {Accart, S.}, {Aerts, C.}, {Aguado,
  J. J.}, {Ajaj, M.}, {Altavilla, G.}, {\'Alvarez, M. A.}, {\'Alvarez
  Cid-Fuentes, J.}, {Alves, J.}, {Anderson, R. I.}, {Anglada Varela, E.},
  {Antoja, T.}, {Audard, M.}, {Baines, D.}, {Baker, S. G.},
  {Balaguer-N\'u\~nez, L.}, {Balbinot, E.}, {Balog, Z.}, {Barache, C.},
  {Barbato, D.}, {Barros, M.}, {Barstow, M. A.}, {Bartolom\'e, S.}, {Bassilana,
  J.-L.}, {Bauchet, N.}, {Baudesson-Stella, A.}, {Becciani, U.}, {Bellazzini,
  M.}, {Bernet, M.}, {Bertone, S.}, {Bianchi, L.}, {Blanco-Cuaresma, S.},
  {Boch, T.}, {Bombrun, A.}, {Bossini, D.}, {Bouquillon, S.}, {Bragaglia, A.},
  {Bramante, L.}, {Breedt, E.}, {Bressan, A.}, {Brouillet, N.}, {Bucciarelli,
  B.}, {Burlacu, A.}, {Busonero, D.}, {Butkevich, A. G.}, {Buzzi, R.}, {Caffau,
  E.}, {Cancelliere, R.}, {C\'anovas, H.}, {Cantat-Gaudin, T.}, {Carballo, R.},
  {Carlucci, T.}, {Carnerero, M. I}, {Carrasco, J. M.}, {Casamiquela, L.},
  {Castellani, M.}, {Castro-Ginard, A.}, {Castro Sampol, P.}, {Chaoul, L.},
  {Charlot, P.}, {Chemin, L.}, {Chiavassa, A.}, {Cioni, M.-R. L.}, {Comoretto,
  G.}, {Cooper, W. J.}, {Cornez, T.}, {Cowell, S.}, {Crifo, F.}, {Crosta, M.},
  {Crowley, C.}, {Dafonte, C.}, {Dapergolas, A.}, {David, M.}, {David, P.}, {de
  Laverny, P.}, {De Luise, F.}, {De March, R.}, {De Ridder, J.}, {de Souza,
  R.}, {de Teodoro, P.}, {de Torres, A.}, {del Peloso, E. F.}, {del Pozo, E.},
  {Delbo, M.}, {Delgado, A.}, {Delgado, H. E.}, {Delisle, J.-B.}, {Di Matteo,
  P.}, {Diakite, S.}, {Diener, C.}, {Distefano, E.}, {Dolding, C.}, {Eappachen,
  D.}, {Edvardsson, B.}, {Enke, H.}, {Esquej, P.}, {Fabre, C.}, {Fabrizio, M.},
  {Faigler, S.}, {Fedorets, G.}, {Fernique, P.}, {Fienga, A.}, {Figueras, F.},
  {Fouron, C.}, {Fragkoudi, F.}, {Fraile, E.}, {Franke, F.}, {Gai, M.},
  {Garabato, D.}, {Garcia-Gutierrez, A.}, {Garc\'{\i}a-Torres, M.}, {Garofalo,
  A.}, {Gavras, P.}, {Gerlach, E.}, {Geyer, R.}, {Giacobbe, P.}, {Gilmore, G.},
  {Girona, S.}, {Giuffrida, G.}, {Gomel, R.}, {Gomez, A.},
  {Gonzalez-Santamaria, I.}, {Gonz\'alez-Vidal, J. J.}, {Granvik, M.},
  {Guti\'errez-S\'anchez, R.}, {Guy, L. P.}, {Hauser, M.}, {Haywood, M.},
  {Helmi, A.}, {Hidalgo, S. L.}, {Hilger, T.}, {Hladczuk, N.}, {Hobbs, D.},
  {Holland, G.}, {Huckle, H. E.}, {Jasniewicz, G.}, {Jonker, P. G.}, {Juaristi
  Campillo, J.}, {Julbe, F.}, {Karbevska, L.}, {Kervella, P.}, {Khanna, S.},
  {Kochoska, A.}, {Kontizas, M.}, {Kordopatis, G.}, {Korn, A. J.},
  {Kostrzewa-Rutkowska, Z.}, {Kruszy\'{}nska, K.}, {Lambert, S.}, {Lanza, A.
  F.}, {Lasne, Y.}, {Le Campion, J.-F.}, {Le Fustec, Y.}, {Lebreton, Y.},
  {Lebzelter, T.}, {Leccia, S.}, {Leclerc, N.}, {Lecoeur-Taibi, I.}, {Liao,
  S.}, {Licata, E.}, {Lindstr\o{}m, E. P.}, {Lister, T. A.}, {Livanou, E.},
  {Lobel, A.}, {Madrero Pardo, P.}, {Managau, S.}, {Mann, R. G.}, {Marchant, J.
  M.}, {Marconi, M.}, {Marcos Santos, M. M. S.}, {Marinoni, S.}, {Marocco, F.},
  {Marshall, D. J.}, {Martin Polo, L.}, {Mart\'{\i}n-Fleitas, J. M.}, {Masip,
  A.}, {Massari, D.}, {Mastrobuono-Battisti, A.}, {Mazeh, T.}, {McMillan, P.
  J.}, {Messina, S.}, {Michalik, D.}, {Millar, N. R.}, {Mints, A.}, {Molina,
  D.}, {Molinaro, R.}, {Moln\'ar, L.}, {Montegriffo, P.}, {Mor, R.},
  {Morbidelli, R.}, {Morel, T.}, {Morris, D.}, {Mulone, A. F.}, {Munoz, D.},
  {Muraveva, T.}, {Murphy, C. P.}, {Musella, I.}, {Noval, L.}, {Ord\'enovic,
  C.}, {Orr\`u, G.}, {Osinde, J.}, {Pagani, C.}, {Pagano, I.}, {Palaversa, L.},
  {Palicio, P. A.}, {Panahi, A.}, {Pawlak, M.}, {Pe\~nalosa Esteller, X.},
  {Penttil\"a, A.}, {Piersimoni, A. M.}, {Pineau, F.-X.}, {Plachy, E.}, {Plum,
  G.}, {Poggio, E.}, {Poretti, E.}, {Poujoulet, E.}, {Prsa, A.}, {Pulone, L.},
  {Racero, E.}, {Ragaini, S.}, {Rainer, M.}, {Raiteri, C. M.}, {Rambaux, N.},
  {Ramos, P.}, {Ramos-Lerate, M.}, {Re Fiorentin, P.}, {Regibo, S.}, {Reyl\'e,
  C.}, {Ripepi, V.}, {Riva, A.}, {Rixon, G.}, {Robichon, N.}, {Robin, C.},
  {Roelens, M.}, {Rohrbasser, L.}, {Romero-G\'omez, M.}, {Rowell, N.}, {Royer,
  F.}, {Rybicki, K. A.}, {Sadowski, G.}, {Sagrist\`a Sell\'es, A.}, {Sahlmann,
  J.}, {Salgado, J.}, {Salguero, E.}, {Samaras, N.}, {Sanchez Gimenez, V.},
  {Sanna, N.}, {Santove\~na, R.}, {Sarasso, M.}, {Schultheis, M.}, {Sciacca,
  E.}, {Segol, M.}, {Segovia, J. C.}, {S\'egransan, D.}, {Semeux, D.}, {Shahaf,
  S.}, {Siddiqui, H. I.}, {Siebert, A.}, {Siltala, L.}, {Slezak, E.}, {Smart,
  R. L.}, {Solano, E.}, {Solitro, F.}, {Souami, D.}, {Souchay, J.}, {Spagna,
  A.}, {Spoto, F.}, {Steele, I. A.}, {Steidelm\"uller, H.}, {Stephenson, C.
  A.}, {S\"uveges, M.}, {Szabados, L.}, {Szegedi-Elek, E.}, {Taris, F.},
  {Tauran, G.}, {Taylor, M. B.}, {Teixeira, R.}, {Thuillot, W.}, {Tonello, N.},
  {Torra, F.}, {Torra, J.}, {Turon, C.}, {Unger, N.}, {Vaillant, M.}, {van
  Dillen, E.}, {Vanel, O.}, {Vecchiato, A.}, {Viala, Y.}, {Vicente, D.},
  {Voutsinas, S.}, {Weiler, M.}, {Wevers, T.}, {Wyrzykowski, L.}, {Yoldas, A.},
  {Yvard, P.}, {Zhao, H.}, {Zorec, J.}, {Zucker, S.}, {Zurbach, C.}, and
  {Zwitter, T.}]{GAIA}
{Gaia Collaboration}.; {Brown, A. G. A.}.; {Vallenari, A.}.; {Prusti, T.}.; {de
  Bruijne, J. H. J.}.; {Babusiaux, C.}.; {Biermann, M.}.; {Creevey, O. L.}.;
  {Evans, D. W.}.; {Eyer, L.}.;  et~al.
\newblock Gaia Early Data Release 3---Summary of the contents and survey
  properties.
\newblock {\em Astron. Astrophys.} {\bf 2021}, {\em 649}.
\newblock {\url{https://doi.org/10.1051/0004-6361/202039657}}.

\bibitem[{Lindegren, L.} \em{et~al.}(2021){Lindegren, L.}, {Bastian, U.},
  {Biermann, M.}, {Bombrun, A.}, {de Torres, A.}, {Gerlach, E.}, {Geyer, R.},
  {Hern\'andez, J.}, {Hilger, T.}, {Hobbs, D.}, {Klioner, S. A.}, {Lammers,
  U.}, {McMillan, P. J.}, {Ramos-Lerate, M.}, {Steidelm\"uller, H.},
  {Stephenson, C. A.}, and {van Leeuwen, F.}]{refId0}
{Lindegren, L.}.; {Bastian, U.}.; {Biermann, M.}.; {Bombrun, A.}.; {de Torres,
  A.}.; {Gerlach, E.}.; {Geyer, R.}.; {Hern\'andez, J.}.; {Hilger, T.}.;
  {Hobbs, D.}.;  et~al.
\newblock Gaia Early Data Release 3 - Parallax bias versus magnitude, colour,
  and position.
\newblock {\em Astronomy \& Astrophysics} {\bf 2021}, {\em 649}, A1.
\newblock {\url{https://doi.org/10.1051/0004-6361/202039653}}.

\bibitem[Wong \em{et~al.}(2019)Wong, Suyu, Chen, Rusu, Millon, Sluse, Bonvin,
  Fassnacht, Taubenberger, Auger, Birrer, Chan, Courbin, Hilbert, Tihhonova,
  Treu, Agnello, Ding, Jee, Komatsu, Shajib, Sonnenfeld, Blandford, Koopmans,
  Marshall, and Meylan]{Wong_2019aa}
Wong, K.C.; Suyu, S.H.; Chen, G.C.F.; Rusu, C.E.; Millon, M.; Sluse, D.;
  Bonvin, V.; Fassnacht, C.D.; Taubenberger, S.; Auger, M.W.;  et~al.
\newblock H0LiCOW - {XIII}. A 2.4 per cent measurement of H0 from lensed
  quasars: 5.3$\sigma$ tension between early- and late-Universe probes.
\newblock {\em Mon. Not. R. Astron. Soc.} {\bf 2019},
  {\em 498}, 1420--1439.
\newblock {\url{https://doi.org/10.1093/mnras/stz3094}}.

\bibitem[Shajib \em{et~al.}(2020)Shajib, Birrer, Treu, Agnello, Buckley-Geer,
  Chan, Christensen, Lemon, Lin, Millon, Poh, Rusu, Sluse, Spiniello, Chen,
  Collett, Courbin, Fassnacht, Frieman, Galan, Gilman, More, Anguita, Auger,
  Bonvin, McMahon, Meylan, Wong, Abbott, Annis, Avila, Bechtol, Brooks, Brout,
  Burke, Rosell, Kind, Carretero, Castander, Costanzi, da~Costa, Vicente,
  Desai, Dietrich, Doel, Drlica-Wagner, Evrard, Finley, Flaugher, Fosalba,
  Garc{\'{\i} }a-Bellido, Gerdes, Gruen, Gruendl, Gschwend, Gutierrez,
  Hollowood, Honscheid, Huterer, James, Jeltema, Krause, Kuropatkin, Li, Lima,
  MacCrann, Maia, Marshall, Melchior, Miquel, Ogando, Palmese,
  Paz-Chinch{\'{o}}n, Plazas, Romer, Roodman, Sako, Sanchez, Santiago,
  Scarpine, Schubnell, Scolnic, Serrano, Sevilla-Noarbe, Smith, Soares-Santos,
  Suchyta, Tarle, Thomas, Walker, and Zhang]{Shajib_2020}
Shajib, A.J.; Birrer, S.; Treu, T.; Agnello, A.; Buckley-Geer, E.J.; Chan,
  J.H.H.; Christensen, L.; Lemon, C.; Lin, H.; Millon, M.;  et~al.
\newblock {STRIDES}: A 3.9 per cent measurement of the Hubble constant from the
  strong lens system {DES} J0408-5354.
\newblock {\em Mon. Not. R. Astron. Soc.} {\bf 2020},
  {\em 494}, 6072--6102.
\newblock {\url{https://doi.org/10.1093/mnras/staa828}}.

\bibitem[Yuan \em{et~al.}(2019)Yuan, Riess, Macri, Casertano, and
  Scolnic]{Yuan_2019}
Yuan, W.; Riess, A.G.; Macri, L.M.; Casertano, S.; Scolnic, D.M.
\newblock Consistent Calibration of the Tip of the Red Giant Branch in the
  Large Magellanic Cloud on the Hubble Space Telescope Photometric System and a
  Re-determination of the Hubble Constant.
\newblock {\em  Astrophys. J.} {\bf 2019}, {\em 886}, 61.
\newblock {\url{https://doi.org/10.3847/1538-4357/ab4bc9}}.

\bibitem[Huang \em{et~al.}(2020)Huang, Riess, Yuan, Macri, Zakamska, Casertano,
  Whitelock, Hoffmann, Filippenko, and Scolnic]{Huang_2020_miras}
Huang, C.D.; Riess, A.G.; Yuan, W.; Macri, L.M.; Zakamska, N.L.; Casertano, S.;
  Whitelock, P.A.; Hoffmann, S.L.; Filippenko, A.V.; Scolnic, D.
\newblock Hubble Space Telescope Observations of Mira Variables in the {SN} Ia
  Host {NGC} 1559: An Alternative Candle to Measure the Hubble Constant.
\newblock {\em  Astrophys. J.} {\bf 2020}, {\em 889}, 5.
\newblock {\url{https://doi.org/10.3847/1538-4357/ab5dbd}}.

\bibitem[Reid \em{et~al.}(2019)Reid, Pesce, and Riess]{Reid_2019}
Reid, M.J.; Pesce, D.W.; Riess, A.G.
\newblock An Improved Distance to {NGC} 4258 and Its Implications for the
  Hubble Constant.
\newblock {\em  Astrophys. J.} {\bf 2019}, {\em 886}, L27.
\newblock {\url{https://doi.org/10.3847/2041-8213/ab552d}}.

\bibitem[Zhang \em{et~al.}(2022)Zhang, d'Amico, Senatore,
  Zhao, and Cai]{Zhang_2022}
Zhang, P.; D'Amico, G.; Senatore, L.; Zhao, C.; Cai, Y.
\newblock {BOSS} Correlation Function analysis from the Effective Field Theory
  of Large-Scale Structure.
\newblock {\em J. Cosmol. Astropart. Phys.} {\bf 2022}, {\em
  2022}, 036.
\newblock {\url{https://doi.org/10.1088/1475-7516/2022/02/036}}.

\bibitem[Wang \em{et~al.}(2017)Wang, Xu, and Zhao]{Wang_2017}
Wang, Y.; Xu, L.; Zhao, G.B.
\newblock A Measurement of the Hubble Constant Using Galaxy Redshift Surveys.
\newblock {\em  Astrophys. J.} {\bf 2017}, {\em 849}, 84.
\newblock {\url{https://doi.org/10.3847/1538-4357/aa8f48}}.

\bibitem[Farren \em{et~al.}(2022)Farren, Philcox, and Sherwin]{Farren_2022}
Farren, G.S.; Philcox, O.H.; Sherwin, B.D.
\newblock Determining the Hubble constant without the sound horizon:
  Perspectives with future galaxy surveys.
\newblock {\em Phys. Rev. D} {\bf 2022}, {\em 105}, 063503.
\newblock {\url{https://doi.org/10.1103/physrevd.105.063503}}.

\bibitem[Philcox \em{et~al.}(2021)Philcox, Sherwin, Farren, and
  Baxter]{Philcox_2021}
Philcox, O.H.; Sherwin, B.D.; Farren, G.S.; Baxter, E.J.
\newblock Determining the Hubble constant without the sound horizon:
  Measurements from galaxy surveys.
\newblock {\em Phys. Rev. D} {\bf 2021}, {\em 103}, 023538.
\newblock {\url{https://doi.org/10.1103/physrevd.103.023538}}.

\bibitem[Anand \em{et~al.}(2022)Anand, Tully, Rizzi, Riess, and
  Yuan]{Anand_2022}
Anand, G.S.; Tully, R.B.; Rizzi, L.; Riess, A.G.; Yuan, W.
\newblock Comparing Tip of the Red Giant Branch Distance Scales: An Independent
  Reduction of the Carnegie-Chicago Hubble Program and the Value of the Hubble
  Constant.
\newblock {\em  Astrophys. J.} {\bf 2022}, {\em 932}, 15.
\newblock {\url{https://doi.org/10.3847/1538-4357/ac68df}}.

\bibitem[de~Jaeger \em{et~al.}(2022)de~Jaeger, Galbany, Riess, Stahl, Shappee,
  Filippenko, and Zheng]{de_Jaeger_2022}
de~Jaeger, T.; Galbany, L.; Riess, A.G.; Stahl, B.E.; Shappee, B.J.;
  Filippenko, A.V.; Zheng, W.
\newblock A 5$\%$ measurement of the Hubble constant from Type {II} supernovae.
\newblock {\em Mon. Not. R. Astron. Soc.} {\bf 2022},
  {\em 514}, 4620--4628.
\newblock {\url{https://doi.org/10.1093/mnras/stac1661}}.

\bibitem[Denzel \em{et~al.}(2020)Denzel, Coles, Saha, and
  Williams]{Denzel_2020}
Denzel, P.; Coles, J.P.; Saha, P.; Williams, L.L.R.
\newblock The Hubble constant from eight time-delay galaxy lenses.
\newblock {\em Mon. Not. R. Astron. Soc.} {\bf 2020},
  {\em 501}, 784--801.
\newblock {\url{https://doi.org/10.1093/mnras/staa3603}}.

\bibitem[Chen \em{et~al.}(2019)Chen, Fassnacht, Suyu, Rusu, Chan, Wong, Auger,
  Hilbert, Bonvin, Birrer, Millon, Koopmans, Lagattuta, McKean, Vegetti,
  Courbin, Ding, Halkola, Jee, Shajib, Sluse, Sonnenfeld, and Treu]{Chen_2019}
Chen, G.C.F.; Fassnacht, C.D.; Suyu, S.H.; Rusu, C.E.; Chan, J.H.H.; Wong,
  K.C.; Auger, M.W.; Hilbert, S.; Bonvin, V.; Birrer, S.;  et~al.
\newblock A {SHARP} view of H0LiCOW: H0 from three time-delay gravitational
  lens systems with adaptive optics imaging.
\newblock {\em Mon. Not. R. Astron. Soc.} {\bf 2019},
  {\em 490}, 1743--1773.
\newblock {\url{https://doi.org/10.1093/mnras/stz2547}}.

\bibitem[Mukherjee \em{et~al.}(2022)Mukherjee, Krolewski, Wandelt, and
  Silk]{Muk_2022}
Mukherjee, S.; Krolewski, A.; Wandelt, B.D.; Silk, J.
\newblock Cross-correlating dark sirens and galaxies: Measurement of $H_0$ from
  GWTC-3 of LIGO-Virgo-KAGRA. \emph{arXiv Prepr.} \textbf{2022}, arXiv:2203.03643.
\newblock {\url{https://doi.org/10.48550/ARXIV.2203.03643}}.

\bibitem[{The LIGO Scientific Collaboration} \em{et~al.}(2021){The LIGO
  Scientific Collaboration}, {The Virgo Collaboration}, {The KAGRA
  Collaboration}, Abbott, Abe, Acernese, Ackley, Adhikari, Adhikari, Adkins,
  Adya, Affeldt, Agarwal, Agathos, Agatsuma, Aggarwal, Aguiar, Aiello, Ain,
  Ajith, Akutsu, Albanesi, Alfaidi, Allocca, Altin, Amato, Anand, Anand,
  Ananyeva, Anderson, Anderson, Ando, Andrade, Andres, Andrés-Carcasona,
  Andrić, Angelova, Ansoldi, Antelis, Antier, Apostolatos, Appavuravther,
  Appert, Apple, Arai, Araya, Araya, Areeda, Arène, Aritomi, Arnaud, Arogeti,
  Aronson, Arun, Asada, Asali, Ashton, Aso, Assiduo, Melo, Aston, Astone,
  Aubin, AultONeal, Austin, Babak, Badaracco, Bader, Badger, Bae, Bae, Baer,
  Bagnasco, Bai, Baird, Bajpai, Baka, Ball, Ballardin, Ballmer, Balsamo,
  Baltus, Banagiri, Banerjee, Bankar, Barayoga, Barbieri, Barbieri, Barish,
  Barker, Barneo, Barone, Barr, Barsotti, Barsuglia, Barta, Bartlett, Barton,
  Bartos, Basak, Bassiri, Basti, Bawaj, Bayley, Bazzan, Becher, Bécsy,
  Bedakihale, Beirnaert, Bejger, Belahcene, Benedetto, Beniwal, Benjamin,
  Bennett, Bentley, BenYaala, Bera, Berbel, Bergamin, Berger, Bernuzzi, Berry,
  Bersanetti, Bertolini, Betzwieser, Beveridge, Bhandare, Bhandari, Bhardwaj,
  Bhatt, Bhattacharjee, Bhaumik, Bianchi, Bilenko, Billingsley, Bini, Birney,
  Birnholtz, Biscans, Bischi, Biscoveanu, Bisht, Biswas, Bitossi, Bizouard,
  Blackburn, Blair, Blair, Blair, Bobba, Bode, Boër, Bogaert, Boldrini,
  Bolingbroke, Bonavena, Bondu, Bonilla, Bonnand, Booker, Boom, Bork, Boschi,
  Bose, Bose, Bossilkov, Boudart, Bouffanais, Bozzi, Bradaschia, Brady,
  Bramley, Branch, Branchesi, Brau, Breschi, Briant, Briggs, Brillet,
  Brinkmann, Brockill, Brooks, Brooks, Brown, Brunett, Bruno, Bruntz, Bryant,
  Bucci, Bulik, Bulten, Buonanno, Burtnyk, Buscicchio, Buskulic, Buy, Byer,
  Davies, Cabras, Cabrita, Cadonati, Caesar, Cagnoli, Cahillane, Bustillo,
  Callaghan, Callister, Calloni, Cameron, Camp, Canepa, Canevarolo,
  Cannavacciuolo, Cannon, Cao, Cao, Capocasa, Capote, Carapella, Carbognani,
  Carlassara, Carlin, Carney, Carpinelli, Carrillo, Carullo, Carver, Diaz,
  Casentini, Castaldi, Caudill, Cavaglià, Cavalier, Cavalieri, Cella,
  Cerdá-Durán, Cesarini, Chaibi, Subrahmanya, Champion, Chan, Chan, Chan,
  Chan, Chan, Chandra, Chang, Chanial, Chao, Chapman-Bird, Charlton, Chase,
  Chassande-Mottin, Chatterjee, Chatterjee, Chatterjee, Chaturvedi, Chaty,
  Chatziioannou, Chen, Chen, Chen, Chen, Chen, Chen, Chen, Chen, Chen, Cheng,
  Cheong, Cheung, Chia, Chiadini, Chiang, Chiarini, Chierici, Chincarini,
  Chiofalo, Chiummo, Choudhary, Choudhary, Christensen, Chu, Chu, Chua, Chung,
  Ciani, Ciecielag, Cieślar, Cifaldi, Ciobanu, Ciolfi, Cipriano, Clara, Clark,
  Clearwater, Clesse, Cleva, Coccia, Codazzo, Cohadon, Cohen, Colleoni,
  Collette, Colombo, Colpi, Compton, Constancio, Conti, Cooper, Corban,
  Corbitt, Cordero-Carrión, Corezzi, Corley, Cornish, Corre, Corsi, Cortese,
  Costa, Cotesta, Cottingham, Coughlin, Coulon, Countryman, Cousins, Couvares,
  Coward, Cowart, Coyne, Coyne, Creighton, Creighton, Criswell, Croquette,
  Crowder, Cudell, Cullen, Cumming, Cummings, Cunningham, Cuoco, Curyło,
  Dabadie, Canton, Dall'Osso, Dálya, Dana, D'Angelo, Danilishin, D'Antonio,
  Danzmann, Darsow-Fromm, Dasgupta, Datrier, Datta, Datta, Dattilo, Dave,
  Davier, Davis, Davis, Daw, de~Alarcón, Dean, DeBra, Deenadayalan, Degallaix,
  De~Laurentis, Deléglise, Del~Favero, De~Lillo, De~Lillo, Dell'Aquila,
  Del~Pozzo, DeMarchi, De~Matteis, D'Emilio, Demos, Dent, Depasse, De~Pietri,
  De~Rosa, De~Rossi, DeSalvo, De~Simone, Dhurandhar, Díaz, Didio, Dietrich,
  Di~Fiore, Di~Fronzo, Di~Giorgio, Di~Giovanni, Di~Giovanni, Di~Girolamo,
  Di~Lieto, Di~Michele, Ding, Di~Pace, Di~Palma, Di~Renzo, Divakarla, Dmitriev,
  Doctor, Donahue, D'Onofrio, Donovan, Dooley, Doravari, Drago, Driggers,
  Drori, Ducoin, Dupej, Dupletsa, Durante, D'Urso, Duverne, Dwyer, Eassa,
  Easter, Ebersold, Eckhardt, Eddolls, Edelman, Edo, Edy, Effler, Eguchi,
  Eichholz, Eikenberry, Eisenmann, Eisenstein, Ejlli, Engelby, Enomoto, Errico,
  Essick, Estellés, Estevez, Etienne, Etzel, Evans, Evans, Evstafyeva, Ewing,
  Fabrizi, Faedi, Fafone, Fair, Fairhurst, Fan, Farah, Farinon, Farr, Farr,
  Fauchon-Jones, Favaro, Favata, Fays, Fazio, Feicht, Fejer, Fenyvesi,
  Ferguson, Fernandez-Galiana, Ferrante, Ferreira, Fidecaro, Figura, Fiori,
  Fiori, Fishbach, Fisher, Fittipaldi, Fiumara, Flaminio, Floden, Fong, Font,
  Fornal, Forsyth, Franke, Frasca, Frasconi, Freed, Frei, Freise, Freitas,
  Frey, Fritschel, Frolov, Fronzé, Fujii, Fujikawa, Fujimoto, Fulda, Fyffe,
  Gabbard, Gadre, Gair, Gais, Galaudage, Gamba, Ganapathy, Ganguly, Gao,
  Gaonkar, Garaventa, Núñez, García-Quirós, Garufi, Gateley, Gayathri, Ge,
  Gemme, Gennai, George, Gerberding, Gergely, Gewecke, Ghonge, Ghosh, Ghosh,
  Ghosh, Ghosh, Ghosh, Giacomazzo, Giacoppo, Giaime, Giardina, Gibson, Gier,
  Giesler, Giri, Gissi, Gkaitatzis, Glanzer, Gleckl, Godwin, Goetz, Goetz,
  Gohlke, Golomb, Goncharov, González, Gosselin, Gouaty, Gould, Goyal, Grace,
  Grado, Graham, Granata, Granata, Grant, Gras, Grassia, Gray, Gray, Greco,
  Green, Green, Gretarsson, Gretarsson, Griffith, Griffiths, Griggs, Grignani,
  Grimaldi, Grimes, Grimm, Grote, Grunewald, Gruning, Gruson, Guerra, Guidi,
  Guimaraes, Guixé, Gulati, Gunny, Guo, Guo, Gupta, Gupta, Gupta, Gupta,
  Gupta, Gustafson, Guzman, Ha, Hadiputrawan, Haegel, Haino, Halim, Hall,
  Hamilton, Hammond, Han, Haney, Hanks, Hanna, Hannam, Hannuksela, Hansen,
  Hansen, Hanson, Harder, Haris, Harms, Harry, Harry, Hartwig, Hasegawa,
  Haskell, Haster, Hathaway, Hattori, Haughian, Hayakawa, Hayama, Hayes, Healy,
  Heidmann, Heidt, Heintze, Heinze, Heinzel, Heitmann, Hellman, Hello,
  Helmling-Cornell, Hemming, Hendry, Heng, Hennes, Hennig, Hennig, Henshaw,
  Hernandez, Vivanco, Heurs, Hewitt, Higginbotham, Hild, Hill, Himemoto, Hines,
  Hirata, Hirose, Ho, Hochheim, Hofman, Hohmann, Holcomb, Holland, Hollows,
  Holmes, Holt, Holz, Hong, Hough, Hourihane, Howell, Hoy, Hoyland, Hreibi,
  Hsieh, Hsieh, Hsiung, Hsu, Huang, Huang, Huang, Huang, Huang, Huang, Hübner,
  Huddart, Hughey, Hui, Hui, Husa, Huttner, Huxford, Huynh-Dinh, Ide,
  Idzkowski, Iess, Inayoshi, Inoue, Iosif, Isi, Isleif, Ito, Itoh, Iyer,
  JaberianHamedan, Jacqmin, Jacquet, Jadhav, Jadhav, Jain, James, Jan, Jani,
  Janquart, Janssens, Janthalur, Jaranowski, Jariwala, Jaume, Jenkins, Jenner,
  Jeon, Jia, Jiang, Jin, Johns, Johnston, Jones, Jones, Jones, Jones, Joshi,
  Ju, Jue, Jung, Jung, Junker, Juste, Kaihotsu, Kajita, Kakizaki, Kalaghatgi,
  Kalogera, Kamai, Kamiizumi, Kanda, Kandhasamy, Kang, Kanner, Kao, Kapadia,
  Kapasi, Karathanasis, Karki, Kashyap, Kasprzack, Kastaun, Kato, Katsanevas,
  Katsavounidis, Katzman, Kaur, Kawabe, Kawaguchi, Kéfélian, Keitel, Key,
  Khadka, Khalili, Khan, Khanam, Khazanov, Khetan, Khursheed, Kijbunchoo, Kim,
  Kim, Kim, Kim, Kim, Kim, Kim, Kimball, Kimura, Kinley-Hanlon, Kirchhoff,
  Kissel, Klimenko, Klinger, Knee, Knowles, Knust, Knyazev, Kobayashi, Koch,
  Koekoek, Kohri, Kokeyama, Koley, Kolitsidou, Kolstein, Komori, Kondrashov,
  Kong, Kontos, Koper, Korobko, Kovalam, Koyama, Kozak, Kozakai, Kringel,
  Krishnendu, Królak, Kuehn, Kuei, Kuijer, Kulkarni, Kumar, Kumar, Kumar,
  Kumar, Kume, Kuns, Kuromiya, Kuroyanagi, Kwak, Lacaille, Lagabbe, Laghi,
  Lalande, Lalleman, Lam, Lamberts, Landry, Lane, Lang, Lange, Lantz, La~Rosa,
  Lartaux-Vollard, Lasky, Laxen, Lazzarini, Lazzaro, Leaci, Leavey, LeBohec,
  Lecoeuche, Lee, Lee, Lee, Lee, Lee, Legred, Lehmann, Lemaître, Lenti,
  Leonardi, Leonova, Leroy, Letendre, Levesque, Levin, Leviton, Leyde, Li, Li,
  Li, Li, Li, Li, Li, Lin, Lin, Lin, Lin, Lin, Lin, Linde, Linker, Linley,
  Littenberg, Liu, Liu, Liu, Liu, Llamas, Lo, Lo, London, Longo, Lopez,
  Portilla, Lorenzini, Loriette, Lormand, Losurdo, Lott, Lough, Lousto,
  Lovelace, Lucaccioni, Lück, Lumaca, Lundgren, Luo, Lynam, Ma'arif, Macas,
  Machtinger, MacInnis, Macleod, MacMillan, Macquet, Hernandez, Magazzù,
  Magee, Maggiore, Magnozzi, Mahesh, Majorana, Maksimovic, Maliakal, Malik,
  Man, Mandic, Mangano, Mansell, Manske, Mantovani, Mapelli, Marchesoni, Pina,
  Marion, Mark, Márka, Márka, Markakis, Markosyan, Markowitz, Maros,
  Marquina, Marsat, Martelli, Martin, Martin, Martinez, Martinez, Martinez,
  Martinovic, Martynov, Marx, Masalehdan, Mason, Massera, Masserot, Masso-Reid,
  Mastrogiovanni, Matas, Mateu-Lucena, Matichard, Matiushechkina, Mavalvala,
  McCann, McCarthy, McClelland, McClincy, McCormick, McCuller, McGhee, McGuire,
  McIsaac, McIver, McRae, McWilliams, Meacher, Mehmet, Mehta, Meijer, Melatos,
  Melchor, Mendell, Menendez-Vazquez, Menoni, Mercer, Mereni, Merfeld, Merilh,
  Merritt, Merzougui, Meshkov, Messenger, Messick, Meyers, Meylahn, Mhaske,
  Miani, Miao, Michaloliakos, Michel, Michimura, Middleton, Mihaylov, Milano,
  Miller, Miller, Miller, Millhouse, Mills, Milotti, Minenkov, Mio, Mir,
  Miravet-Tenés, Mishkin, Mishra, Mishra, Mistry, Mitra, Mitrofanov,
  Mitselmakher, Mittleman, Miyakawa, Miyo, Miyoki, Mo, Modafferi, Moguel,
  Mogushi, Mohapatra, Mohite, Molina, Molina-Ruiz, Mondin, Montani, {S},
  {More}, Moore, Moragues, Moraru, Morawski, More, Moreno, Moreno, Mori,
  Morisaki, Morisue, Moriwaki, Mours, Mow-Lowry, Mozzon, Muciaccia, Mukherjee,
  Mukherjee, Mukherjee, Mukherjee, Mukherjee, Mukund, Mullavey, Munch, Muñiz,
  Murray, Musenich, Muusse, Nadji, Nagano, Nagar, Nakamura, Nakano, Nakano,
  Nakayama, Napolano, Nardecchia, Narikawa, Narola, Naticchioni, Nayak, Nayak,
  Neil, Neilson, Nelson, Nelson, Nery, Neubauer, Neunzert, Ng, Ng, Nguyen,
  Nguyen, Nguyen, Quynh, Ni, Ni, Nichols, Nishimoto, Nishizawa, Nissanke,
  Nitoglia, Nocera, Norman, North, Nozaki, Nurbek, Nuttall, Obayashi, Oberling,
  O'Brien, O'Dell, Oelker, Ogaki, Oganesyan, Oh, Oh, Oh, Ohashi, Ohashi,
  Ohkawa, Ohme, Ohta, Okada, Okutani, Olivetto, Oohara, Oram, O'Reilly,
  Ormiston, Ormsby, O'Shaughnessy, O'Shea, Oshino, Ossokine, Osthelder, Otabe,
  Ottaway, Overmier, Pace, Pagano, Pagano, Page, Pagliaroli, Pai, Pai, Pal,
  Palamos, Palashov, Palomba, Pan, Pan, Panda, Pang, Pankow, Pannarale, Pant,
  Panther, Paoletti, Paoli, Paolone, Pappas, Parisi, Park, Park, Parker,
  Pascucci, Pasqualetti, Passaquieti, Passuello, Patel, Pathak, Patricelli,
  Patron, Paul, Payne, Pedraza, Pedurand, Pegoraro, Pele, Arellano, Penano,
  Penn, Perego, Pereira, Pereira, Perez, Périgois, Perkins, Perreca, Perriès,
  Pesios, Petermann, Petterson, Pfeiffer, Pham, Pham, Phukon, Phurailatpam,
  Piccinni, Pichot, Piendibene, Piergiovanni, Pierini, Pierro, Pillant, Pillas,
  Pilo, Pinard, Pineda-Bosque, Pinto, Pinto, Piotrzkowski, Piotrzkowski,
  Pirello, Pitkin, Placidi, Placidi, Planas, Plastino, Pluchar, Poggiani,
  Polini, Pong, Ponrathnam, Porter, Poulton, Poverman, Powell, Pracchia,
  Pradier, Prajapati, Prasai, Prasanna, Pratten, Principe, Prodi, Prokhorov,
  Prosposito, Prudenzi, Puecher, Punturo, Puosi, Puppo, Pürrer, Qi, Quartey,
  Quetschke, Quinonez, Quitzow-James, Raab, Raaijmakers, Radkins, Radulesco,
  Raffai, Rail, Raja, Rajan, Ramirez, Ramirez, Ramos-Buades, Rana, Rapagnani,
  Ray, Raymond, Raza, Razzano, Read, Rees, Regimbau, Rei, Reid, Reid, Reitze,
  Relton, Renzini, Rettegno, Revenu, Reza, Rezac, Ricci, Richards, Richardson,
  Richardson, Riemenschneider, Riles, Rinaldi, Rink, Robertson, Robie, Robinet,
  Rocchi, Rodriguez, Rolland, Rollins, Romanelli, Romano, Romel, Romero,
  Romero-Shaw, Romie, Ronchini, Rosa, Rose, Rosińska, Ross, Rowan, Rowlinson,
  Roy, Roy, Roy, Rozza, Ruggi, Ruiz-Rocha, Ryan, Sachdev, Sadecki, Sadiq, Saha,
  Saito, Sakai, Sakellariadou, Sakon, Salafia, Salces-Carcoba, Salconi, Saleem,
  Salemi, Samajdar, Sanchez, Sanchez, Sanchez, Sanchis-Gual, Sanders, Sanuy,
  Saravanan, Sarin, Sassolas, Satari, Sathyaprakash, Sauter, Savage, Savant,
  Sawada, Sawant, Sayah, Schaetzl, Scheel, Scheuer, Schiworski, Schmidt,
  Schmidt, Schnabel, Schneewind, Schofield, Schönbeck, Schulte, Schutz,
  Schwartz, Scott, Scott, Seglar-Arroyo, Sekiguchi, Sellers, Sengupta,
  Sentenac, Seo, Sequino, Sergeev, Setyawati, Shaffer, Shahriar, Shaikh, Shams,
  Shao, Sharma, Sharma, Shawhan, Shcheblanov, Sheela, Shikano, Shikauchi,
  Shimizu, Shimode, Shinkai, Shishido, Shoda, Shoemaker, Shoemaker,
  ShyamSundar, Sieniawska, Sigg, Silenzi, Singer, Singh, Singh, Singh, Singha,
  Sintes, Sipala, Skliris, Slagmolen, Slaven-Blair, Smetana, Smith, Smith,
  Smith, Soldateschi, Somala, Somiya, Song, Soni, Sordini, Sorrentino,
  Sorrentino, Soulard, Souradeep, Sowell, Spagnuolo, Spencer, Spera,
  Spinicelli, Srivastava, Srivastava, Staats, Stachie, Stachurski, Steer,
  Steinlechner, Steinlechner, Stergioulas, Stops, Stover, Strain, Strang,
  Stratta, Strong, Strunk, Sturani, Stuver, Suchenek, Sudhagar, Sudhir,
  Sugimoto, Suh, Sullivan, Summerscales, Sun, Sunil, Sur, Suresh, Sutton,
  Suzuki, Suzuki, Suzuki, Swinkels, Szczepańczyk, Szewczyk, Tacca, Tagoshi,
  Tait, Takahashi, Takahashi, Takano, Takeda, Takeda, Talbot, Talbot, Tamanini,
  Tanaka, Tanaka, Tanaka, Tanasijczuk, Tanioka, Tanner, Tao, Tao, Tapia,
  Martín, Taranto, Taruya, Tasson, Tenorio, Terhune, Terkowski,
  Thirugnanasambandam, Thomas, Thomas, Thompson, Thompson, Thondapu, Thorne,
  Thrane, Tiwari, Tiwari, Tiwari, Toivonen, Tolley, Tomaru, Tomura, Tonelli,
  Tornasi, Torres-Forné, Torrie, Melo, Töyrä, Trapananti, Travasso, Traylor,
  Trevor, Tringali, Tripathee, Troiano, Trovato, Trozzo, Trudeau, Tsai, Tsang,
  Tsang, Tsao, Tse, Tso, Tsuchida, Tsukada, Tsuna, Tsutsui, Turbang, Turconi,
  Turski, Tuyenbayev, Ubhi, Uchikata, Uchiyama, Udall, Ueda, Uehara, Ueno,
  Ueshima, Unnikrishnan, Urban, Ushiba, Utina, Vajente, Vajpeyi, Valdes,
  Valentini, Valsan, van Bakel, van Beuzekom, van Dael, Brand, Broeck,
  Vander-Hyde, van Haevermaet, van Heijningen, van Putten, van Remortel,
  Vardaro, Vargas, Varma, Vasúth, Vecchio, Vedovato, Veitch, Veitch,
  Venneberg, Venugopalan, Verkindt, Verma, Verma, Vermeulen, Veske, Vetrano,
  Viceré, Vidyant, Viets, Vijaykumar, Villa-Ortega, Vinet, Virtuoso, Vitale,
  Vocca, von Reis, von Wrangel, Vorvick, Vyatchanin, Wade, Wade, Wagner, Walet,
  Walker, Wallace, Wallace, Wang, Wang, Wang, Ward, Warner, Was, Washimi,
  Washington, Watchi, Weaver, Weaving, Webster, Weinert, Weinstein, Weiss,
  Weller, Weller, Wellmann, Wen, Weßels, Wette, Whelan, White, Whiting,
  Whittle, Wilken, Williams, Williams, Williamson, Willis, Willke, Wilson,
  Wipf, Wlodarczyk, Woan, Woehler, Wofford, Wong, Wong, Wright, Wu, Wu, Wu,
  Wysocki, Xiao, Yamada, Yamamoto, Yamamoto, Yamamoto, Yamashita, Yamazaki,
  Yang, Yang, Yang, Yang, Yang, Yang, Yap, Yeeles, Yeh, Yelikar, Ying,
  Yokoyama, Yokozawa, Yoo, Yoshioka, Yu, Yu, Yuzurihara, Zadrożny, Zanolin,
  Zeidler, Zelenova, Zendri, Zevin, Zhan, Zhang, Zhang, Zhang, Zhang, Zhang,
  Zhang, Zhao, Zhao, Zhao, Zhao, Zhou, Zhou, Zhu, Zhu, Zimmerman, Zucker, and
  Zweizig]{Abb_2021_b}
{The LIGO Scientific Collaboration}; {The Virgo Collaboration}; {The KAGRA
  Collaboration}; Abbott, R.; Abe, H.; Acernese, F.; Ackley, K.; Adhikari, N.;
  Adhikari, R.X.; Adkins, V.K.;  et~al.
\newblock Constraints on the cosmic expansion history from GWTC-3. \emph{arXiv Prepr.} \textbf{2021}, arXiv:2111.03604.
\newblock {\url{https://doi.org/10.48550/ARXIV.2111.03604}}.

\bibitem[Gayathri \em{et~al.}(2020)Gayathri, Healy, Lange, O'Brien,
  Szczepanczyk, Bartos, Campanelli, Klimenko, Lousto, and
  O'Shaughnessy]{Gayat_GW}
Gayathri, V.; Healy, J.; Lange, J.; O'Brien, B.; Szczepanczyk, M.; Bartos, I.;
  Campanelli, M.; Klimenko, S.; Lousto, C.; O'Shaughnessy, R.
\newblock Hubble Constant Measurement with GW190521 as an Eccentric Black Hole
  Merger. \emph{arXiv Prepr.}, \textbf{2020},  arXiv:2009.14247.
\newblock {\url{https://doi.org/10.48550/ARXIV.2009.14247}}.

\bibitem[Pesce \em{et~al.}(2020)Pesce, Braatz, Reid, Riess, Scolnic, Condon,
  Gao, Henkel, Impellizzeri, Kuo, and Lo]{Pesce_2020}
Pesce, D.W.; Braatz, J.A.; Reid, M.J.; Riess, A.G.; Scolnic, D.; Condon, J.J.;
  Gao, F.; Henkel, C.; Impellizzeri, C.M.V.; Kuo, C.Y.;  et~al.
\newblock The Megamaser Cosmology Project. {XIII}. Combined Hubble Constant
  Constraints.
\newblock {\em  Astrophys. J.} {\bf 2020}, {\em 891}, L1.
\newblock {\url{https://doi.org/10.3847/2041-8213/ab75f0}}.

\bibitem[Kourkchi \em{et~al.}(2020)Kourkchi, Tully, Anand, Courtois, Dupuy,
  Neill, Rizzi, and Seibert]{Kourkchi_2020}
Kourkchi, E.; Tully, R.B.; Anand, G.S.; Courtois, H.M.; Dupuy, A.; Neill, J.D.;
  Rizzi, L.; Seibert, M.
\newblock Cosmicflows-4: The Calibration of Optical and Infrared
  Tully---Fisher Relations.
\newblock {\em  Astrophys. J.} {\bf 2020}, {\em 896}, 3.
\newblock {\url{https://doi.org/10.3847/1538-4357/ab901c}}.

\bibitem[Blakeslee \em{et~al.}(2021)Blakeslee, Jensen, Ma, Milne, and
  Greene]{Blakeslee_2021}
Blakeslee, J.P.; Jensen, J.B.; Ma, C.P.; Milne, P.A.; Greene, J.E.
\newblock The Hubble Constant from Infrared Surface Brightness Fluctuation
  Distances.
\newblock {\em  Astrophys. J.} {\bf 2021}, {\em 911}, 65.
\newblock {\url{https://doi.org/10.3847/1538-4357/abe86a}}.

\bibitem[Dialektopoulos \em{et~al.}(2019)Dialektopoulos, Borka, Capozziello,
  Borka~Jovanovi\'c, and Jovanovi\'c]{Dialektopoulos_2018iph}
Dialektopoulos, K.F.; Borka, D.; Capozziello, S.; Borka~Jovanovi\'c, V.;
  Jovanovi\'c, P.
\newblock {Constraining nonlocal gravity by S2 star orbits}.
\newblock {\em Phys. Rev. D} {\bf 2019}, {\em 99}, 044053.
\newblock {\url{https://doi.org/10.1103/PhysRevD.99.044053}}.

\bibitem[Bouch\`e \em{et~al.}(2022)Bouch\`e, Capozziello, Salzano, and
  Umetsu]{Bouche:2022jts}
Bouch\`e, F.; Capozziello, S.; Salzano, V.; Umetsu, K.
\newblock {Testing non-local gravity by clusters of galaxies}.
\newblock {\em Eur. Phys. J. C} {\bf 2022}, {\em 82},~652,
\newblock {\url{https://doi.org/10.1140/epjc/s10052-022-10586-5}}.

\bibitem[Capozziello \em{et~al.}(1996)Capozziello, De~Ritis, Rubano, and
  Scudellaro]{Capozziello:1996bi}
Capozziello, S.; De~Ritis, R.; Rubano, C.; Scudellaro, P.
\newblock {Noether symmetries in cosmology}.
\newblock {\em La Rivista del Nuovo Cimento} {\bf 1996}, {\em 19N4},~1--114.
\newblock {\url{https://doi.org/10.1007/BF02742992}}.

\bibitem[Capozziello \em{et~al.}(2022)Capozziello, D'Agostino, and
  Luongo]{Capozziello:2022rac}
Capozziello, S.; D'Agostino, R.; Luongo, O.
\newblock {The phase-space view of non-local gravity cosmology}.
\newblock {\em Phys. Lett. B} {\bf 2022}, {\em 834},~137475.
\newblock {\url{https://doi.org/10.1016/j.physletb.2022.137475}}.

\bibitem[Capozziello and Lambiase(2022)]{NLicecube}
Capozziello, S.; Lambiase, G.
\newblock PeV IceCube signals and $H_0$ tension in the framework of Non-Local
  Gravity. \emph{arXiv Prepr.} \textbf{2022}, arXiv:2206.03690.
\newblock {\url{https://doi.org/10.48550/ARXIV.2206.03690}}.

\bibitem[Dialektopoulos and Capozziello(2018)]{Dialektopoulos_2018_NS}
Dialektopoulos, K.F.; Capozziello, S.
\newblock Noether symmetries as a geometric criterion to select theories of
  gravity.
\newblock {\em Int. J. Geom. Methods Mod. Phys.}
  {\bf 2018}, {\em 15}, 1840007.
\newblock {\url{https://doi.org/10.1142/s0219887818400078}}.

\bibitem[Gillessen \em{et~al.}(2009)Gillessen, Eisenhauer, Trippe, Alexander,
  Genzel, Martins, and Ott]{Gillessen_2009}
Gillessen, S.; Eisenhauer, F.; Trippe, S.; Alexander, T.; Genzel, R.; Martins,
  F.; Ott, T.
\newblock Monitoring stellar orbits around the massive black hole in the
  galactic centre.
\newblock {\em  Astrophys. J.} {\bf 2009}, {\em 692}, 1075.
\newblock {\url{https://doi.org/10.1088/0004-637x/692/2/1075}}.

\bibitem[Postman \em{et~al.}(2012)Postman, Coe, Benítez, Bradley, Broadhurst,
  Donahue, Ford, Graur, Graves, Jouvel, Koekemoer, Lemze, Medezinski, Molino,
  Moustakas, Ogaz, Riess, Rodney, Rosati, Umetsu, Zheng, Zitrin, Bartelmann,
  Bouwens, Czakon, Golwala, Host, Infante, Jha, Jimenez-Teja, Kelson, Lahav,
  Lazkoz, Maoz, McCully, Melchior, Meneghetti, Merten, Moustakas, Nonino,
  Patel, Regös, Sayers, Seitz, and der Wel]{Postman_2012}
Postman, M.; Coe, D.; Benítez, N.; Bradley, L.; Broadhurst, T.; Donahue, M.;
  Ford, H.; Graur, O.; Graves, G.; Jouvel, S.;  et~al.  The Cluster Lensing And Supernova survey with Hubble: An overview.
{\em  Astrophys. J. Suppl. Ser.} {\bf 2012}, {\em
  199}, 25.
\newblock {\url{https://doi.org/10.1088/0067-0049/199/2/25}}.

\bibitem[Umetsu \em{et~al.}(2016)Umetsu, Zitrin, Gruen, Merten, Donahue, and
  Postman]{Umetsu_2016}
Umetsu, K.; Zitrin, A.; Gruen, D.; Merten, J.; Donahue, M.; Postman, M.
\newblock CLASH: Joint analysis of strong-lensing, weak-lensing shear, and
  magnification data for 20 galaxy clusters.
\newblock {\em  Astrophys. J.} {\bf 2016}, {\em 821}, 116.
\newblock {\url{https://doi.org/10.3847/0004-637X/821/2/116}}.

\bibitem[Burstein \em{et~al.}(1997)Burstein, Bender, Faber, and
  Nolthenius]{Burstein_1997}
Burstein, D.; Bender, R.; Faber, S.; Nolthenius, R.
\newblock Global Relationships Among the Physical Properties of Stellar
  Systems.
\newblock {\em  Astron. J.} {\bf 1997}, {\em 114}. \newblock {\url{https://doi.org/10.1086/118570}}.

\bibitem[Belgacem \em{et~al.}(2019)Belgacem, Finke, Frassino, and
  Maggiore]{Belgacem_2019_LLR}
Belgacem, E.; Finke, A.; Frassino, A.; Maggiore, M.
\newblock Testing nonlocal gravity with Lunar Laser Ranging.
\newblock {\em J. Cosmol. Astropart. Phys.} {\bf 2019}, {\em
  2019}, 035.
\newblock {\url{https://doi.org/10.1088/1475-7516/2019/02/035}}.

\bibitem[Vainshtein(1972)]{VAINSHTEIN1972393}
Vainshtein, A.
\newblock To the problem of nonvanishing gravitation mass.
\newblock {\em Phys. Lett. B} {\bf 1972}, {\em 39}, 393--394.
\newblock {\url{https://doi.org/https://doi.org/10.1016/0370-2693(72)90147-5}}.

\bibitem[Belgacem \em{et~al.}(2018{\natexlab{a}})Belgacem, Dirian, Foffa, and
  Maggiore]{Belgacem_2018_DW1}
Belgacem, E.; Dirian, Y.; Foffa, S.; Maggiore, M.
\newblock Gravitational-wave luminosity distance in modified gravity theories.
\newblock {\em Phys. Rev. D} {\bf 2018}, {\em 97}, 104066.
\newblock {\url{https://doi.org/10.1103/physrevd.97.104066}}.

\bibitem[Belgacem \em{et~al.}(2018{\natexlab{b}})Belgacem, Dirian, Foffa, and
  Maggiore]{Belgacem_2018_DW2}
Belgacem, E.; Dirian, Y.; Foffa, S.; Maggiore, M.
\newblock Modified gravitational-wave propagation and standard sirens.
\newblock {\em Phys. Rev. D} {\bf 2018}, {\em 98}, 023510.
\newblock {\url{https://doi.org/10.1103/physrevd.98.023510}}.

\bibitem[Abbott \em{et~al.}(2017)Abbott, Abbott, Abbott, Acernese, Ackley,
  Adams, Adams, Addesso, Adhikari, Adya, Affeldt, Afrough, Agarwal, Agathos,
  Agatsuma, Aggarwal, Aguiar, Aiello, Ain, Ajith, Allen, Allen, Allocca, Aloy,
  Altin, Amato, Ananyeva, Anderson, Anderson, Angelova, Antier, Appert, Arai,
  Araya, Areeda, Arnaud, Arun, Ascenzi, Ashton, Ast, Aston, Astone, Atallah,
  Aufmuth, Aulbert, AultONeal, Austin, Avila-Alvarez, Babak, Bacon, Bader, Bae,
  Baker, Baldaccini, Ballardin, Ballmer, Banagiri, Barayoga, Barclay, Barish,
  Barker, Barkett, Barone, Barr, Barsotti, Barsuglia, Barta, Bartlett, Bartos,
  Bassiri, Basti, Batch, Bawaj, Bayley, Bazzan, B{\'{e} }csy, Beer, Bejger,
  Belahcene, Bell, Berger, Bergmann, Bero, Berry, Bersanetti, Bertolini,
  Betzwieser, Bhagwat, Bhandare, Bilenko, Billingsley, Billman, Birch, Birney,
  Birnholtz, Biscans, Biscoveanu, Bisht, Bitossi, Biwer, Bizouard, Blackburn,
  Blackman, Blair, Blair, Blair, Bloemen, Bock, Bode, Boer, Bogaert, Bohe,
  Bondu, Bonilla, Bonnand, Boom, Bork, Boschi, Bose, Bossie, Bouffanais, Bozzi,
  Bradaschia, Brady, Branchesi, Brau, Briant, Brillet, Brinkmann, Brisson,
  Brockill, Broida, Brooks, Brown, Brown, Brunett, Buchanan, Buikema, Bulik,
  Bulten, Buonanno, Buskulic, Buy, Byer, Cabero, Cadonati, Cagnoli, Cahillane,
  Bustillo, Callister, Calloni, Camp, Canepa, Canizares, Cannon, Cao, Cao,
  Capano, Capocasa, Carbognani, Caride, Carney, Diaz, Casentini, Caudill,
  Cavagli{\`{a}}, Cavalier, Cavalieri, Cella, Cepeda, Cerd{\'{a}}-Dur{\'{a}}n,
  Cerretani, Cesarini, Chamberlin, Chan, Chao, Charlton, Chase,
  Chassande-Mottin, Chatterjee, Chatziioannou, Cheeseboro, Chen, Chen, Chen,
  Cheng, Chia, Chincarini, Chiummo, Chmiel, Cho, Cho, Chow, Christensen, Chu,
  Chua, Chua, Chung, Chung, Ciani, Ciolfi, Cirelli, Cirone, Clara, Clark,
  Clearwater, Cleva, Cocchieri, Coccia, Cohadon, Cohen, Colla, Collette,
  Cominsky, Jr., Conti, Cooper, Corban, Corbitt, Cordero-Carri{\'{o}}n, Corley,
  Cornish, Corsi, Cortese, Costa, Coughlin, Coughlin, Coulon, Countryman,
  Couvares, Covas, Cowan, Coward, Cowart, Coyne, Coyne, Creighton, Creighton,
  Cripe, Crowder, Cullen, Cumming, Cunningham, Cuoco, Canton, D{\'{a}}lya,
  Danilishin, D'Antonio, Danzmann, Dasgupta, Costa, Dattilo, Dave, Davier,
  Davis, Daw, Day, De, DeBra, Degallaix, Laurentis, Del{\'{e}}glise, Pozzo,
  Demos, Denker, Dent, Pietri, Dergachev, Rosa, DeRosa, Rossi, DeSalvo,
  de~Varona, Devenson, Dhurandhar, D{\'{\i}}az, Fiore, Giovanni, Girolamo,
  Lieto, Pace, Palma, Renzo, Doctor, Dolique, Donovan, Dooley, Doravari,
  Dorrington, Douglas, {\'{A}}lvarez, Downes, Drago, Dreissigacker, Driggers,
  Du, Ducrot, Dupej, Dwyer, Edo, Edwards, Effler, Eggenstein, Ehrens, Eichholz,
  Eikenberry, Eisenstein, Essick, Estevez, Etienne, Etzel, Evans, Evans,
  Factourovich, Fafone, Fair, Fairhurst, Fan, Farinon, Farr, Farr,
  Fauchon-Jones, Favata, Fays, Fee, Fehrmann, Feicht, Fejer, Fernandez-Galiana,
  Ferrante, Ferreira, Ferrini, Fidecaro, Finstad, Fiori, Fiorucci, Fishbach,
  Fisher, Fitz-Axen, Flaminio, Fletcher, Fong, Font, Forsyth, Forsyth,
  Fournier, Frasca, Frasconi, Frei, Freise, Frey, Frey, Fries, Fritschel,
  Frolov, Fulda, Fyffe, Gabbard, Gadre, Gaebel, Gair, Gammaitoni, Ganija,
  Gaonkar, Garcia-Quiros, Garufi, Gateley, Gaudio, Gaur, Gayathri, Gehrels,
  Gemme, Genin, Gennai, George, George, Gergely, Germain, Ghonge, Ghosh, Ghosh,
  Ghosh, Giaime, Giardina, Giazotto, Gill, Glover, Goetz, Goetz, Gomes,
  Goncharov, Gonz{\'{a}}lez, Castro, Gopakumar, Gorodetsky, Gossan, Gosselin,
  Gouaty, Grado, Graef, Granata, Grant, Gras, Gray, Greco, Green, Gretarsson,
  Groot, Grote, Grunewald, Gruning, Guidi, Guo, Gupta, Gupta, Gushwa,
  Gustafson, Gustafson, Halim, Hall, Hall, Hamilton, Hammond, Haney, Hanke,
  Hanks, Hanna, Hannam, Hannuksela, Hanson, Hardwick, Harms, Harry, Harry,
  Hart, Haster, Haughian, Healy, Heidmann, Heintze, Heitmann, Hello, Hemming,
  Hendry, Heng, Hennig, Heptonstall, Heurs, Hild, Hinderer, Hoak, Hofman, Holt,
  Holz, Hopkins, Horst, Hough, Houston, Howell, Hreibi, Hu, Huerta, Huet,
  Hughey, Husa, Huttner, Huynh-Dinh, Indik, Inta, Intini, Isa, Isac, Isi, Iyer,
  Izumi, Jacqmin, Jani, Jaranowski, Jawahar, Jim{\'{e}}nez-Forteza, Johnson,
  Johnson-McDaniel, Jones, Jones, Jonker, Ju, Junker, Kalaghatgi, Kalogera,
  Kamai, Kandhasamy, Kang, Kanner, Kapadia, Karki, Karvinen, Kasprzack,
  Kastaun, Katolik, Katsavounidis, Katzman, Kaufer, Kawabe,
  K{\'{e}}f{\'{e}}lian, Keitel, Kemball, Kennedy, Kent, Key, Khalili, Khan,
  Khan, Khan, Khazanov, Kijbunchoo, Kim, Kim, Kim, Kim, Kim, Kim, Kimbrell,
  King, King, Kinley-Hanlon, Kirchhoff, Kissel, Kleybolte, Klimenko, Knowles,
  Koch, Koehlenbeck, Koley, Kondrashov, Kontos, Korobko, Korth, Kowalska,
  Kozak, Krämer, Kringel, Krishnan, Kr{\'{o}}lak, Kuehn, Kumar, Kumar, Kumar,
  Kuo, Kutynia, Kwang, Lackey, Lai, Landry, Lang, Lange, Lantz, Lanza,
  Lartaux-Vollard, Lasky, Laxen, Lazzarini, Lazzaro, Leaci, Leavey, Lee, Lee,
  Lee, Lee, Lee, Lehmann, Lenon, Leonardi, Leroy, Letendre, Levin, Li, Linker,
  Littenberg, Liu, Lo, Lockerbie, London, Lord, Lorenzini, Loriette, Lormand,
  Losurdo, Lough, Lousto, Lovelace, Lück, Lumaca, Lundgren, Lynch, Ma, Macas,
  Macfoy, Machenschalk, MacInnis, Macleod, Hernandez, Maga{\~{n}}a-Sandoval,
  Zertuche, Magee, Majorana, Maksimovic, Man, Mandic, Mangano, Mansell, Manske,
  Mantovani, Marchesoni, Marion, M{\'{a}}rka, M{\'{a}}rka, Markakis, Markosyan,
  Markowitz, Maros, Marquina, Martelli, Martellini, Martin, Martin, Martynov,
  Mason, Massera, Masserot, Massinger, Masso-Reid, Mastrogiovanni, Matas,
  Matichard, Matone, Mavalvala, Mazumder, McCarthy, McClelland, McCormick,
  McCuller, McGuire, McIntyre, McIver, McManus, McNeill, McRae, McWilliams,
  Meacher, Meadors, Mehmet, Meidam, Mejuto-Villa, Melatos, Mendell, Mercer,
  Merilh, Merzougui, Meshkov, Messenger, Messick, Metzdorff, Meyers, Miao,
  Michel, Middleton, Mikhailov, Milano, Miller, Miller, Miller, Millhouse,
  Milovich-Goff, Minazzoli, Minenkov, Ming, Mishra, Mitra, Mitrofanov,
  Mitselmakher, Mittleman, Moffa, Moggi, Mogushi, Mohan, Mohapatra, Montani,
  Moore, Moraru, Moreno, Morriss, Mours, Mow-Lowry, Mueller, Muir, Mukherjee,
  Mukherjee, Mukherjee, Mukund, Mullavey, Munch, Mu{\~{n}}iz, Muratore, Murray,
  Napier, Nardecchia, Naticchioni, Nayak, Neilson, Nelemans, Nelson, Nery,
  Neunzert, Nevin, Newport, Newton, Ng, Nguyen, Nichols, Nielsen, Nissanke,
  Nitz, Noack, Nocera, Nolting, North, Nuttall, Oberling, O'Dea, Ogin, Oh, Oh,
  Ohme, Okada, Oliver, Oppermann, Oram, O'Reilly, Ormiston, Ortega,
  O'Shaughnessy, Ossokine, Ottaway, Overmier, Owen, Pace, Page, Page, Pai, Pai,
  Palamos, Palashov, Palomba, Pal-Singh, Pan, Pan, Pang, Pang, Pankow,
  Pannarale, Pant, Paoletti, Paoli, Papa, Parida, Parker, Pascucci,
  Pasqualetti, Passaquieti, Passuello, Patil, Patricelli, Pearlstone, Pedraza,
  Pedurand, Pekowsky, Pele, Penn, Perez, Perreca, Perri, Pfeiffer, Phelps,
  Piccinni, Pichot, Piergiovanni, Pierro, Pillant, Pinard, Pinto, Pirello,
  Pitkin, Poe, Poggiani, Popolizio, Porter, Post, Powell, Prasad, Pratt,
  Pratten, Predoi, Prestegard, Prijatelj, Principe, Privitera, Prodi,
  Prokhorov, Puncken, Punturo, Puppo, Pürrer, Qi, Quetschke, Quintero,
  Quitzow-James, Raab, Rabeling, Radkins, Raffai, Raja, Rajan, Rajbhandari,
  Rakhmanov, Ramirez, Ramos-Buades, Rapagnani, Raymond, Razzano, Read,
  Regimbau, Rei, Reid, Reitze, Ren, Reyes, Ricci, Ricker, Rieger, Riles, Rizzo,
  Robertson, Robie, Robinet, Rocchi, Rolland, Rollins, Roma, Romano, Romel,
  Romie, Rosi{\'{n}}ska, Ross, Rowan, Rüdiger, Ruggi, Rutins, Ryan, Sachdev,
  Sadecki, Sadeghian, Sakellariadou, Salconi, Saleem, Salemi, Samajdar, Sammut,
  Sampson, Sanchez, Sanchez, Sanchis-Gual, Sandberg, Sanders, Sassolas,
  Sathyaprakash, Saulson, Sauter, Savage, Sawadsky, Schale, Scheel, Scheuer,
  Schmidt, Schmidt, Schnabel, Schofield, Schönbeck, Schreiber, Schuette,
  Schulte, Schutz, Schwalbe, Scott, Scott, Seidel, Sellers, Sengupta, Sentenac,
  Sequino, Sergeev, Shaddock, Shaffer, Shah, Shahriar, Shaner, Shao, Shapiro,
  Shawhan, Sheperd, Shoemaker, Shoemaker, Siellez, Siemens, Sieniawska, Sigg,
  Silva, Singer, Singh, Singhal, Sintes, Slagmolen, Smith, Smith, Smith,
  Somala, Son, Sonnenberg, Sorazu, Sorrentino, Souradeep, Spencer, Srivastava,
  Staats, Staley, Steinke, Steinlechner, Steinlechner, Steinmeyer, Stevenson,
  Stone, Stops, Strain, Stratta, Strigin, Strunk, Sturani, Stuver,
  Summerscales, Sun, Sunil, Suresh, Sutton, Swinkels, Szczepa{\'{n}}czyk,
  Tacca, Tait, Talbot, Talukder, Tanner, T{\'{a}}pai, Taracchini, Tasson,
  Taylor, Taylor, Tewari, Theeg, Thies, Thomas, Thomas, Thomas, Thorne, Thorne,
  Thrane, Tiwari, Tiwari, Tokmakov, Toland, Tonelli, Tornasi,
  Torres-Forn{\'{e}}, Torrie, Töyrä, Travasso, Traylor, Trinastic, Tringali,
  Trozzo, Tsang, Tse, Tso, Tsukada, Tsuna, Tuyenbayev, Ueno, Ugolini,
  Unnikrishnan, Urban, Usman, Vahlbruch, Vajente, Valdes, van Bakel, van
  Beuzekom, van~den Brand, Broeck, Vander-Hyde, van~der Schaaf, van Heijningen,
  van Veggel, Vardaro, Varma, Vass, Vas{\'{u}}th, Vecchio, Vedovato, Veitch,
  Veitch, Venkateswara, Venugopalan, Verkindt, Vetrano, Vicer{\'{e}}, Viets,
  Vinciguerra, Vine, Vinet, Vitale, Vo, Vocca, Vorvick, Vyatchanin, Wade, Wade,
  Wade, Walet, Walker, Wallace, Walsh, Wang, Wang, Wang, Wang, Wang, Ward,
  Warner, Was, Watchi, Weaver, Wei, Weinert, Weinstein, Weiss, Wen, Wessel,
  We{\ss}els, Westerweck, Westphal, Wette, Whelan, Whitcomb, Whiting, Whittle,
  Wilken, Williams, Williams, Williamson, Willis, Willke, Wimmer, Winkler,
  Wipf, Wittel, Woan, Woehler, Wofford, Wong, Worden, Wright, Wu, Wysocki,
  Xiao, Yamamoto, Yancey, Yang, Yap, Yazback, Yu, Yu, Yvert, Zadro{\.{z}}ny,
  Zanolin, Zelenova, Zendri, Zevin, Zhang, Zhang, Zhang, Zhang, Zhao, Zhou,
  Zhou, Zhu, Zhu, Zimmerman, Zucker, Zweizig, Burns, Veres, Kocevski, Racusin,
  Goldstein, Connaughton, Briggs, Blackburn, Hamburg, Hui, von Kienlin,
  McEnery, Preece, Wilson-Hodge, Bissaldi, Cleveland, Gibby, Giles, Kippen,
  McBreen, Meegan, Paciesas, Poolakkil, Roberts, Stanbro, Savchenko, Ferrigno,
  Kuulkers, Bazzano, Bozzo, Brandt, Chenevez, Courvoisier, Diehl, Domingo,
  Hanlon, Jourdain, Laurent, Lebrun, Lutovinov, Mereghetti, Natalucci, Rodi,
  Roques, Sunyaev, and Ubertini]{Abbott_2017_GW}
Abbott, B.P.; Abbott, R.; Abbott, T.D.; Acernese, F.; Ackley, K.; Adams, C.;
  Adams, T.; Addesso, P.; Adhikari, R.X.; Adya, V.B.;  et~al.
\newblock Gravitational Waves and Gamma-Rays from a Binary Neutron Star Merger:
  {GW}170817 and {GRB} 170817A.
\newblock {\em  Astrophys. J.} {\bf 2017}, {\em 848}, L13.
\newblock {\url{https://doi.org/10.3847/2041-8213/aa920c}}.

\end{thebibliography}
\end{document}